\documentclass[11pt] {article}
\usepackage{latexsym,amssymb}
\usepackage{epsfig}
\usepackage{epic,eepic}
\usepackage{amsmath}

\newcommand{\beq}{\begin{equation}} 
\newcommand{\eeq}{\end{equation}}
\newcommand{\beqs}{\begin{eqnarray}} 
\newcommand{\eeqs}{\end{eqnarray}}
\newcommand{\lag}{\mathcal{L}}
\newcommand{\tr}{\mathrm{tr\,}}
\newcommand{\wholesp}{\int \mathrm{d}^4x \mathrm{d}^2\theta
                      \mathrm{d}^2\bar{\theta}\; }
\newcommand{\halfsp}{\int \mathrm{d}^4x \mathrm{d}^2\theta\;}
\newcommand{\halfspbar}{\int \mathrm{d}^4x \mathrm{d}^2\bar{\theta}\;}
\newcommand{\oone}{\mathcal{O}_1(x)}
\newcommand{\otwo}{\mathcal{O}_2(y)}

\newcommand{\sh} {\hat S}
\newcommand{\refe}[1]{(\ref{#1})}
\newcommand{\ff}{{\mathcal{F}}}
\newcommand{\ad}{{\mathrm{Ad}}}
\newcommand{\measure}[2]{\displaystyle \int\!\! \mathrm{d}^{#1}\!#2\,} 
\newcommand{\pd}[2]{\frac{\partial #1}{\partial #2}}
\newcommand{\nn} {\nonumber}

\begin{document}
\begin{titlepage}
\vskip 2.5cm
\begin{center}
{\LARGE \bf An Introduction to \phantom{y}}\\
\smallskip
{\LARGE \bf Supersymmetric Gauge Theories}\\
\smallskip
{\LARGE \bf and Matrix Models}\\
\vspace{2.71cm}
{\Large
Riccardo Argurio,}
\vskip 0.5cm
{\large \it Physique Th\'eorique et Math\'ematique \\
\smallskip
Universit\'e Libre de Bruxelles, C.P. 231, 1050 Bruxelles, Belgium}
\vskip 0.8cm
{\Large Gabriele Ferretti and 
Rainer Heise}
\vskip 0.5cm
{\large \it Institute for Theoretical Physics - G\"oteborg University and \\
\smallskip
Chalmers University of Technology, 412 96 G\"oteborg, Sweden}
\end{center}
\vspace{2cm}
\begin{abstract}

\end{abstract}
We give an introduction to the recently established connection between
supersymmetric gauge theories and matrix models. We begin by reviewing 
previous material that is required in order to follow 
the latest developments. This includes the superfield formulation of 
gauge theories, holomorphy, the chiral ring, the Konishi anomaly and
the large $N$ limit. 
We then present both the diagrammatic proof of the connection 
and the one based on the anomaly. Our discussion is entirely 
field theoretical and self contained.
\end{titlepage}

\tableofcontents

\section{Introduction}

The scope of this review is to give a pedagogical introduction to some
new techniques developed in the context of supersymmetric gauge theories,
enabling one to compute systematically non-perturbative quantities 
in the low energy effective physics. 
These new developments were initiated by a
conjecture put forward by Dijkgraaf and Vafa~\cite{dv} where it was
argued that non-perturbative quantities such as the glueball 
superpotential can be computed by means of a matrix model.

The main motivation for the study of supersymmetric (SUSY)
gauge theories is of course that they are expected to be directly
relevant to the next round of high energy experiments, where evidence for 
the supersymmetric partners of the standard model particles 
should hopefully be found. To be phenomenologically viable,
supersymmetry must of course be broken in some way. While this is the
main phenomenological context for the study of supersymmetry, 
there are further theoretical
reasons for studying SUSY gauge theories. These theories, even when 
preserving supersymmetry (to be precise, minimal $\mathcal{N} = 1$ SUSY),
are believed to share many key dynamical features 
with QCD such as confinement,
dynamical generation of a mass gap in the gauge sector, chiral
symmetry breaking and the related non-trivial vacuum structure. 
All these features are non-perturbative in nature
and notoriously difficult to address in real life QCD. 
With unbroken SUSY there are a  number of powerful tools at one's
disposal, such as non-renormalization theorems and holomorphy.
While important dynamical issues such as confinement remain out of reach
and have to be assumed, it is possible to make 
exact statements concerning the vacuum structure of the theory and the
value of the vacuum condensates.

The quantitative investigation of the non-perturbative aspects of SUSY
gauge theories began in the early 80's, as reviewed in \cite{amati},
and received new impetus during the 90's, see \cite{nonpert90}. 
Subsequently, in the second part of the 90's it was realized
that there are a huge number of possibilities to embed gauge
theories within string theory, using D-branes and dualities.
This led to an enormous amount of work in many directions, one of
which \cite{previousDV} culminated in the proposal of \cite{dv}.
This conjecture, originally motivated in the above string theory
context, was subsequently given a purely field theoretical 
interpretation in various different approaches. In \cite{ferrari1} 
the emphasis was on the relation with the Seiberg-Witten solution
of $\mathcal{N}=2$ pure gauge theory \cite{sw}, broken to 
$\mathcal{N}=1$ by the presence of a tree level superpotential
for the adjoint matter field. In \cite{dglvz} a purely perturbative
argument was given, in which the glueball superfield was treated as
a classical background field. Finally, in \cite{cdsw} the connection
with a matrix model was made using generalized Konishi anomaly
relations~\cite{konishiorig}. 
 
In this review we will concentrate on the purely field 
theoretical \hbox{$\mathcal{N}\!=\!1$} 
approach to these new techniques, very much in the
spirit of \cite{dglvz} and \cite{cdsw}. This choice is partly dictated
by the wish to provide a self contained introduction using the least
amount of advanced results. Hopefully this will equip the reader
with the basic knowledge to address the more advanced current
literature on the subject. In this spirit, we have made some effort
to discuss earlier results in $\mathcal{N}=1$ gauge theories to
put the new developments into context.

The review is organized as follows. In Section~2 we briefly present the
superspace formulation of $\mathcal{N}=1$ gauge theory. This is the
language in which all the results will be derived. We introduce chiral
and vector superfields and show how they combine into the most
general gauge theory with matter. 
Section~3 discusses some basic dynamical
properties of these theories such as the perturbative 
non-renormalization theorem, the holomorphy 
of the Wilsonian superpotential, 
the linearity principle and the formulation of the low energy physics
in terms of the glueball superfield. 
The review of the latest developments begins in Section~4 where the 
conjecture of Dijkgraaf and Vafa~\cite{dv} is presented and derived in
a perturbative context. This section describes the
computation of \cite{dglvz} by considering the simplest example,
namely that of a cubic superpotential for matter in the adjoint
representation of $U(N)$. We show
explicitly the equivalence between the field theory computation and
the matrix model concentrating on a simple diagram. 
Furthermore, we show that only planar diagrams are relevant to
the problem. 
Section~5 introduces the concept of the chiral ring,
which provides a useful tool to analyze the vacuum
structure of the theory. To do so,
we define chiral operators and derive their
basic properties. We then introduce an equivalence relation between chiral
operators leading to the concept of the chiral ring and discuss its
quantum corrections. 
Section~6 describes the Konishi anomaly~\cite{konishiorig}
and its generalizations~\cite{cdsw}. Finally, in Section~7 we show how
one can determine the
effective superpotential, using the results of the two previous
sections, and how this is related to the matrix model.
This last section is closely based on~\cite{cdsw}.

Appendix A fixes the notation. Appendix B reviews the one cut solution
to the matrix model of~\cite{bipz}, needed for the complete
solution of the example in Section~4. 
In Appendix C we make a humble attempt at reviewing the literature 
that uses the new techniques discussed here.
With this we hope to partly correct for
not having touched many interesting new developments in the main text.
A second apology is due for omitting many important results in 
SUSY gauge theories previous to the 
Dijkgraaf and Vafa conjecture, let alone for not mentioning extended 
supersymmetry. Fortunately there are many extensive reviews covering these 
subjects.

\section{Superspace formulation of gauge theories}

The purpose of this section is to familiarize the reader with the
techniques of superspace which has the advantage of making SUSY
manifest, see e.g.~\cite{wb} for a list of references to the original papers.
We begin by showing what a SUSY Lagrangian looks like in a non
manifestly SUSY notation (the so called component notation) and then explain
how it can be rewritten in a way so as to make SUSY manifest. The reader
familiar with the formalism can skip directly to Section 3.

The two typical examples of four dimensional SUSY Lagrangians 
are the so called Wess-Zumino (WZ)
model (no gauge interactions, only matter fields) and the pure gauge
theory (only gauge interaction, no matter fields).

The WZ model can be written as~\cite{Wess:1973kz}:
\beq
     \lag = \partial_\mu\bar\phi\partial^\mu\phi - 
     i \bar{\psi}\bar{\sigma}^\mu
     \partial_\mu\psi + \bar f f +  \hbox{interactions}, \label{WZ}
\eeq
where $\phi$ is a complex scalar, $\psi$ a Weyl fermion\footnote{Weyl
spinors are ubiquitous in SUSY Lagrangians, and the conventions used
are slightly different from the more familiar ones involving Dirac
spinors. We review the notation in Appendix~A.} and $f$ an
auxiliary field (a field without a kinetic term) that can be 
integrated out by its own (algebraic) equation of motion. The 
interaction terms can also be written explicitly
and we will discuss them at length in the following. For the purpose of 
introducing the SUSY transformations however the kinetic terms
(\ref{WZ}) are enough. 

The pure gauge theory on the other hand has a 
Lagrangian~\cite{nonabsusy}

\beq
      \lag = \tr\left(-\frac{1}{2} F^{\mu\nu}F_{\mu\nu} - 
      2 i \bar{\lambda}\bar{\sigma}^\mu
      D_\mu\lambda + D^2\right),   \label{puregauge}
\eeq
where $F_{\mu\nu}$ stands for the usual field strength, $\lambda$ is a
Weyl fermion (the ``gluino'' or ``gaugino'') and $D$ is, again, an auxiliary
field. All fields in (\ref{puregauge})
are matrix valued in some Lie algebra and the trace is the 
trace over the generators: $\tr(T^a T^b) = \frac{1}{2}\delta^{ab}$. 
One could also add a so called 
``topological term'' proportional to 
$\epsilon_{\mu\nu\rho\sigma}\tr F^{\mu\nu}F^{\rho\sigma}$ or, 
for $U(1)$ factors in the gauge group, a so called Fayet-Iliopoulos 
(FI) term proportional to $D$
but otherwise gauge invariance prevents any other type of 
renormalizable interaction.

The most general four 
dimensional gauge theory with $\mathcal{N}=1$ SUSY is, loosely speaking, 
based on a combination of (\ref{WZ}) and  (\ref{puregauge}) as we are going
to show. 

The SUSY transformations can be 
written explicitly as transformations on the fields. 
In the context of the WZ model, one can show that
\beqs
     \delta \phi &=& \xi\psi \nonumber \\
     \delta \psi &=& i\sigma^\mu\bar{\xi}\partial_\mu\phi +
                     \xi f  \label{SUSYWZ} \\
     \delta f &=& i\bar{\xi}\bar{\sigma}^\mu\partial_\mu\psi
                     \nonumber
\eeqs
in terms of one\footnote{Thus the name $\mathcal{N}=1$.} constant Grassmann 
(i.e. anticommuting) Weyl spinor $\xi_\alpha$
($\bar{\xi}_{\dot{\alpha}} = (\xi_\alpha)^*$), changes (\ref{WZ}) 
by at most a total derivative, thus leaving the action 
$S=\int \mathrm{d}^4 x \lag$ invariant.
A similar transformation can be shown to leave (\ref{puregauge})
invariant, again up to a total spacetime derivative:
\beqs
     \delta A^\mu &=& \frac{1}{\sqrt{2}}(-i\bar\lambda\bar\sigma^\mu\xi + 
               i\bar\xi\bar\sigma^\mu\lambda) \nonumber \\
     \delta \lambda &=&  \frac{1}{\sqrt{2}}(\sigma^{\mu\nu}\xi F_{\mu\nu} 
        + i\xi D) \label{SUSYYM} \\
     \delta D &=&  \frac{1}{\sqrt{2}}(-\xi \sigma^\mu D_\mu\bar\lambda - 
          D_\mu \lambda \sigma^\mu\bar\xi). \nonumber
\eeqs
While the invariance of (\ref{WZ}) under (\ref{SUSYWZ}) is
straightforward, in checking the invariance of (\ref{puregauge}) under 
(\ref{SUSYYM}) a more involved four Fermi term arises from
the variation of the gauge potential $A_\mu$ inside $D_\mu$,
which can be shown to vanish using Fierz identities. 
For later convenience, in (\ref{SUSYWZ}) and
(\ref{SUSYYM}) we rescaled the SUSY parameter by a factor of 
$1/\sqrt{2}$ from that which is commonly used.

The formulation briefly sketched above is manifestly Lorentz and gauge
invariant, but lacks manifest supersymmetry invariance. 
If we had such a manifest SUSY formulation we would not need to go
through the explicit computations discussed in the previous paragraph.
This task is accomplished by noticing that the SUSY transformations 
can be written in terms of ``translations in 
superspace''~\cite{Salam:1974yz}. To
understand  what that means, let us consider an ordinary translation of a
scalar field $\phi(x)$.
We know that there exist an operator $\mathcal{P}^\mu$ such that
\beq
     \phi(x+a) = e^{-i a \mathcal{P}} \phi(x)  e^{ i a \mathcal{P}}.
\eeq

When $a$ is small we can expand both sides. On the one hand: 
\beq
          \phi(x+a) = (1 - i a^\mu \mathcal{P}_\mu + \dots) \phi(x) 
                     (1 + i a^\mu \mathcal{P}_\mu + \dots) =
         \phi(x) - i a^\mu [\mathcal{P}_\mu, \phi(x)] + \dots
\eeq
on the other hand
\beq
          \phi(x+a) = \phi(x) + a^\mu \partial_\mu \phi(x) + \dots
\eeq
Thus:
\beq
      [\mathcal{P}_\mu, \phi(x)] = i \partial_\mu \phi(x) \equiv
        P_\mu \phi(x). 
\eeq
Hence, we have two  objects playing the role of ``momentum'':
$\mathcal{P}_\mu$ which is the generator of translations and 
$P_\mu \equiv i \partial_\mu$ which gives the \emph{representation}
of $\mathcal{P}_\mu$ in field space. These two objects are often
written using the same symbol.

Supersymmetry is the generalization of this to
``superspace'', that is space parameterized by the usual coordinates
$x^\mu$ and new complex Grassmann (anti-commuting) 
coordinates $\theta^\alpha$ and 
$\bar{\theta}^{\dot{\alpha}} = (\theta^\alpha)^*$. A ``superfield'' is a
function in superspace: $Y(x, \theta, \bar{\theta})$.
Since $(\theta_1)^2 = (\theta_2)^2 = 0$ and similarly for $\bar\theta$
due to anticommutativity, 
the Taylor expansion of $Y$ in $\theta$ and $\bar{\theta}$ terminates:
\beqs
     Y(x, \theta, \bar{\theta}) &=& \phi(x) + \theta\eta(x) +
     \bar{\theta}\bar{\chi}(x) + \theta^2 m(x) + \bar{\theta}^2 n(x) + 
     \theta\sigma^\mu\bar{\theta} A_\mu(x) + \nonumber \\
     && \theta^2\bar{\theta}\bar{\lambda}(x) + 
     \bar{\theta}^2\theta\psi(x) +\theta^2\bar{\theta}^2 d(x).
\eeqs
We have not indicated it explicitly but $Y$ could carry some spin
indices that would reflect in the component structure. The above
formula refers to a Lorentz scalar for simplicity.

In analogy with usual translations, we now define translations in
superspace by a quantity $(a, \xi, \bar{\xi})$ as:
\beqs
     \theta_\alpha &\to&  \theta_\alpha +  \xi_\alpha \nonumber \\
     \bar{\theta}_{\dot{\alpha}} &\to&  \bar{\theta}_{\dot{\alpha}} 
       +  \bar{\xi}_{\dot{\alpha}} \\
     x^\mu &\to& x^\mu + a^\mu + \frac{i}{2} \theta\sigma^\mu\bar{\xi} - 
       \frac{i}{2} \xi\sigma^\mu \bar\theta \nonumber,
\eeqs
where one should note the crucial addition of the fermionic bilinears
to the shift of $x^\mu$. They are of course needed so that the
composition of two fermionic translations yields a translation in $x^\mu$.

The operators generating such transformation on the superfields are
$\mathcal{P}$,  $\mathcal{Q}$ and $\bar{\mathcal{Q}}$ obeying
\beqs
    &&Y(x^\mu + a^\mu + \frac{i}{2} \theta\sigma^\mu\bar{\xi} - 
       \frac{i}{2} \xi\sigma^\mu \bar\theta, \theta_\alpha +  \xi_\alpha, 
       \bar{\theta}_{\dot{\alpha}} + \bar{\xi}_{\dot{\alpha}}) = 
      \nonumber\\
    && e^{-i a\mathcal{P} + \xi\mathcal{Q} - \bar{\xi}\bar{\mathcal{Q}}}
     Y(x^\mu, \theta_\alpha, \bar{\theta}_{\dot{\alpha}})
    e^{i a\mathcal{P} - \xi\mathcal{Q} + \bar{\xi}\bar{\mathcal{Q}}}.
\eeqs
Not surprisingly, the operators $\mathcal{Q}$ and $\bar{\mathcal{Q}}$
are also represented by differential operators acting on superfields, as
one can check:
\beq
    [\mathcal{Q}_\alpha, Y] =  Q_\alpha Y, \quad
    [\bar{\mathcal{Q}}_{\dot{\alpha}}, Y] = \bar{Q}_{\dot{\alpha}} Y,
    \label{qoper}
\eeq
where on the left hand side we have a commutator or anticommutator
depending on the spin of $Y$.
Explicitly:
\beq
    Q_\alpha = \partial_\alpha - 
    \frac{i}{2} \sigma^\mu_{\alpha {\dot{\alpha}}}
    \bar{\theta}^{\dot{\alpha}}\partial_\mu, \quad
    \bar{Q}_{\dot{\alpha}} = \bar{\partial}_{\dot{\alpha}} - 
    \frac{i}{2}\theta^\alpha \sigma^\mu_{\alpha {\dot{\alpha}}}\partial_\mu.
    \label{theQs}
\eeq
One can show that $P$, $Q$ and $\bar{Q}$ obey the SUSY algebra
\beq
     \{Q_\alpha, \bar{Q}_{\dot{\alpha}}\} = 
      -\sigma^\mu_{\alpha {\dot{\alpha}}} P_\mu. \label{SUSYalg}
\eeq

Having seen that SUSY is a translation in superspace, any
integral over superspace of a
superfield $Y$:
\beq
        \wholesp Y,  \label{wholesuper}
\eeq 
will be manifestly supersymmetric
because the integration measure is translational invariant by construction 
(see Appendix~A). Of course, $Y$ has to be a scalar superfield and 
can arise as the composition of other superfields.

In the case of translations we were done at this point. Here we have
to perform an extra step because the generic superfield $Y$ above
forms a \emph{reducible} representation of the SUSY algebra
(\ref{SUSYalg}), that is, it is possible to restrict the form of $Y$ to
contain only a subset of the original fields that are still mapped
into each other by SUSY. It should be obvious that $Y$ is ``too big''
because we would like, for instance, to formulate the WZ model
(\ref{WZ}) which does not have any vector fields or the pure gauge
theory (\ref{puregauge}) which does not have any scalars.
The two ways of reducing the field content in $Y$, 
that are going to be used are: \emph{(i)} The chiral (or antichiral)
projection and \emph{(ii)} The real projection.

\subsection{Chiral superfields}

A chiral superfield can be constructed by noticing that there exist
two differential operators:
\beq
     D_\alpha = \partial_\alpha
          +\frac{i}{2}\sigma^\mu_{\alpha {\dot{\alpha}}}
     \bar{\theta}^{\dot{\alpha}}\partial_\mu, \quad
     \bar{D}_{\dot{\alpha}} = \bar{\partial}_{\dot{\alpha}} +
     \frac{i}{2}\theta^\alpha \sigma^\mu_{\alpha {\dot{\alpha}}}\partial_\mu
     \label{theDs}
\eeq
that anticommute with all the supercharges (\ref{theQs}). 
This means that if $Y$ is a superfield, so are $D_\alpha Y$ and
$\bar{D}_{\dot{\alpha}} Y$ and it is consistent with SUSY to set one of them
to zero. (Setting both would result in requiring $Y=$ const.)
We call chiral superfield a superfield $\Phi$ 
obeying $\bar{D}_{\dot{\alpha}} \Phi = 0$. (Analogously, 
$D_\alpha \bar{\Phi} = 0$
defines an antichiral superfield.)

The solution to $\bar{D}_{\dot{\alpha}} \Phi = 0$ can be obtained by
observing that the (complex)
combination $y^\mu = x^\mu + \frac{i}{2}\theta\sigma^\mu\bar{\theta}$ obeys
$\bar{D}_{\dot{\alpha}} y^\mu =0$, and so does trivially
$\theta_\alpha$. Thus 
\beq
     \Phi = \phi(y) + \theta\psi(y)+ \theta^2 f(y). \label{easyphi}
\eeq
is the general expression for a chiral superfield. 
Expanding (\ref{easyphi}) around $x$ yields
\beq
    \Phi = \phi(x) + 
    \frac{i}{2}\theta\sigma^\mu\bar{\theta}\partial_\mu\phi(x) -
     \frac{1}{4}\theta^2 \bar{\theta}^2 \Box \phi(x) + 
     \theta\psi(x) - \frac{i}{2}\theta^2 
     \partial_\mu\psi(x)\sigma^\mu\bar{\theta} + \theta^2 f(x) 
    \label{chiralfield}
\eeq

Let us begin by considering the action of $\mathcal{Q}$ on $\Phi$.
Recalling that SUSY can be written as a
translation in superspace, we have, on the one hand\footnote{The SUSY
transformation of the coordinate $y^\mu$ is independent of $\xi$. We take
formally $\bar\xi = 0$ since we are only interested in $Q$ and not $\bar Q$.}:
\beq
    \delta_\xi\Phi = \xi Q \Phi = \xi \psi(y) + \xi\theta f(y),
\eeq
on the other hand:
\beq
    \delta_\xi\Phi = [\xi \mathcal{Q}, \Phi] =  [\xi \mathcal{Q},
     \phi] + \theta^\alpha  [\xi \mathcal{Q}, \psi_\alpha] + 
     \theta^2 [\xi \mathcal{Q}, f].
\eeq
Thus, the transformation properties (\ref{SUSYWZ}) are reproduced, now 
written in terms of the SUSY generators
\beqs
     {[ \mathcal{Q}_\alpha,\phi]} &=& \psi_\alpha \nonumber \\
     {\{ \mathcal{Q}_\alpha,\psi_\beta \}} &=&
     - \epsilon_{\alpha\beta} f  \label{QPsi} \\
     {[ \mathcal{Q}_\alpha, f]} &=& 0.  \nonumber
\eeqs
Similarly, one can check that
\beqs
     {[ \bar{\mathcal{Q}}_{\dot{\alpha}},\phi]} &=& 0 \nonumber \\
     {\{ \bar{\mathcal{Q}}_{\dot{\alpha}},\psi_\beta\}}&=&
      -i \sigma^\mu_{\beta \dot{\alpha}} \partial_\mu \phi \label{QbarPsi}\\
     {[ \bar{\mathcal{Q}}_{\dot{\alpha}}, f]} &=& 
      i \sigma^\mu_{\alpha\dot{\alpha}}\partial_\mu\psi^\alpha.  \nonumber
\eeqs

In order to write a SUSY action we recall that the superfield $Y$ in 
(\ref{wholesuper}) can be the composition of other superfields.
For instance, given a chiral superfield $\Phi$ one can write
\beq
     \wholesp \bar\Phi \Phi, \label{ek}
\eeq
which has been chosen as an example because by expanding $\Phi$ and
$\bar\Phi$ and
integrating over $\theta$ and $\bar{\theta}$ we obtain precisely the kinetic
terms in the WZ model (\ref{WZ}).

It is often stated that there is an exception to (\ref{wholesuper}) when the
integrand happens to be a chiral superfield $\Sigma$ (or a product of such).
In this case, the integral over the whole superspace vanishes,
however, the integral over \emph{chiral} superspace
\beq
     \halfsp \Sigma \label{halfsuper}
\eeq
is still manifestly SUSY (translation invariant) as can readily be
seen in the formalism where $x^\mu$ is replaced by $y^\mu$ so that
$\bar{\theta}$ never appears. Equivalently, it can be noticed from eqs.
(\ref{QPsi}) and (\ref{QbarPsi}) that the highest component $f$ of a
chiral superfield (which is singled out by the $\theta^2$ integral)
transforms as a total spacetime derivative.

However, it should be appreciated that (\ref{halfsuper}) is actually 
\emph{more general} than (\ref{wholesuper}) in the sense that any
integral of type (\ref{wholesuper}) can be written as (\ref{halfsuper}).
Remember that integrals over a Grassmann variable are equivalent 
to derivatives and that total spacetime
derivatives $\partial_\mu$ do not contribute to the bosonic 
integral\footnote{We need not worry about boundary terms at this stage.} 
so that one can write
\beq
     \wholesp Y = \halfsp \bar{D}^2 Y. \label{fake}
\eeq
The superfield $\bar{D}^2 Y$ is manifestly chiral, regardless of
the form of $Y$ because three $\bar{D}_{\dot{\alpha}}$ always give zero
since they anticommute and there are only two possible values
${\dot{\alpha}}$ can take. We call such terms ``chirally exact''.
On the other hand, not all terms of the form (\ref{halfsuper}) 
can be written as (\ref{wholesuper}) in terms of a local integral.
For instance, the quantity
\beq
        \halfsp \Phi^n,  \label{honest}
\eeq
for some chiral superfield $\Phi$, cannot be expressed as (\ref{wholesuper})
because we do not have any
derivative that can be reconverted into a $\mathrm{d}^2\bar{\theta}$. 
We call terms such as (\ref{honest}) that cannot be written as local
integrals over the whole superspace ``F-terms'' and all the others,
including (\ref{fake}), ``D-terms''. 

The above expression (\ref{ek}) is so compact that 
it may be a bit confusing at
first -- for instance, if one tried to
obtain the equations of motion by varying $\bar\Phi$
one would get the nonsensical result $\Phi = 0$. 
The error, of course, is that the field $\Phi$ obeys 
$\bar{D}_{\dot{\alpha}} \Phi = 0$ but the action (\ref{ek}) doesn't know
that. An easy way to enforce this is to use the above tricks and write
\beq
     \wholesp \bar\Phi \Phi = \int \mathrm{d}^4 x 
       \mathrm{d}^2\bar{\theta} \bar\Phi(D^2 \Phi).
\eeq
Now the fields are unconstrained in (anti)-chiral superspace 
and varying with respect to $\bar\Phi$
yields $ D^2 \Phi = 0$ which implies the free equations of motion for
$\phi$, $\psi$ and $f$. 

Interactions can be introduced in the same
way. Since the simplest non-vanishing integral over all of superspace
(\ref{ek}) already involves derivatives, we must turn to terms
like (\ref{honest}). In fact any \emph{holomorphic} function $W(\Phi)$
(i.e. such that ${\partial W}/{\partial\bar\Phi} = 0$)
to be called the superpotential, will be acceptable in the sense that 
if $\Phi$ is chiral, so is $W(\Phi)$:
\beq
     \bar{D}_{\dot{\alpha}}W(\Phi) = \frac{\partial W(\Phi)}{\partial \Phi}
     \bar{D}_{\dot{\alpha}}\Phi + \frac{\partial W(\Phi)}{\partial \bar{\Phi}}
     \bar{D}_{\dot{\alpha}}\bar\Phi = 0.
\eeq
Thus
\beq
    \halfsp W(\Phi) + \int \mathrm{d}^4x \mathrm{d}^2\bar\theta\;
    \bar{W}(\bar{\Phi})
\eeq
gives a supersymmetric interaction term that can be added to the
kinetic terms in (\ref{ek}). If renormalizability is an
issue (and it will not be in many following cases since we will be
dealing with \emph{effective theories}) we must restrict $W$ to
be at most a cubic polynomial.

\subsection{Vector superfields}

To introduce gauge interactions we need a second type of superfield --
the \emph{vector} superfield. This is just the original $Y$ 
(conventionally called $V$ in this case and Lie algebra valued) 
on which we impose the reality condition
\beq
            V = V^\dagger.
\eeq
This condition is trivially supersymmetrically covariant because if $V$ is the
most arbitrary superfield, so is $V^\dagger$ and equating the two
is a superfield equation. In components, displaying the Lie algebra index: 
\beqs
    V^a &=& C^a(x) + i\theta\chi^a(x) -i\bar{\theta}\bar{\chi}^a(x) +
     i \theta^2 m^a(x) - i \bar{\theta}^2 \bar m^a(x)
     - \theta\sigma^\mu\bar{\theta}A^a_\mu(x) \nonumber \\
     &&+\sqrt{2} i \theta^2\bar{\theta}
     (\bar{\lambda}^a(x) + \frac{i}{2\sqrt{2}}\bar{\sigma}^\mu\partial_\mu
     \chi^a(x)) - \sqrt{2} i \bar{\theta}^2\theta 
     (\lambda^a(x)\cr &&+ 
     \frac{i}{2\sqrt{2}}\sigma^\mu\partial_\mu\bar{\chi}^a(x))
     + \theta^2\bar{\theta}^2(D^a(x) - \frac{1}{4}\Box C^a(x)),
\eeqs
where $C^a$, $A_\mu^a$ and $D^a$ are real and the shifts in
$\lambda^a$ and $D^a$ by derivatives of $\chi^a$ and $C^a$ are for
later convenience.
If $T^a$ are the generators of the Lie algebra ($[T^a, T^b] = i
f^{abc} T^c$), one can write a matrix valued object $V = V^a T^a$
obeying $V = V^\dagger$ because $V^a$ are real and $T^a$ hermitian.

Let us now see the relation between the vector superfield and gauge 
invariance. Consider matter described by a chiral superfield $\Phi$ 
transforming in some representation (reducible or irreducible) of the
Lie algebra. The usual gauge transformation
\beq
   \Phi \to \Phi^\prime = e^{- i \Lambda} \Phi
\eeq
keeps this form but now $\Lambda\equiv \Lambda^a T^a$ must 
also be promoted to a chiral
superfield because otherwise $\Phi^\prime$ would no longer be chiral. 
However there is a problem with the kinetic term because, being
$\Lambda$ complex
\beq
   \bar\Phi\Phi \to  \bar\Phi e^{ i \bar\Lambda}
    e^{- i \Lambda}\Phi \not= \bar\Phi \Phi.
\eeq
Thus, we must compensate by inserting 
$e^V$ between $\bar\Phi$ and $\Phi$ and postulate the
transformation law:
\beq
     e^V \to e^{V^\prime} =  e^{- i \bar\Lambda} e^V e^{i \Lambda}
       \label{gaugeV}
\eeq
so that $ \bar\Phi  e^V \Phi$ is gauge invariant.

At first sight, it is not obvious that (\ref{gaugeV}) implies an 
infinitesimal gauge transformation like
\beq
    \delta A_\mu = \partial_\mu \alpha + i{[A_\mu, \alpha]}
    \label{goodold}
\eeq 
but, in fact, (\ref{goodold}) is included in the transformation 
(\ref{gaugeV}) of the components of $V$ as one should check
at least for an Abelian group, where  (\ref{gaugeV})
simplifies to $V \to {V^\prime} =  V - i \bar\Lambda + i \Lambda$.

Using (\ref{gaugeV}) it is possible to show that one can
gauge away many of the fields in the original definition of $V$ and go
to the so called Wess-Zumino gauge where
\beq
    V^a = 
     - \theta\sigma^\mu\bar{\theta}A^a_\mu(x) + 
     \sqrt{2} i \theta^2\bar{\theta}
     \bar{\lambda}^a(x) -  \sqrt{2} i \bar{\theta}^2\theta\lambda^a(x) +
     \theta^2\bar{\theta}^2 D^a(x). \label{VinWZ}
\eeq
In this form, $V$ is no longer a superfield -- this is not a SUSY
gauge. Indeed acting with $Q$ or
$\bar Q$ would regenerate the terms that have been set to zero in the
WZ gauge. However, these new terms can be once again removed by a new
(field dependent) gauge transformation. In other words, in this gauge
the SUSY algebra closes up to a gauge transformation.

As for the superpotential, no modification is required. 
As long as $W$ is holomorphic and gauge invariant in the usual
sense, the gauge parameter can be promoted to a full chiral superfield
because its complex conjugate is never used. This is often stated by
saying that the F-terms are invariant under the complexified gauge
group.

We now turn to finding the
kinetic term for $V$, i.e. the manifestly SUSY way of writing
(\ref{puregauge}). From the transformation law (\ref{gaugeV}) and the fact
that $A_\mu$ appears explicitly in $V$ it is clear that we must act with 
some derivative to construct an object to be identified with the
supersymmetric field strength. 

It turns out that the right expression is
\beq
   W_\alpha = \bar{D}^2 (e^{-V}(D_\alpha e^V)) \label{susyfieldstrength}
\eeq
where we have defined the manifestly chiral superfield 
$W_\alpha$\footnote{Not
to be confused with the superpotential, also conventionally denoted
with a $W$ but without any spinor indices.}.
The antichiral superfield $\bar W_{\dot \alpha}$ has a similar definition,
obtained from $\bar W_{\dot \alpha} = (W_\alpha)^*$.

It is a nice exercise to check that, under a gauge transformation 
(\ref{gaugeV}), $W_\alpha$ transforms as
\beq
     W_\alpha \to  W_\alpha^\prime = e^{-i\Lambda} W_\alpha
     e^{i\Lambda}.  \label{gaugeW}
\eeq
Note the difference between (\ref{gaugeW}) and (\ref{gaugeV})
-- in (\ref{gaugeW}) there is no $\bar{\Lambda}$ appearing and so it is
possible to make a gauge invariant combination e.g. by taking traces.
In checking (\ref{gaugeW}) one has to notice that, first
(\ref{gaugeV}) implies $e^{-V} \to
e^{-i\Lambda}e^{-V}e^{i\bar{\Lambda}}$, second that the
$\bar{\Lambda}$ terms commute through $D_\alpha$ and cancel and,
finally, that the spurious term
$\bar{D}^2 D_\alpha(e^{i\Lambda})$ vanishes.

$W_\alpha$ can be easily computed in the WZ gauge to give
\beq
    W_\alpha = -\sqrt{2} i\lambda_\alpha(y) + \theta_\alpha D(y) - 
    i  {\sigma^{\mu\nu}}_\alpha^\beta \theta_\beta F_{\mu\nu}(y)
    + \sqrt{2} \theta^2 \sigma^\mu_{\alpha\dot{\beta}}
     D_\mu\bar{\lambda}^{\dot{\beta}}(y),
\eeq
where, as usual, expanding in $y$ would introduce 
additional spacetime derivatives. Since $W_\alpha$ starts off with 
the gaugino, it is often called the gaugino (or gluino) superfield.

The pure gauge theory action (\ref{puregauge}) can be written in
a manifestly SUSY way using the field $W_\alpha$:
\beq
   \halfsp \tr W^\alpha W_\alpha + 
   \int \mathrm{d}^4x \mathrm{d}^2\bar\theta\; 
   \tr \bar{W}_{\dot\alpha} \bar{W}^{\dot\alpha}.
   \label{manifestgauge}
\eeq

For later purpose, it is convenient to reformulate expression 
(\ref{susyfieldstrength}) for the field strength $W_\alpha$ in terms of
(anti)commutators of some new ``covariant'' derivatives.
Define
\beq
    \bar{\nabla}_{\dot{\alpha}} \equiv  \bar{D}_{\dot{\alpha}} 
    \quad\hbox{and} \quad \nabla_\alpha \equiv 
    e^{-V}D_\alpha e^V, \label{nablas}
\eeq
the second equation being intended as an operator, that is
$D_\alpha$ acts on $e^V$ and whatever follows it.
Given the two new derivatives, one can define ``covariantly chiral/antichiral''
fields $\Phi$ (as before) and $\tilde\Phi = \bar\Phi e^V$ obeying 
$\bar{\nabla}_{\dot{\alpha}}\Phi = 0$ and 
${\nabla}_\alpha \tilde\Phi  = 0$ (this last expression to be interpreted as
a right action.).
Evaluating the anticommutator yields:
\beq
    \{\nabla_\alpha, \bar{\nabla}_{\dot{\alpha}} \} = 
     i\sigma^\mu_{\alpha\dot{\alpha}}\partial_\mu +
     \bar{D}_{\dot{\alpha}}( e^{-V}(D_\alpha e^V)) \equiv
     i \nabla_{\alpha\dot{\alpha}}.
\eeq
The last equality defines the new
object $\nabla_{\alpha\dot{\alpha}}$, whose lowest component is the
usual covariant derivative. Taking one more commutator we obtain:
\beq
    W_\alpha = \frac{i}{2}{[{\bar\nabla}^{\dot\alpha}, 
    \nabla_{\alpha\dot\alpha}]}.
\eeq

\subsection{The complete Lagrangian}

Let us summarize all the results thus far by writing the most general 
$\mathcal{N}= 1$ gauge theory for a simple gauge group $G$.
It is convenient to combine the usual Yang-Mills coupling $g$ and the 
vacuum angle $\Theta$ into a complex parameter
\beq
      \tau = \frac{4\pi}{g^2} - i \frac{\Theta}{2\pi}. \label{coupl}
\eeq

Let $\Phi$ be the matter chiral superfield transforming in a 
(generically reducible) representation of $G$. The complete Lagrangian
takes the form
\beqs
     S &=& \frac{\tau}{16\pi}\halfsp \tr W^\alpha W_\alpha +
           \frac{\bar\tau}{16\pi}\halfspbar 
           \tr \bar{W}_{\dot\alpha} \bar{W}^{\dot\alpha} + \nonumber\\ 
     && \wholesp \bar\Phi e^V \Phi + \halfsp W(\Phi) +
        \halfspbar \bar{W}(\bar\Phi). \label{ich}
\eeqs 
The superpotential $W(\Phi)$ is invariant under $G$ and all the
coupling constants inside it are complex. 
In fact, in string theory it is
quite natural to think of these couplings as the vacuum expectation values of
some background chiral superfields as we will discuss later.

It is instructive to write (\ref{ich}) in components in the WZ
gauge. The gauge kinetic Lagrangian (first line in 
eq.~(\ref{ich})) becomes
\beq
      \frac{1}{g^2}\tr\left(-\frac{1}{2} F^{\mu\nu}F_{\mu\nu} - 
      2 i \bar{\lambda}\bar{\sigma}^\mu
      D_\mu\lambda + D^2\right) + \frac{\Theta}{32\pi^2}
      \epsilon^{\mu\nu\rho\lambda}\tr F_{\mu\nu} F_{\rho\lambda}.
\eeq
The matter ``kinetic'' Lagrangian (first term in the second line
of (\ref{ich})) reads
\beq
     D_\mu\bar\phi D^\mu\phi - i \bar{\psi}\bar{\sigma}^\mu
     D_\mu\psi + \bar{f} f + 
     \sqrt{2} i \left(\bar\phi \lambda \psi 
     -\bar\psi\bar\lambda \phi\right) + \bar\phi D \phi,
\eeq
where the indices\footnote{The index $a=1\dots dim(G)$ runs over the 
adjoint representation of $G$. The indices $i, j = 1\dots dim(R)$ label the
representation $R$ of $\Phi$ and its conjugate representation $\bar R$.} 
are contracted in the only possible way, i.e.
$\bar\phi \lambda \psi \equiv
\bar{\phi}^i \lambda^{\alpha a} {T^a}^j_i \psi_{\alpha j}$. 
Lastly, the contribution from the superpotential reads
\beq
   -\frac{1}{2}\frac{\partial^2 W}{\partial\phi_i\partial\phi_j}\psi_i\psi_j+
   \frac{\partial W}{\partial\phi_i} f_i + c.c.
\eeq

Let us collect all terms contributing to the bosonic potential
\beq
    \mathcal{V} = - \frac{1}{2 g^2} {D^a}^2 - \bar\phi T^a \phi D^a -
      \bar{f}^i f_i - \frac{\partial W}{\partial\phi_i} f_i - 
     \frac{\partial \bar{W}}{\partial\bar{\phi}^i} \bar{f}^i. \label{poti}
\eeq
The auxiliary fields $D$ and $f$ appear only in these terms and we can
solve for them as
\beq
    D^a = - g^2 \bar\phi T^a \phi,
   \quad f_i = - \frac{\partial \bar{W}}{\partial\bar{\phi}^i},
    \quad \bar{f}^i = - \frac{\partial W}{\partial\phi_i}. \label{bruce}
\eeq
Substituting the solution (\ref{bruce}) in the potential (\ref{poti}) gives
\beqs
   \mathcal{V} &=& \frac{g^2}{2}(\bar\phi T^a \phi)^2 + 
   \frac{\partial \bar{W}}{\partial\bar{\phi}^i}
   \frac{\partial W}{\partial\phi_i} \nonumber \\
    &=& \frac{1}{2 g^2} {D^a}^2 + \bar{f}^i f_i,
\eeqs
the last line to be interpreted as evaluated at the solution.

The potential $\mathcal{V}$ is obviously non-negative, and 
the classical SUSY vacua are described by $\mathcal{V} = 0$, i.e. by
the simultaneous vanishing of the $D$ and $f$ terms, referred to as
D-flatness and F-flatness conditions.

The space of solutions to the D-flatness conditions $\bar\phi T^a\phi
= 0$ up to gauge transformations is an important object 
known as the \emph{classical moduli space} 
of the theory. An extremely useful mathematical fact\footnote{For
a careful explanation accessible to physicists and a list of earlier
references see~\cite{Luty:1995sd}.} is that such
space can always be parameterized in terms of a set of independent
holomorphic gauge invariants $X_r(\phi)$.
One of the most common examples is a field $\Phi$ in the adjoint
representation of $SU(N)$. In this case the D-flatness
condition is $\bar\phi T^a \phi = i f^{abc} \bar{\phi}^b \phi^c = 0$,
that is, ${[\phi^\dagger, \phi]} = 0$ as a matrix equation. The
classical moduli space is parameterized by $X_k = \tr \phi^k$ for 
$k = 2,\dots N$.  

As for the F-flatness condition, it implies that when a superpotential
is present classical SUSY vacua are found at its extrema. 
Thus, in the presence of a superpotential, some of the ``moduli'' $X_r$
defined above may be subjected to further restrictions.

So far our discussion has been in terms of the classical theory and one
must ask how this picture is modified in the quantum theory. 
We will see that, in general, the superpotential will be modified by
(non-perturbative) quantum corrections but the 
property of the superpotential of identifying the SUSY vacua
carries over to the low energy effective theory as long as one does not 
encounter any singularity, typically due to the presence of extra 
massless fields. In this review we will discuss at length how 
the superpotential is modified by quantum corrections but unfortunately
we will not have time to discuss the many applications to the search
for SUSY vacua.

\section{Basic Dynamical Facts}

Having seen the kinematic (classical) structure of SUSY gauge
theories we now review some basic facts about their (quantum)
dynamics. The aim of this section is to present enough information to
be able to put the latest developments into context, without any
ambition of being exhaustive.

It is a familiar fact that in a quantum theory the
couplings appearing in the physical quantities 
(observables) are
actually running parameters depending on the renormalization scale
$\mu$ and that they appear in the observables in such a way that the total
$\mu$ dependence cancels. 
All of this is still true in SUSY gauge theories but SUSY implies some
simplifications in the running of the coupling constants. 

The simplest (and very important) example of this fact is the perturbative
\emph{non-renormalization} of the superpotential. 
This can already be seen in the WZ model, defined by the bare
action\footnote{We consider a cubic superpotential to ensure that the 
full theory is renormalizable in the ordinary sense.}: 
\beq
     S =  \wholesp \bar{\Phi}_0  \Phi_0 
     + \halfsp\; { \frac{m_0}{2}}\Phi_0^2 
     + { \frac{\lambda_0}{3}} \Phi_0^3
     + c.c.  
     \label{WZbare}
\eeq
Loop computations and renormalization can be done in a variety of
equivalent ways -- we use the so called ``renormalized perturbation
theory'' and write action (\ref{WZbare}) in terms of
the renormalized quantities: 
\beqs
     S &=&  \wholesp Z_\Phi\bar\Phi \Phi 
     + \halfsp\; Z_m \frac{m}{2}\Phi^2 
     + Z_\lambda \frac{\lambda}{3} \Phi^3
     + c.c. = \nonumber \\
     &=& \wholesp \bar\Phi \Phi 
     + \halfsp\; \frac{m}{2}\Phi^2
     +  \frac{\lambda}{3} \Phi^3
     + c.c. +  \nonumber \\
     && \wholesp\; \delta_\Phi \bar\Phi \Phi 
     + \halfsp  \frac{\delta_m}{2}\Phi^2 
     +  \frac{\delta_\lambda}{3} \Phi^3
     + c.c. \label{WZren}
\eeqs

It is possible to develop Feynman rules directly in superspace without
having to expand into component fields.
Without presenting the details that are explained very clearly in the 
literature, the main point~\cite{noren} is that 
the most general term that can be generated in loop diagrams has
\emph{only one} Grassmann integral over \emph{all four} $\theta$:
\beq
    \int\;\mathrm{d}^4 x_1\dots\mathrm{d}^4 x_n 
    \mathrm{d}^2\theta \mathrm{d}^2\bar\theta
     G(x_1 \dots x_n)F_1(x_1,\theta, \bar\theta)\dots 
     F_n(x_n,\theta, \bar\theta), 
    \label{loca}
\eeq
where the $F_i$ stand for products of superfields and their derivatives. 
We notice in passing that the expression would be valid if we also had
the gauge superfield $V$ with the possibility of the $F_i$ being equal
to $V$, $D_\alpha V$ etc. in (\ref{loca}).

We know that the theory is renormalizable because it is so in components.
Thus the only possible primitively ultraviolet (UV) 
divergent graphs are those with an external field structure already
present in the bare Lagrangian.
Suppose now that there were a UV 
divergent graph with only two (or three) external chiral fields
$\Phi$. The local counterterm needed to cancel
it would be of the form (say in dimensional
regularization, where $\epsilon = 4 - d$ and $l$ is the degree of the pole)
\beq
     \frac{1}{\epsilon^l}\wholesp \Phi^2(x,\theta, \bar\theta)
\eeq
but this diagram is identically zero since it is the integral of
a chiral quantity over all of superspace. This means that diagrams with 
two or three chiral fields are always UV convergent and thus
$\delta_m = \delta_\lambda = 0$ implying
$Z_m = Z_\lambda = 1$. At this stage, we do not need to worry about
terms like
\beq
    \wholesp \Phi\frac{D^2}{\Box}\Phi, \label{badguy}
\eeq
which could be turned into a local F-term by replacing 
$\mathrm{d}^2\bar{\theta} \to \bar{D}^2$ and noting that 
$\bar{D}^2 D^2\Phi = - \Box \Phi$,
because such non-local behavior can never arise in the UV. 

\begin{figure}[htb]
  \begin{center}

    \setlength{\unitlength}{2pt}
    \begingroup\makeatletter\ifx\SetFigFont\undefined%
    \gdef\SetFigFont#1#2#3#4#5{%
      \reset@font\fontsize{#1}{#2pt}%
      \fontfamily{#3}\fontseries{#4}\fontshape{#5}%
      \selectfont}%
    \fi\endgroup%
    {\renewcommand{\dashlinestretch}{30}
      \begin{picture}(72,72)(0,0)
        \thicklines
        \path(00,72)(36,36)(72,36)(36,36)(00,00)
        \path(12,60)(24,24)
        \path(12,12)(18.41,31.26) 
        \path(21.58,40.74)(24,48)

        \put(0,64){\makebox(0,0)[lb]{\smash{{{\SetFigFont{12}{0}{\rmdefault}{\mddefault}{\updefault}$\phi$}}}}}
        \put(13,60){\makebox(0,0)[lb]{\smash{{{\SetFigFont{12}{0}{\rmdefault}{\mddefault}{\updefault}$\lambda$}}}}}
        \put(25,48){\makebox(0,0)[lb]{\smash{{{\SetFigFont{12}{0}{\rmdefault}{\mddefault}{\updefault}$\bar\lambda$}}}}}
        \put(36,30){\makebox(0,0)[lb]{\smash{{{\SetFigFont{12}{0}{\rmdefault}{\mddefault}{\updefault}$\lambda$}}}}}
        \put(69,30){\makebox(0,0)[lb]{\smash{{{\SetFigFont{12}{0}{\rmdefault}{\mddefault}{\updefault}$\phi$}}}}}
        \put(0,6){\makebox(0,0)[lb]{\smash{{{\SetFigFont{12}{0}{\rmdefault}{\mddefault}{\updefault}$\phi$}}}}}
        \put(13,6){\makebox(0,0)[lb]{\smash{{{\SetFigFont{12}{0}{\rmdefault}{\mddefault}{\updefault}$\lambda$}}}}}
        \put(25,18){\makebox(0,0)[lb]{\smash{{{\SetFigFont{12}{0}{\rmdefault}{\mddefault}{\updefault}$\bar\lambda$}}}}}
        \thicklines
        \put(20,36){\arc{10}{1.8925}{5.0359}}
        
      \end{picture}
      }
    \caption{The two-loop contribution to the one-particle effective
    superpotential.} 
  \label{peter}
\end{center}
\end{figure}
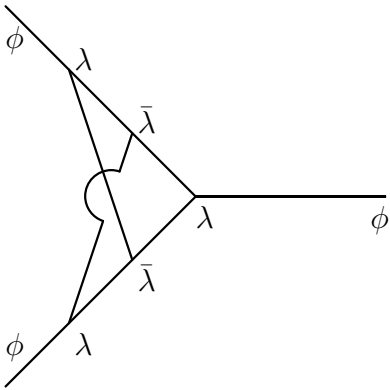

The situation changes if we are interested in the effective action
generating the one particle irreducible diagrams 
(1PI action)~\cite{Jack:pd}. 
In this context, a UV convergent but
IR divergent D-term of the type (\ref{badguy}) can arise giving a
term that \emph{looks} like a finite contribution to the
superpotential. One explicit example comes from the massless WZ model,
where a diagram like
Fig.~\ref{peter} 
gives a finite contribution to the 1PI
effective superpotential:
\beq
    W_{1PI} = \halfsp \;  \frac{\lambda}{3}\Phi^3 + \zeta(3)
     \frac{\lambda^3\bar{\lambda}^2}{(16\pi^2)^2}\Phi^3 + \dots \label{onePI}
\eeq 
and the dots indicate that there could be more
contributions.

Thus we see that the 1PI superpotential is (by definition) a
holomorphic quantity in the fields but \emph{not necessarily} in the
coupling constants. This is an unpleasant state of affairs that
suggests that we are not looking at the ``right'' quantity for our
purposes. First of all, since we are interested in gauge theories, the
1PI action constructed via perturbation theory is not well defined
since the fields appearing in the Lagrangian are not the correct
degrees of freedom at low energy. Even more basically, if we think
of the coupling constants as the vacuum expectation values (v.e.v.) of
some very heavy chiral superfield \cite{Seiberg:1994bp}
(as is natural in string
theory), then a term like the last one in 
(\ref{onePI}) is \emph{not} supersymmetric and should be written
instead as:
\beqs
    &&\wholesp \lambda^3 \bar{\lambda}^2 \Phi^2\frac{D^2}{\Box}\Phi =
    \halfsp \bar{D}^2\left(\lambda^3 \bar{\lambda}^2\Phi^2
    \frac{D^2}{\Box}\Phi \right) \\
    &=&\halfsp \left( \lambda^3 \Phi^2 \bar{D}^2  \bar{\lambda}^2
    \frac{D^2}{\Box}\Phi + \lambda^3 \Phi^2 \bar{D}_{\dot{\alpha}}
    \bar{\lambda}^2 \bar{D}^{\dot{\alpha}} \frac{D^2}{\Box}\Phi -
    \lambda^3 \bar{\lambda}^2 \Phi^3 \right), \nonumber
\eeqs
where in the last line only the sum of all three terms is manifestly SUSY.
In this framework, the whole expression should thus be thought of as 
a singular D-term.

All this suggests that we forget about the 1PI action and look instead
at the Wilsonian action defined as follows: If $S_{\mu_0}$ denotes the
(bare) action at a scale $\mu_0$, then $S_{\mu}$  (for
$\mu<\mu_0$) denotes the action describing the same physics but where
the degrees of freedom with momenta between $\mu$ and $\mu_0$ have
been integrated out. 
We can be more precise if we restrict our
attention to the Wilsonian superpotential because any non-holomorphic
dependence on the couplings must be regarded as a (now properly IR
regularized) D-term~\cite{Seiberg:1994bp}.
The Wilsonian superpotential $W_{\mathrm{eff}}$ is thus a 
holomorphic quantity in both
fields and coupling constants and it is \emph{not} 
perturbatively renormalized. Thus we write:
\beq
    W_{\mathrm{eff}} = W_{\mathrm{tree}} + W_{\mathrm{non-pert}},
    \label{weff}
\eeq
where $W_{\mathrm{tree}}$ is the tree level superpotential and 
$W_{\mathrm{non-pert}}$ a non-perturbative contribution that cannot
be ruled out by the perturbative analysis above. Such term is actually
absent in the pure WZ model but will be present in many gauge theories and
will indeed play a crucial role in all that follows.

Let us make two more remarks. In the case where interacting 
massless fields are absent, the Wilsonian and the 1PI effective 
superpotentials coincide. Furthermore, the perturbative non-renormalization
of the superpotential can also be proven using holomorphy and the
symmetries of the theory as discussed in~\cite{Seiberg:1994bp}. 

Let us now turn to the renormalization of the 
gauge coupling constant $\tau$ introduced in (\ref{coupl}). Since the
gauge kinetic term is written as an integral over chiral
superspace one might be tempted to argue that the 
non-renormalization theorem would apply to this case too. However, this
conclusion is wrong because one can write
\beq
    \halfsp \tr W^\alpha W_\alpha = 
    \wholesp \tr (e^{-V} D^\alpha e^V W_\alpha).
\eeq
The integrand is not gauge invariant, however the
integral is (since it is just a rewriting of a manifestly gauge
invariant one) and it is local. Thus, UV divergences of this type are
allowed by the non-renormalization theorem and the coupling $\tau$ is
not protected from running. 

Having established that the coupling constant runs,
we can use a piece of information well known from
non-supersymmetric theories that states that the $\Theta$ term, being
the coefficient of a boundary term, does not get renormalized. Hence,
the beta-function for $\tau$ only involves $Re (\tau)$ and
thus cannot be holomorphic. The only exception would be if it were a
constant, which would imply that $g$ is renormalized only to one loop.
Indeed,
\beq
      \mu\frac{d}{d\mu}\tau = \frac{\beta}{2\pi}, \label{arf}
\eeq
where $\beta$ is a real constant, implies
\beq
    \mu\frac{d}{d\mu} \Theta = 0\quad\hbox{and}\quad 
    \mu\frac{d}{d\mu} g = - \frac{\beta}{16 \pi^2} g^3. \label{arff}
\eeq
An explicit computation shows that this is \emph{not} the case for the 
standard coupling constant that in general gets renormalized to all orders.
Once again, this suggests that we should introduce a different 
coupling constant, to be referred to as the Wilsonian coupling, obeying
(\ref{arf}), (\ref{arff}). The relation between the ordinary (1PI) and 
the Wilsonian coupling is of course non-holomorphic~\cite{Shifman:1991dz}
and this explains why 
the two beta-functions disagree already at two loops. Even in this
context, it can be shown that
the lack of holomorphy comes from an IR effect~\cite{Shifman:1991dz}
and it is thus consistent
to exclude it from the Wilsonian beta-function.
From now on $\tau$ will always denote the Wilsonian coupling.

A standard computation then gives:
\beq
    \beta = \frac{1}{2}\left( 3\; T(\hbox{Adjoint})-\sum_\rho T(\rho)\right),
\eeq
where $T(\rho)$ denotes the index of the irreducible representations $\rho$
in which $\Phi$ transforms, normalized to $T(\hbox{Fundamental}) = 1$.

Eq. (\ref{arf}) allows one to introduce a perturbatively RG invariant scale
known as the holomorphic scale:
\beq
  \Lambda^\beta = \mu^\beta e^{-2 \pi \tau(\mu)} =
   \mu^\beta e^{-\frac{8 \pi^2}{g^2(\mu)} + i \Theta}. \label{biglambda}
\eeq 

Consider two theories related to each other by
integrating out some field of mass $m$.
Because the running is perturbatively only one-loop, the
holomorphic scales are easily matched as
\beq
       \Bigg(\frac{\tilde{\Lambda}}{m}\Bigg)^{\tilde{\beta}} = 
       \Bigg(\frac{\Lambda}{m}\Bigg)^{\beta}, \label{matching}
\eeq
where the tilde quantities refer to the IR theory where the field of
mass $m$ has been integrated out.

The importance of the holomorphic scale $\Lambda$ is that it controls
the quantum corrections of the observables in the theory, with
$\Lambda \to 0$ corresponding to the classical limit.

Let us now return to the Wilsonian superpotential (\ref{weff}).
Notice that one can always write 
$W_{\mathrm{tree}} = \sum \lambda_r X_r(\Phi_i)$ for
the gauge invariant quantities $X_r(\Phi_i)$ introduced in the
discussion of the D-terms\footnote{In some cases, some of the $X_r$
could also be dependent.}.
The question now is what is the form of $W_{\mathrm{non-pert}}$. 
One would expect that $W_{\mathrm{non-pert}}$ depends (holomorphically,
of course) on the scale $\Lambda$, the couplings $\lambda_r$ and the
invariants $X_r(\Phi_i)$. But, in fact, $W_{\mathrm{non-pert}}$ is
\emph{independent} of the couplings $\lambda_r$. This fact is
sometimes referred to as the ``linearity 
principle''~\cite{linearityprinciple} because 
it implies that the full superpotential is linear in the
couplings.

The linearity principle is less surprising than it seems at first when
one realizes that, having ruled out any perturbative dependence on
$\lambda_r$, any non-perturbative dependence would have to have a
singularity as $\lambda_r\to 0$ in some direction in the complex
plane. For instance, note that this rules out terms of the type
$e^{-1/\lambda}$ that go to zero only if the origin is approached 
for $Re(\lambda)>0$.
The gauge coupling $\tau$ is a different story because 
in this case the real part and the imaginary part enter differently
(the real part is always positive)
and thus the combination (\ref{biglambda})
appearing in $\Lambda$ is allowed and it can
be generated e.g. by instanton effects.

Having established that 
$W_{\mathrm{non-pert}} = W_{\mathrm{non-pert}}(X_r, \Lambda)$ it is
sometimes possible to completely fix the functional dependence up to
some numerical constant by dimensional analysis and symmetry
considerations. The most celebrated example of this is the
Affleck-Dine-Seiberg superpotential for SQCD, i.e. Supersymmetric
$SU(N_c)$ gauge theory with $N_f$ quarks ($1\leq N_f < N_c$) in the fundamental
representation \cite{ads}. The new techniques which are the object of this 
review allow a more systematic derivation of these non-perturbative
corrections to $W_{\mathrm{eff}}$ applicable in more general situations.
 
What we have seen is that one can write
\beq
     W_{\mathrm{eff}} = \lambda_1 X_1 + \lambda_2 X_2 +\dots +
     W_{\mathrm{non-pert}}(X_1, X_2, \dots, \Lambda_1, \Lambda_2,
    \dots),  \label{inter}
\eeq
where we have introduced more than one scale $\Lambda_s$ to allow for
the case where the gauge group is not simple. Let us focus on, say, 
the dependence on $\lambda_1$. Clearly, we can integrate out the field 
$X_1$ at low energy (where the superpotential piece dominates) by
solving
\beq
     \frac{\partial}{\partial X_1}W_{\mathrm{eff}} = 0.
\eeq
This is tantamount to doing a Legendre transform:
\beq
    \lambda_1 = - \frac{\partial}{\partial X_1}W_{\mathrm{non-pert}}. 
    \label{la1x1}
\eeq
Solving (\ref{la1x1}) for $X_1$ in terms of $\lambda_1$ and the other 
variables and substituting into (\ref{inter}) one obtains
$W_{\mathrm{eff}}$ where now the dependence on $\lambda_1$ will be
complicated because we have integrated out its partner.
Thus we can think of $\lambda_1$ and $X_1$ as forming a canonical
pair. One could integrate out all the composite fields $X_r$ to give 
$W_{\mathrm{eff}}$ solely in terms of the couplings $\lambda_r$ and
the scales $\Lambda_s$.

The Legendre transform is clearly invertible and thus, contrary to our
``Wilsonian'' intuition, as long as we are only interested in F-terms
we do not lose any information integrating out a field and, in fact,
we can integrate it back in by reversing the procedure:
\beq
    \langle X_r\rangle = \frac{\partial}{\partial\lambda_r}
    W_{\mathrm{eff}}. \label{xla}
\eeq
Note that this equation will be relevant later in a slightly different way:
if we happen to know the v.e.v. of $X_r$ in some other way, 
eq. (\ref{xla}) can be used
to determine $W_{\mathrm{eff}}$ as we will discuss in Section 7.

So far the scales $\Lambda_s$ have played a passive
role. However, the recent developments depend crucially on the fact
that we can also introduce a canonical conjugate for each $\Lambda_s$.
We can use the perturbatively exact definition of the holomorphic
scale (\ref{biglambda}) to rewrite (for each $s$) the gauge kinetic 
term as a (tree level plus one loop) contribution to 
the Wilsonian superpotential:
\beq
      \frac{\tau(\mu)}{16 \pi} \tr W^\alpha W_\alpha \equiv 
      \beta\log(\Lambda/\mu) S,
\eeq
where $S$ is the glueball superfield
\beq
     S = - \frac{1}{32 \pi^2} \tr(W^\alpha W_\alpha) =
           \frac{1}{16 \pi^2} \tr(\lambda^\alpha \lambda_\alpha + \dots). 
           \label{glueS}
\eeq
The field $S$ is a chiral superfield whose lowest component is
proportional to the gluino bilinear $\tr \lambda^\alpha \lambda_\alpha$.

The linearity principle still applies
in this case. While $S$ does not appear in the original superpotential
one can ``integrate it in'' by solving for $\Lambda$:
\beq
     \langle S\rangle = 
     \frac{1}{\beta} \Lambda \frac{d}{d\Lambda} W_{\mathrm{eff}}.  
     \label{soflambda}
\eeq

Notice that even without matter fields there is a superpotential for
$S$: the so-called Veneziano-Yankielowicz (VY) superpotential~\cite{vy}.
Consider, as an example, pure $SU(N)$ SYM theory for which $\beta =
3N$. By the assumption of confinement, we expect that all degrees of 
freedom are massive and thus the effective superpotential at low energy
can only be a constant\footnote{A constant superpotential does not break
rigid supersymmetry.}.
Dimensional analysis shows that $W_{\mathrm{eff}} = a \Lambda^3$
for some numerical constant $a$. Using (\ref{soflambda}) yields 
$\langle S \rangle= (a/N) \Lambda^3$. 
From an explicit instanton computation we know that
\beq
      \langle S\rangle = \Lambda^3,  \label{corr}
\eeq
that is, $a=N$.\footnote{The computation of the exact numerical 
coefficient is a subtle issue. To derive the correct result
(\ref{corr}) one needs to perform a computation at weak coupling. We
refer to the extensive literature on the subject reviewed in 
\cite{instant}. (See also \cite{Finnell:1995dr}.)} 

We can now express the non-perturbative 
superpotential as a function of $S$ as
\beq
     W_{\mathrm{non-pert}} = - NS\log\frac{S}{\mu^3} + NS.
\eeq

Sometimes one also writes in a somewhat mixed notation
\beq
     W_{\mathrm{eff}} =  W_{\mathrm{non-pert}} + 
      3NS \log\frac{\Lambda}{\mu} =   \label{WVY}
      NS \left(1 - \log\frac{S}{\Lambda^3} \right)  \equiv W_{VY},
\eeq
usually referred to as the VY superpotential.
The advantage of this seemingly roundabout notation is that upon
minimizing $W_{VY}$ with respect to $S$ one recovers (\ref{corr}).
Thus in the following we will consider $W_{\mathrm{eff}}$ as dependent
on $S$ although its ``natural'' variable should be $\Lambda$.

The general situation is summarized in Table~\ref{tab:1}
\begin{table}[htbp]
  \begin{center}
    \begin{tabular}{|l|c|c|c|c|c|c|}
      \hline
      Couplings & $\beta_1\log\Lambda_1/\mu$ & $\beta_2\log\Lambda_2/\mu$ & 
      \dots & $\lambda_1$ & $\lambda_2$ & \dots\\
      \hline
      Fields & $S_1$ & $S_2$ &  \dots & $X_1$ & $X_2$ & \dots\\
      \hline
    \end{tabular}
    \caption{Fields and their corresponding couplings}
    \label{tab:1}
  \end{center}
\end{table}

Given the F-terms as functions of one of the variables for
each column, one can obtain the
dependence on the other variables by performing a Legendre transform.

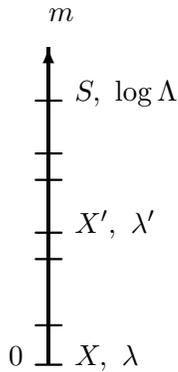
\begin{figure}
  \begin{center}
    \begin{picture}(60,60)(0,0)
      \thicklines
      \put(0,0){\vector(0,1){120}}
      \put(0,130){\makebox{$m$}}      
      \path(-5,0)(5,0)\path(-5,15)(5,15)
      \put(10,0){\makebox{$X,\ \lambda$}}
      \put(-15,0){\makebox{$0$}}
      \path(-5,50)(5,50)\path(-5,40)(5,40)
      \put(10,50){\makebox{$X',\ \lambda'$}}
      \path(-5,100)(5,100)\path(-5,70)(5,70)\path(-5,80)(5,80)
      \put(10,100){\makebox{$S,\ \log\Lambda$}}
    \end{picture}
    \end{center}
  \caption{The different energy scales}
  \label{fig:sca}
\end{figure}

Let us assume that the mass spectrum for these fields is as given in
Fig.~\ref{fig:sca}.
Generically, all glueball fields $S$ will be massive whereas we can
split the remaining invariants into massive ones, schematically denoted
by  $X^\prime$ with couplings $\lambda^\prime$, and massless ones $X$
with couplings $\lambda$.
The ``most Wilsonian'' thing to do would be to express
the F-terms as functions of the massless fields $X$, the couplings
$\lambda^\prime$ and the holomorphic scale $\Lambda$.

What we will see in the following section however is that the new
techniques allow us to determine the F-terms as functions of
$\lambda$, $\lambda^\prime$ and $S$ instead. 
Curiously, expressing the
F-terms in this set of variables reveals simple dependences on other
parameters of the theory such as the number of colors or the number of
flavors that are obscured in the initial description. 

\section{Diagrammatic computation of the glueball superpotential}

In this section we will describe the first of the two main approaches to
the calculation of the glueball superpotential, originally presented
in \cite{dglvz}. This approach is essentially perturbative in nature,
and we will illustrate it by means of an example. We will follow the 
discussion of \cite{dglvz} very closely and make some additional
simplifications along the way for the sake of clarity.

The computation of \cite{dglvz} is based on the ideas
put forward by Dijkgraaf and Vafa in \cite{dv}, where it was conjectured
that the effective superpotential of a gauge theory with matter in
the adjoint of $U(N)$ and an arbitrary tree level 
superpotential $W_{\mathrm{tree}}$ 
could be fully captured by a bosonic matrix model with a \emph{potential} 
coinciding with $W_{\mathrm{tree}}$. More specifically, one is instructed
to compute the \emph{planar} contribution\footnote{We will see in detail 
how this is done.} $\ff$ to the free energy of the matrix model:
\beq
     e^{-\frac{1}{\epsilon^2}\ff}= 
     \int\! d M\; e^{-\frac{1}{\epsilon}W_{\mathrm{tree}}(M)},
\eeq
where $M$ is a $\hat N\times \hat N$ hermitian matrix. Taking the
large $\hat N$ limit while keeping $S = \epsilon \hat N$ constant 
singles out the planar diagrams and yields
$\ff$ as a function of $S$ and the couplings in $W_{\mathrm{tree}}$.
The effective superpotential for the gauge theory\footnote{For simplicity we
limit ourselves to discuss the unbroken phase at this stage.}
expressed in terms of these variables is then
\beq
     W_{\mathrm{eff}} = W_{VY} + W_{DV},
\eeq
where $W_{VY}$ is the same as (\ref{WVY}) and
\beq
    W_{DV} = N \pd{}{S} \ff.  \label{bigdeal}
\eeq
While the original conjecture was
based on a chain of string dualities, the computation of \cite{dglvz}
is purely field theoretic, the reduction to a matrix model computation
being a direct consequence of ${\cal N}=1$ SUSY in four dimensions, as 
we will show below. 

Consider the gauge theory described above and for simplicity 
take $W_{\mathrm{tree}}$ to be a cubic superpotential:
\beq 
      W_{\mathrm{tree}} = \frac{m}{2} \tr \Phi^2 + \frac{g}{3} \tr \Phi^3.
      \label{cubic} 
\eeq
The reader should not confuse $g$ with the gauge coupling that will
always be ``hidden'' in $\Lambda$ in the following.
Note that renormalizability is not an issue here, since on the one
hand the computation we will review is not sensitive to the degree
of the superpotential, and on the other hand the theory we consider
could be thought of as an effective field theory of an underlying
renormalizable theory. Thus, the generalization to an arbitrary
superpotential is straightforward.

Thought as a polynomial in a complex variable $z$, the superpotential
$W_{\mathrm{tree}}(z) = \frac{m}{2} z^2 + \frac{g}{3}z^3$ 
has two extrema\footnote{Here and in the following we make an abuse of 
notation replacing $\Phi$ by $z$ in $W_{\mathrm{tree}}$ but refraining from 
taking the trace.}: $z=0$ and $z=-m/g$. The
classical vacua are constructed by setting each (complex) eigenvalue of the
$N\times N$ matrix $\langle \Phi \rangle$ equal to one of the two extrema. This
generically breaks the gauge symmetry from $U(N)$ to $U(p)\times
U(N-p)$. In this section we will study only the unbroken $U(N)$ phase,
i.e. we set the v.e.v. of $\Phi$ identically to zero.  

The goal is to compute the effective superpotential
$W_{\mathrm{eff}}(S,g,m)$, as discussed at the end of Section 3.
Apart from the non-perturbative VY piece (\ref{WVY}) already present
in the pure gauge theory, the only contributions to 
$W_{\mathrm{eff}}$ come from the perturbative evaluation of the path
integral over the chiral superfield in a constant background for the
vector superfield. 

Note that this is \emph{not} in contradiction
to the non-renormalization theorem, since when one integrates out $S$,
all terms in $W_{\mathrm{eff}}$ appear as powers of the holomorphic scale
$\Lambda$. Thus it is true that $W_{\mathrm{eff}}$ is a perturbative
series in the coupling $g$, but each term is accompanied by a 
positive power of
the essentially non-perturbative object $\Lambda$ coming from the
gauge sector. Were we to decouple the gauge sector from the chiral
superfield, we would have to take $\Lambda$ to zero, which would make
$W_{\mathrm{eff}}$ vanish.

We choose the simplest possible background for our purposes, i.e. a 
constant gaugino $\lambda_\alpha$ and a vanishing gauge field, $A_\mu=0$.
In order for the above background to satisfy trivially the equations
of motion, we can set to zero the complex conjugate gaugino, 
$\bar{\lambda}_{\dot\alpha}=0$, employing the usual trick of thinking
of $\lambda_\alpha$ and $\bar{\lambda}_{\dot\alpha}$ as independent
objects. This is justified as long as we are interested in 
$\bar{\lambda}_{\dot\alpha}$ independent quantities such as the 
F-terms.\footnote{It is possible to take a different approach and keep
all generality in the background and in the superpotential, while using
the full power of supergraph calculus (R.A. thanks D.~Zanon for an
interesting discussion on this issue). This
would certainly be appropriate in an exhaustive review, however
here we prefer not to introduce the more advanced material required
for this treatment and instead we stick to a more intuitive approach
as in \cite{dglvz}.}

In superfield terms, the background simply reads:
\beq
     V = - \sqrt{2} i \bar{\theta}^2 \theta \lambda, \qquad \qquad 
     W_\alpha = - \sqrt{2} i \lambda_\alpha.
\eeq
There is a further restriction to be obeyed by the constant 
gluino background -- a SUSY variation
of the gauge field, $\delta A_\mu= i \bar{\xi} \bar{\sigma}_\mu \lambda$,
is required to be pure gauge, thus implying that the matrix valued 
$\lambda_\alpha$ satisfy:
\beq
    \left\{ \lambda_\alpha, \lambda_\beta \right\} =0. \label{bkgnd}
\eeq 

In this background, covariant spacetime derivatives reduce to ordinary 
partial derivatives, the gluino superfield $W_\alpha$ is constant (and thus
covariantly constant) and, most importantly, traces of more than two
$W_\alpha$ vanish. This comes about because eq. (\ref{bkgnd}) implies that,
for instance, $\tr W_\alpha W_\beta W_\gamma$ must be totally antisymmetric
in the spinorial indices, which take only two values. A further and less
obvious consequence of (\ref{bkgnd}) is that the glueball superfield $S$, 
as given in (\ref{glueS}), satisfies $S^h=0$, where $h$ is the dual
Coxeter number of the gauge group. This issue will be discussed in more
detail in the next section, but it should be clear already at this stage 
that $S$ is nilpotent because there are only a finite number of 
Grassmann variables in $W_\alpha$.

We want to compute the holomorphic contribution to the partition function:
\beq
      Z=\int {\cal D}\Phi{\cal D}\bar\Phi \; e^{i S}, \label{pathtot}
\eeq
where the action is:
\beq
      S=\wholesp \tr \left(e^{-V}\bar\Phi e^V \Phi \right)
      + \halfsp W_{\mathrm{tree}}(\Phi) + \halfspbar \bar{W}_{\mathrm{tree}}
      (\bar{\Phi}).
\eeq
In the kinetic term we have written the adjoint action on $\Phi$ explicitly, 
i.e. in terms of $N\times N$ matrices.
Equivalently, one can define the adjoint action of $V$ on $\Phi$ as
$\ad V \Phi = {[V, \Phi]}$. One can then exponentiate it by formal 
power series and check that:
\beqs
   e^{\ad V} \Phi & = & 1 + \ad V \Phi + \frac{1}{2} (\ad V)^2 \Phi + \dots
   \nonumber \\
    & = & 1 + {[V, \Phi]} + \frac{1}{2} {[V, [V, \Phi]]}+ \dots = 
   e^V \Phi e^{-V}.
\eeqs
Both ways of writing this expression will be used.

As far as the computation of the F-terms is concerned, we can set
$\bar g=0$ and $\bar m=1$ independently of $g$ and $m$, 
since they are not going to enter any holomorphic quantity\footnote{
Notice that setting $\bar m=1$ will imply that some of the intermediate 
formulas that follow will not be manifestly dimensionally correct. The final
(holomorphic) result will of course be.}.
In this way the overall $\bar \Phi$ dependence will be quadratic.
We can thus write:
\beq
      Z=\int {\cal D}\Phi \; e^{i \halfsp W_{\mathrm{tree}}(\Phi)}
      \int {\cal D}\bar\Phi e^{i S_0}, \label{zeta}
\eeq
where
\beq
      S_0 = \wholesp \tr \left(e^{-V}\bar\Phi e^V \Phi \right)
       + \halfspbar  \frac{1}{2} \tr \bar{\Phi}^2.
\eeq
We start by evaluating the (Gaussian) path integral over $\bar \Phi$.

The action $S_0$ can be reformulated by defining:
\beq
       \tilde \Phi \equiv e^{-V}\bar\Phi e^V = e^{-\ad V}\bar\Phi =
       \bar\Phi e^{\ad V}.
\eeq
Using the covariant fermionic derivatives defined in (\ref{nablas}), where
$V$ should be intended in the adjoint representation
(i.e. $\nabla_\alpha =  e^{-\ad V} D_\alpha  e^{\ad V}$) the fields now
obey $\bar{\nabla}_{\dot{\alpha}}\Phi = 0$ and 
${\nabla}_\alpha \tilde\Phi  = 0$. More importantly, the path integral
over $\bar \Phi$ can be safely replaced by a path integral over $\tilde \Phi$,
since no anomalies can arise, $\Phi$ being in the adjoint representation.

Finally, to complete the square, we need to turn the integral over antichiral
superspace into an integral over full superspace:
\beqs
S_0 & = & \wholesp \tr (\tilde \Phi \Phi) + \halfspbar \frac{1}{2}
          \tr {\tilde \Phi}^2 \nonumber \\
    & = & \wholesp \tr \left(\tilde \Phi \Phi + \frac{1}{2} \tilde\Phi
          \frac{1}{\nabla^2} \tilde \Phi \right),
\eeqs
where we have used the fact that $\int \mathrm{d}^2 \theta$
can be replaced by $\nabla^2$ when 
inside the spacetime integral and the trace. 

Note that $\nabla^2$ is not invertible in all of superspace (for instance
it vanishes on covariantly antichiral superfields like $\tilde\Phi$), 
but we can define its inverse 
up to an element of its kernel. In the literature, it is also often written
equivalently either as $({\bar\nabla}^2 \nabla^2)^{-1} {\bar\nabla}^2$
or as ${\bar\nabla}^2 (\nabla^2 {\bar\nabla}^2)^{-1}$. The first expression
is obviously the left-inverse of $\nabla^2$, while the second is the
right-inverse, and again they coincide up to an element of the kernel
as can be checked by conjugation.

Completing the square, the action can thus be rewritten as:
\beq
      S_0 = \wholesp \frac{1}{2}\left [\tr (\tilde \Phi +\nabla^2 \Phi)
              \frac{1}{\nabla^2}  (\tilde \Phi +\nabla^2 \Phi)  
              -  \tr (\Phi\nabla^2 \Phi) \right].
\eeq
Now the first piece is trivially a quadratic integral on antichiral
superspace for the shifted
antichiral superfield ${\tilde\Phi}'=\tilde \Phi +\nabla^2 \Phi$, 
while the second piece can be turned into an integral over chiral superspace:
\beq
     S_0= \halfspbar \frac{1}{2} \tr ({\tilde\Phi}')^2 
         - \halfsp  \frac{1}{2} \tr (\Phi{\bar\nabla}^2 \nabla^2 \Phi ).
\eeq
The path integral over ${\tilde\Phi}$ can be harmlessly turned into 
a path integral over ${\tilde\Phi}'$, which is Gaussian and independent
of the background, thus contributing just a constant to the partition
function $Z$.

Returning to (\ref{zeta}), we are left with a chiral path 
integral over $\Phi$. Contrary to the 
integral over $\bar \Phi$, the action here contains interactions, forcing
us to a perturbative evaluation. We will however encounter a remarkable
simplification due to supersymmetry which will lead us to consider
a matrix model.

To obtain the explicit form of the propagator for $\Phi$ we need to
expand the expression ${\bar\nabla}^2 \nabla^2 \Phi$. Recalling the
definitions, we have that (in this simple background):
\beq
   {\bar\nabla}^2 \nabla^2 \Phi  =  {\bar D}^2 e^{-\ad V} D^2 e^{\ad V}\Phi
    = - \Box \Phi - \ad W^\alpha D_\alpha \Phi = 
    - \Box \Phi - \{ W^\alpha,  D_\alpha \Phi\},
\eeq 
where in this case the adjoint action gives rise to an anticommutator.

The path integral (\ref{zeta}) is thus reduced to the following
chiral path integral:
\beq
     Z'=\int {\cal D}\Phi \; e^{i \halfsp \frac{1}{2} \tr \Phi
       (\Box + \ad W^\alpha D_\alpha + m )\Phi + \frac{g}{3} \tr \Phi^3}.
       \label{pathchi}
\eeq
The quantity $Z'$ defined above is the purely holomorphic part of the
full partition function (\ref{pathtot}), i.e. the piece that depends
only on the holomorphic couplings and on the holomorphic background
$\lambda_\alpha$. We can thus associate it with the coupling dependent
part of the effective superpotential:
\beq 
     Z'\equiv e^{i\halfsp W_{{DV}}(S,g,m)}. \label{zedpr}
\eeq

The procedure outlined above is quite general, and it applies to any
group, matter content, and tree level superpotential.

There are however two important points to make. The first one is that, since
$Z'$ is the partition function obtained by 
integrating over the matter fields,
there is no hope of getting the piece of the effective
superpotential pertaining to the pure gauge low-energy dynamics,
i.e. the Veneziano-Yankielowicz piece (\ref{WVY}), in this way. 
This piece is 
independent of the matter couplings $g$ and $m$ and cannot arise from a
perturbative evaluation of (\ref{pathchi}). 

The second point is that the perturbative evaluation of (\ref{pathchi})
has to take into account the fact that the classical gluino background
satisfies (\ref{bkgnd}). As said before, this implies that the gauge invariant
gluino bilinear $S$ is nilpotent, i.e. $S^h=0$. For instance, in the $U(N)$
case considered above, $W_{{DV}}(S,g,m)$ would have a truncated
Taylor expansion which stops at $S^N$. 
Of course, this makes no sense when $S$ is considered as an effective
chiral superfield, and we will indeed find superpotentials 
$W_{{DV}}(S,g,m)$ with an infinite Taylor expansion, of which 
the first $h-1$ terms coincide with the ones computed by the method
above. 

In some cases however, one can compute
$W_{{DV}}(S,g,m)$ to arbitrary powers of $S$ without any
obstacle. This is possible for instance in the case considered here
because the dependence on $N$ factorizes in front
of $W_{{DV}}$, as stated in (\ref{bigdeal}). 
Since $\ff$ is independent of $N$ we can compute its power series 
expansion to arbitrary order by formally considering $N$ large and then
substitute into (\ref{bigdeal}) with the actual value of $N$.
In other cases discrepancies
with exact field theory results can show up at order $h$, 
as discussed in~\cite{contro}. 

We now proceed to evaluate perturbatively (\ref{pathchi}) and to show
that in the case of $U(N)$ with adjoint matter, it reduces to the
evaluation of a matrix model integral.

First of all, it will be useful to go to momentum space. Following 
\cite{dglvz}, we introduce not only the usual (bosonic) 4-momentum $p_\mu$, 
but also a fermionic momentum $\pi^\alpha$ conjugate to $\theta_\alpha$.
In the quadratic part of the action in (\ref{pathchi}), we Wick rotate
to Euclidean space and Fourier transform in both the bosonic and
fermionic coordinates by substituting:
\beq
     \Box \rightarrow -\Box_{Eucl} \rightarrow p^2, 
     \qquad \qquad D_\alpha \rightarrow \pi_\alpha.
\eeq
It may be surprising that we do not 
include a factor of $i$ in the last definition but this is consistent with our
choice of hermiticity for $D_\alpha$.

We now turn to the propagator 
$\langle\Phi^a_c(p,\pi) \Phi^b_d(-p,-\pi)\rangle$ for $\Phi$. 
In the following $\Phi$ is thought of as a $N\times N$ matrix 
and the roman indices $a, b, \dots$ run from $1$ to $N$.
Looking at the kinetic term in (\ref{pathchi}) we read 
off the inverse propagator:
\beq
     \Gamma^{a b}_{c d}(p,\pi) = (p^2 + m)\delta^a_d \delta^b_c +
     ({W^\alpha}^a_d   \delta^b_c  - {W^\alpha}^b_c \delta^a_d )\pi_\alpha.
\eeq

We can write the propagator itself using a Schwinger parameter $s$:
\beq
     \Delta(p,\pi) = \int_0^{\infty}\! d s\; e^{-s (p^2+m+\ad W^\alpha
                     \pi_\alpha)}, \label{proppy}
\eeq
where, as usual, the exponential
is defined by power series 
expansion\footnote{Notice that $\{W^\alpha, \pi_\alpha \Phi\} =
{[W^\alpha\pi_\alpha, \Phi]}$.}. 
Actually, since $\pi_\alpha$ is a Grassmann two-component
spinor, the expansion stops at the second order. Thus we have:
\beq
    e^{-s G}=1-s G +\frac{1}{2}s^2 G^2,  \label{expanprop}
\eeq
with
\beq
     1^{a b}_{c d}\equiv\delta^a_d \delta^b_c, \qquad  G^{a b}_{c d}\equiv
     ({W^\alpha}^a_d   \delta^b_c  - {W^\alpha}^b_c \delta^a_d )\pi_\alpha.
\eeq

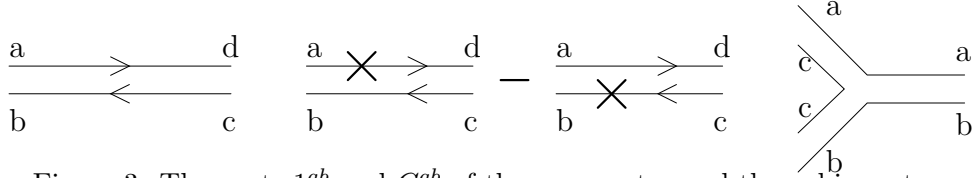
\begin{figure}[tb]
  \begin{center}
    \setlength{\unitlength}{3.5pt}
    \begingroup\makeatletter\ifx\SetFigFont\undefined%
    \gdef\SetFigFont#1#2#3#4#5{%
      \reset@font\fontsize{#1}{#2pt}%
      \fontfamily{#3}\fontseries{#4}\fontshape{#5}%
      \selectfont}%
    \fi\endgroup%
    {\renewcommand{\dashlinestretch}{30}
      \begin{picture}(24,10)(5,0)
        \thinlines
        \path(0,4)(24,4)
        \path(0,7)(24,7)
        \path(11,8)(13,7)(11,6)
        \path(13,5)(11,4)(13,3)
        \put(0,8){\makebox(0,0)[lb]{\smash{{{\SetFigFont{12}{0}{\rmdefault}{\mddefault}{\updefault}a}}}}}
        \put(23,8){\makebox(0,0)[lb]{\smash{{{\SetFigFont{12}{0}{\rmdefault}{\mddefault}{\updefault}d}}}}}
        \put(0,0){\makebox(0,0)[lb]{\smash{{{\SetFigFont{12}{0}{\rmdefault}{\mddefault}{\updefault}b}}}}}
        \put(23,0){\makebox(0,0)[lb]{\smash{{{\SetFigFont{12}{0}{\rmdefault}{\mddefault}{\updefault}c}}}}}
      \end{picture}
      }
    {\renewcommand{\dashlinestretch}{30}
      \begin{picture}(45,10)(0,0)
        \thicklines
        \path(7.5,8.5)(4.5,5.5)
        \path(7.5,5.5)(4.5,8.5)
        \thinlines
        \path(0,4)(18,4)
        \path(0,7)(18,7) 
        \path(11,8)(13,7)(11,6)
        \path(13,3)(11,4)(13,5)
        \put(0,8){\makebox(0,0)[lb]{\smash{{{\SetFigFont{12}{0}{\rmdefault}{\mddefault}{\updefault}a}}}}}
        \put(17,8){\makebox(0,0)[lb]{\smash{{{\SetFigFont{12}{0}{\rmdefault}{\mddefault}{\updefault}d}}}}}
        \put(0,0){\makebox(0,0)[lb]{\smash{{{\SetFigFont{12}{0}{\rmdefault}{\mddefault}{\updefault}b}}}}}
        \put(17,0){\makebox(0,0)[lb]{\smash{{{\SetFigFont{12}{0}{\rmdefault}{\mddefault}{\updefault}c}}}}}
        \thicklines
        \path(31.5,5.5)(34.5,2.5)
        \path(31.5,2.5)(34.5,5.5)
        \thinlines
        \path(27,4)(45,4)
        \path(27,7)(45,7) 
        \path(38,8)(40,7)(38,6)
        \path(40,3)(38,4)(40,5)
        \put(27,8){\makebox(0,0)[lb]{\smash{{{\SetFigFont{12}{0}{\rmdefault}{\mddefault}{\updefault}a}}}}}
        \put(44,8){\makebox(0,0)[lb]{\smash{{{\SetFigFont{12}{0}{\rmdefault}{\mddefault}{\updefault}d}}}}}
        \put(27,0){\makebox(0,0)[lb]{\smash{{{\SetFigFont{12}{0}{\rmdefault}{\mddefault}{\updefault}b}}}}}
        \put(44,0){\makebox(0,0)[lb]{\smash{{{\SetFigFont{12}{0}{\rmdefault}{\mddefault}{\updefault}c}}}}}
        \thicklines
        \path(21,5.5)(24,5.5)
      \end{picture}
      }
    {\renewcommand{\dashlinestretch}{30}
      \begin{picture}(18,18)(-5,4.5)
        \thinlines
        \path(0,18)(7.5,10.5)(18,10.5)
        \path(0,0)(7.5,7.5)(18,7.5)
        \path(0,4.24)(5,9)(0,13.76)
        \put(17,12){\makebox(0,0)[lb]{\smash{{{\SetFigFont{12}{0}{\rmdefault}{\mddefault}{\updefault}a}}}}}
        \put(17,4){\makebox(0,0)[lb]{\smash{{{\SetFigFont{12}{0}{\rmdefault}{\mddefault}{\updefault}b}}}}}
        \put(0,11){\makebox(0,0)[lb]{\smash{{{\SetFigFont{12}{0}{\rmdefault}{\mddefault}{\updefault}c}}}}}
        \put(0,6){\makebox(0,0)[lb]{\smash{{{\SetFigFont{12}{0}{\rmdefault}{\mddefault}{\updefault}c}}}}}
        \put(3,17){\makebox(0,0)[lb]{\smash{{{\SetFigFont{12}{0}{\rmdefault}{\mddefault}{\updefault}a}}}}}
        \put(3,0){\makebox(0,0)[lb]{\smash{{{\SetFigFont{12}{0}{\rmdefault}{\mddefault}{\updefault}b}}}}}
      \end{picture}
      }
    \caption{The parts $1^{a b}_{c d}$ and $G^{ab}_{cd}$ of the
    propagator and  the cubic vertex.}
    \label{lines}
  \end{center}
\end{figure}

In order to compute $Z'$ we have to evaluate vacuum diagrams
of the theory defined by the action in (\ref{pathchi}). Because the matter 
field $\Phi$ is in the adjoint, the diagrams can be written using
the double line notation~\cite{thooft} (see Fig. \ref{lines}). 
The identity in (\ref{expanprop}) is 
represented by two lines with opposite orientations, while the $G$
and $G^2$ terms are represented by the same lines with one or two $W^\alpha$
insertions respectively. The cubic vertex, being a trace, is represented
by three lines joining three different two-line propagators. 

Having established the above, one can assign a topology 
to any given graph in the 
usual way. The graph represents a discretization of a two-dimensional
closed surface over which it can be drawn without any twisting or
intersection. Call $V$ the number of vertices, $P$ the number of 
propagators and $F$ the number of faces i.e. regions encircled by a 
line, including the external one. Then the Euler characteristic of
the surface is given by:
\beq
      \chi = V-P+F, \label{euler}
\eeq
and it is a topological invariant. For instance, a sphere has Euler
characteristic $\chi=2$ and a torus has $\chi=0$. Actually, every handle
one ``adds'' to the surface reduces $\chi$ by $2$, so that the following
formula is valid for closed orientable surfaces:
\beq
     \chi = 2 - 2h,
\eeq 
where $h$ here is the number of handles, also referred to as
the genus of the surface.

Allowing for open and/or unoriented surfaces, the Euler characteristic
can take odd values (but always keeping $\chi\leq 2$ for a single
surface). For instance, $\chi=1$ for a disk (open, oriented surface) 
and for the projective plane (closed, unoriented surface).

An important result that is obtained even before starting any computation
is that \emph{only planar graphs contribute to the superpotential}. 
This can be understood
as follows. A generic (connected)
vacuum diagram has $V$ vertices and $P$ propagators.
Every vertex carries with it an integral over chiral (coordinate) superspace, 
while every propagator, after Fourier transforming it, carries 
an integral over chiral momentum superspace. All but one of the coordinate
superspace integrals give delta functions enforcing momentum conservation
(both bosonic and fermionic) at every vertex, the remaining overall
chiral superspace integral being the one which appears in the effective
action in (\ref{zedpr}). 
Solving for the delta functions, one is left with $P-V+1\equiv L$
momentum superspace (loop) integrals. 
So far this is just ordinary loop counting applied also to 
the Grassmann momenta.
Thus one has to integrate
over $2L$ Grassmannian momenta $\pi_\alpha$. 
In the integrand, these momenta necessarily
appear in bilinears together with the gluino background $W^\alpha$
because of the specific form of the propagator (\ref{proppy}).
Since traces of more than two $W^\alpha$ vanish due to (\ref{bkgnd}),
the result of the integration is non-zero only if we are able to write
at least $L$ traces. Remembering now that the number of traces coincides
with the number of closed lines, or ``faces'' of the graph, 
we get the following constraint on the graph's topology:
\beq
     F\geq L= P-V+1 \qquad  \Rightarrow \qquad \chi \geq 1.
\eeq
This means that in the case of the matter field in the adjoint of
$U(N)$, we are concerned only with graphs with the topology of a sphere, 
that is planar graphs. All the other ones are exactly zero
already at finite $N$.
In other words, planarity is an exact consequence of
supersymmetry, and not a leading order 
approximation due to a large $N$ limit.

When open or unoriented graphs have to be taken into account, 
also graphs with one boundary or one cross cap give a non-vanishing 
result. Graphs with boundaries arise in theories with matter
fields in the fundamental representation, 
whose propagator has only one color line (possibly supplemented
by a flavor line), while unoriented graphs arise in theories 
where the gauge group is $SO(N)$ or $Sp(N)$.
Note also that graphs of higher genera become important when the
theory is coupled to gravity, in the background of which they
no longer vanish.

Let us now compute in detail one diagram in order to give a feeling
of how a field theory computation can be reduced to
a matrix model one.

Let us compute the ``stop sign'' diagram of Fig. \ref{fig1}:
\begin{figure}[htb]
  \begin{center}
    \setlength{\unitlength}{2pt}
    {\renewcommand{\dashlinestretch}{30}
      \begin{picture}(48,48)(0,0)
        \put(24,24){\circle{48}}
        \put(24,24){\arc{42}{3.2156}{6.2091}}
        \path(45,25.5)(3,25.5)
        \put(24,24){\arc{42}{0.0740}{3.0676}}
        \path(45,22.5)(3,22.5)
      \end{picture}
      }
    
    \caption{The stop sign diagram.}
    \label{fig1}
  \end{center}
\end{figure}
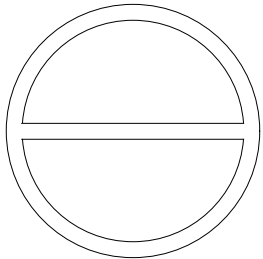

This diagram yields:
\beq
    \frac{g^2}{6}\int\!\! \frac{\mathrm{d}^4p_1}{(2\pi)^4}
    \frac{\mathrm{d}^4p_2}{(2\pi)^4}
    \mathrm{d}^2\pi_1\mathrm{d}^2\pi_2 
    \Delta^{a b}_{c d}(p_1,\pi_1) \Delta^{e d}_{a f}(p_2,\pi_2)
    \Delta^{c f}_{e b}(-p_1-p_2,-\pi_1-\pi_2) 
    \label{stopsign}
\eeq
The overall factor of $\frac{g^2}{6}$ is understood as follows. 
We have a $\frac{1}{2}$ from the Taylor expansion to second order 
of the exponential of the interaction, a $(g/3)^2$ from the two  
interaction vertices, and a $3$ from the number of independent ways one
can contract the $\Phi$ to give a planar graph\footnote{There are also 
three more contractions that do not yield a planar graph and thus vanish
according to the above argument. This explains the discrepancy between 
(\ref{stopsign}) and the factor that would arise 
in a single component $\phi^3$ theory.}.

In terms of the three Schwinger parameters, we can rewrite the diagram as:
\beq
     \frac{g^2}{6}\int\! d s_1 d s_2 d s_3\; e^{-m(s_1+s_2+s_3)} 
     I_{\mathrm{b o s e}} I_{\mathrm{f e r m i}}. \label{schwint}
\eeq
The bosonic integral is straightforward, since it does not imply any 
group theory factors. It is given by:
\beqs
     I_{\mathrm{b o s e}} &=& 
     \int\!  \frac{\mathrm{d}^4p_1}{(2\pi)^4}
    \frac{\mathrm{d}^4p_2}{(2\pi)^4}\;
     e^{-s_1 p_1^2 -s_2 p_2^2 -s_3 (p_1 + p_2)^2} \nonumber \\
     &= & \frac{1}{(4\pi)^4} \frac{1}{(s_1 s_2 + s_1 s_3 + s_2 s_3)^2}.
     \label{ibose}
\eeqs
The computation of the fermionic integral is more involved, since
the integrand involves traces over the gauge group:
\beq
     I_{\mathrm{fermi}} = \!\int\!\! d^2\pi_1 d^2\pi_2
     \left(e^{-s_1 \ad W^\alpha \pi_{1 \alpha }}\right)^{a b}_{c d}\!
     \left(e^{-s_2 \ad W^\alpha \pi_{2 \alpha}}\right)^{ed}_{a f}\!
     \left(e^{s_3 \ad W^\alpha (\pi_{1 \alpha }+
     \pi_{2 \alpha })}\right)^{cf}_{e b}
\eeq
The integral over the fermionic momenta is saturated only if
we bring down two $\pi_1$ and two $\pi_2$:
\beq
     \int\! d^2\pi_1 d^2\pi_2 \; \pi_1^2 \pi_2^2 =1.
\eeq
The result thus gives
an overall fourth power of the gluino $W^\alpha$. These four gluinos
have to be inserted at some point of the three color loops 
of the diagram. Since we cannot put more than two gluinos on the same line 
(otherwise the trace would vanish),
the two choices we are left with are either to put
two $W^\alpha$ on two color loops and leave the third loop without insertion,
or to put two $W^\alpha$ on one loop and one each in the two remaining
color loops. 

The latter combination leads to a term including
$\tr  W^\alpha \tr W_\alpha$, which is the effective kinetic term for
the decoupled $U(1)$. Terms of this type are interesting in the case
of broken gauge group but 
here we focus only on the effective superpotential
for the gluino bilinear $S$, i.e. we are concerned only with those
combinations where either two $W^\alpha$ or none are inserted in each
color loop. The $U(1)$ couplings will be briefly discussed in Section 7.

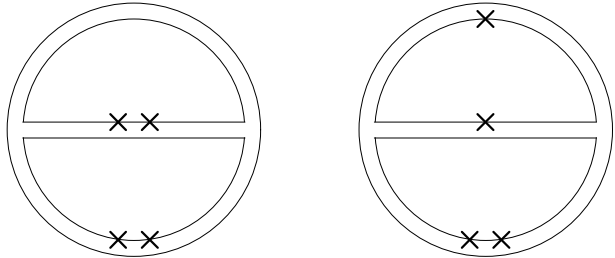
\begin{figure}[htb]
  \begin{center}
 \setlength{\unitlength}{2pt}
    {\renewcommand{\dashlinestretch}{30}
      \begin{picture}(48,48)(-60,0)
        \put(24,24){\circle{48}}
        \put(24,24){\arc{42}{3.21308}{-0.07148}}
        \path(45,25.5)(3,25.5)
        \put(24,24){\arc{42}{0.07148}{3.0701}}
        \path(45,22.5)(3,22.5)
        \thicklines
        \path(22.5,24)(25.5,27)
        \path(22.5,27)(25.5,24)
        \path(22.5,46.5)(25.5,43.5)
        \path(22.5,43.5)(25.5,46.5)
        \path(25.5,4.75)(28.5,1.75)
        \path(28.5,4.75)(25.5,1.75)
        \path(22.5,4.75)(19.5,1.75)
        \path(19.5,4.75)(22.5,1.75)
      \end{picture}
      }  
  {\renewcommand{\dashlinestretch}{30}
    \begin{picture}(48,48)(60,0)
      \put(24,24){\circle{48}}
      \put(24,24){\arc{42}{3.21308}{-0.07148}}
      \path(45,25.5)(3,25.5)
      \put(24,24){\arc{42}{0.07148}{3.0701}}
      \path(45,22.5)(3,22.5)
      \thicklines
      \path(25.5,24)(28.5,27)
      \path(28.5,24)(25.5,27)
      \path(22.5,24)(19.5,27)
      \path(19.5,24)(22.5,27)
      \path(25.5,4.75)(28.5,1.75)
      \path(28.5,4.75)(25.5,1.75)
      \path(22.5,4.75)(19.5,1.75)
      \path(19.5,4.75)(22.5,1.75)
    \end{picture}
    }      
    \caption{The stop sign with two $W_\alpha$ insertions at two of the
      propagators (left) and the stop sign with one $W_\alpha$
      insertion at two of the propagators and one insertion of two
      $W_\alpha$ at the third propagator (right).}
    \label{insertions}
  \end{center}
\end{figure}

It is easy to work out all the insertions pictorially, instead of 
developing the propagators by brute force. One can then realize
that two different kinds of diagrams contribute. The first, depicted 
on the left of Fig.~\ref{insertions},
arises when 
two propagators have two insertions and the third has none\footnote{At this
point it is important not to confuse propagators and color loops -- 
propagators can carry at most two insertions because their expansion
stops at second order.}. 
Diagrams of these type contribute to $I_{\mathrm{fermi}}$ a quantity:
\beq
     3 N (\tr W^\alpha W_\alpha)^2  \left(\frac{1}{2}s_1^2 \frac{1}{2}s_2^2
     + \frac{1}{2}s_1^2 \frac{1}{2}s_3^2 + \frac{1}{2}s_2^2 \frac{1}{2}s_3^2
     \right),
\eeq
where the factor of $N$ is given by the color loop without insertions,
$\tr 1= N$, while the factor of $3$ comes from the choice of this color
loop.

The second kind of diagram, depicted on the right of Fig. \ref{insertions}, 
arises when one propagator carries two insertions
and the other two one each. The contribution from these diagrams is:
\beq
     3 N (\tr W^\alpha W_\alpha)^2  \left( \frac{1}{2}s_1^2 s_2 s_3 
     + s_1 \frac{1}{2}s_2^2 s_3 + s_1 s_2 \frac{1}{2}s_3^2\right),
\eeq
the numerical prefactor having the same origin as above.

Summing these contributions, we get, using (\ref{glueS}):
\beq
    I_{\mathrm{fermi}} = 3N (4\pi)^4 S^2 (s_1 s_2 + s_1 s_3 + s_2 s_3)^2.
\eeq
Comparing now with the result of the bosonic integral (\ref{ibose}),
we observe the striking fact that the $s_i$ dependent part of the
numerator and the denominator
exactly cancel\footnote{Curiously, this works only in four spacetime 
dimensions, as the overall powers of $s_i$ in the numerator are 
determined by the dimension of the Weyl spinor, whereas the ones of
the denominator are determined by the dimension of spacetime.}, 
leaving:
\beq
    I_{\mathrm{b o s e}} I_{\mathrm{f e r m i}} = 3N S^2
\eeq
that is, a result independent of the Schwinger parameters $s_i$ !

Plugging this back into (\ref{schwint}), the integral over $s_1$, $s_2$
and $s_3$ becomes trivial:
\beq
    \int\! d s_1 d s_2 d s_3\; e^{-m(s_1+s_2+s_3)} =\frac{1}{m^3}.
\eeq
In other words, all that is left from the propagators is a contribution
of $\frac{1}{m}$ each, exactly what one would expect from a zero-dimensional
field theory, i.e. from a matrix model.
In particular, note that the result is finite. This is non-trivial since
after all we are dealing with a two-loop computation. 

The final result for the stop sign diagram is thus:
\beq
    \frac{1}{6} \frac{g^2 }{m^3} 3N S^2.     \label{setsun}
\eeq

\begin{figure}[htbp]
  \begin{center} 
    \setlength{\unitlength}{2pt}
    {\renewcommand{\dashlinestretch}{30}
      \begin{picture}(48,48)(0,0)
        \put(63,24){\circle{35}} 
        \put(-21,24){\circle{35}}
        \put(63,24){\arc{42}{3.21308}{3.0701}}
        \path(42,25.5)(0,25.5)
        \put(-21,24){\arc{42}{0.07148}{-0.07148}}
        \path(42,22.5)(0,22.5)
      \end{picture}
      }  
    \caption{The drum-bell diagram.}
    \label{fig:drum}
  \end{center}
\end{figure}
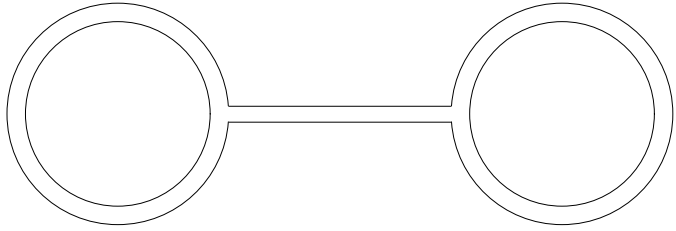

To order $g^2$ there is only another diagram, shown in Fig.~\ref{fig:drum}, 
contributing to the effective superpotential.
We leave it as an exercise to check that, after all the $W_\alpha$ insertions
are made, it contributes a quantity
\beq
    \frac{1}{2} \frac{g^2 }{m^3} 3N S^2. \label{drube}
\eeq
Summing (\ref{setsun}) and (\ref{drube}) and recalling the overall 
minus sign coming from the Wick rotation we get the full contribution 
to the effective superpotential to this order:
\beq
    W_{DV} = - 2 N \frac{g^2 }{m^3} S^2 + \dots
\eeq

In all generality, we can argue as above that for a 
planar diagram with $F$
color loops, the $S^L\equiv S^{F-1}$ term will have a prefactor of $N$
from the color loop without insertions and a factor of $F$ from the
choice of such a loop:
\beq
    I_{\mathrm{b o s e}} I_{\mathrm{f e r m i}}=F N S^{F-1} = N \pd{}{S}
    S^F. \label{partial} 
\eeq
The numerical and coupling dependent prefactor will depend on the
interaction vertex and on the combinatorics.
That the $s_i$ dependence completely cancels in the above expression
is less trivial to see, but can actually be proven in all generality
introducing a new set of $2L$ auxiliary Grassmann coordinates to enforce
the fact that one allows for at most two insertions of $W^\alpha$ in
any color loop. We preferred here to present a sample computation,
and we refer the interested reader to the original derivation \cite{dglvz}.

The relevant question now is whether we can define a matrix model
from which to extract all the information we need to compute
the effective superpotential of the gauge theory. The answer turns
out to be quite straightforward.

Let us look at the following matrix model:
\beq
     {\cal Z} = \int\! d M\; e^{-\frac{1}{\epsilon} \tr \left( \frac{m}{2} M^2
     +\frac{g}{3} M^3\right)}. \label{matmod}
\eeq
Here $M$ is an $\hat N \times \hat N$ hermitian matrix, where $\hat N$
is not related to the $N$ of the gauge theory. Actually, for the time
being the only explicit link between the matrix model and the gauge theory is 
that the potential of the matrix model is given by the tree level
superpotential $W_{\mathrm{tree}}$ of the gauge theory, after 
substituting $\Phi$ for $M$. 
Otherwise, note for instance that the matrix model is purely bosonic.
The parameter $\epsilon$ is introduced for later convenience, and plays
the role of an overall coupling constant.

Strictly speaking the potential in (\ref{matmod}) is not
well defined because it does not have a minimum. This should not be however
a matter of worry, since we will use (\ref{matmod}) to define a (formal)
perturbative expansion around its local minimum (recall that we are 
interested in the case where, on the gauge theory side, the v.e.v. of $\Phi$
does not break the gauge group). 

Indeed, we write:
\beqs
     {\cal Z}& = &\int\! d M\; e^{-\frac{1}{\epsilon} \tr \left
         ( \frac{m}{ 2} M^2 
     \right)}\sum_{n=0}^\infty \frac{(-1)^n}{n!}\left(\frac{1}{\epsilon}
     \frac{g}{3} \tr M^3\right)^n  \nonumber \\
      & = &    \sum_{n=0}^\infty\frac{(-1)^n}{n!} \left\langle 
      \left(\frac{1}{\epsilon}\frac{g}{3} \tr M^3\right)^n \right\rangle .
      \label{matexpan}
\eeqs
In the second line above, we have used the correlation functions
of the ``free'' matrix model, that is the one given by a purely quadratic
potential. The free propagator is given by:
\beq
     \langle M^a_b M^c_d \rangle = \frac{\epsilon}{ m} \delta^a_d
     \delta^c_b,
\eeq
and can also be represented by a double line as in the previous set up.
Note however that here the roman indices run between $1$ and $\hat N$.

The vacuum diagrams of the interacting theory are represented by the
correlation functions (\ref{matexpan}) of the free theory.
Every vertex is multiplied by a factor of $g/ \epsilon$. Hence a generic
diagram will give a contribution of:
\beq
     c \left(\frac{\epsilon}{ m}\right)^P \left(\frac{g}{\epsilon}
     \right)^V {\hat N}^F, \label{matgraph}
\eeq
where $c$ is a numerical coefficient including the powers of $\frac{1}{3}$
coming from the vertices, the coefficients of the Taylor expansion of 
the exponential and the various combinatorics. Again, $F$ is the number 
of index loops, i.e. traces.

The matrix model simplifies greatly when one takes the large $\hat N$ limit.
In order to do this properly, let us take simultaneously the limit
$\epsilon \rightarrow 0$, while keeping the combination:
\beq
     \epsilon \hat N = S 
\eeq
constant (it is of course \emph{not} an accident that the constant is called
$S$, we will shortly see below why).

We now solve for $\hat N$ in (\ref{matgraph}), and recalling the expression
for the Euler characteristic (\ref{euler}), we obtain for the graph:
\beq
    c\; \epsilon^{-\chi} \left(m^{-P} g^V S^F \right).
\eeq
The expression between parenthesis is finite in the limit, and 
when  $\epsilon \rightarrow 0$ the planar diagrams ($\chi=2$) will clearly 
dominate, the other ones being suppressed by powers of $1/ {\hat N}^2$.

Reexpressing the partition function for the matrix model as:
\beq
     {\cal Z} = e^{-\frac{1}{\epsilon^2} {\cal F}_{\chi=2} - 
     {\cal F}_{\chi=0} - \epsilon^2 {\cal F}_{\chi=-2} + \dots},
\eeq
we observe that the result of computing all planar diagrams is 
contained in the planar free energy $\ff \equiv {\cal F}_{\chi=2}(S,g,m)$.
For instance, the contribution to $\ff$ from the
``stop sign'' diagram considered before, is:
\beq
     - \frac{1}{ 6} \frac{1}{ m^3} g^2 S^3,
\eeq
where the power of $S$ is now simply the number of color index loops.
The cubic matrix model can be solved exactly in the planar limit, as
first shown in \cite{bipz}
and in Appendix B we indeed show how to obtain the complete  
$\ff$.

Recalling the expression (\ref{partial}) derived on the gauge theory side,
the only non-trivial pieces of
informations to compute are the numerical factors in front of every
term in the perturbative expansion in $S$. 
These are precisely the numbers contained in the function $\ff$. 
Hence the effective superpotential of
the gauge theory is given in terms of the matrix model planar free energy by:
\beq
     W_{{DV}}(S,g,m) = N \pd{}{S} \ff (S,g,m) .
     \label{wdv} 
\eeq
Let us underscore again the important point -- planarity is
achieved through the large $\hat N$ limit on the matrix model side, 
but is an exact feature at finite $N$ on the field theory side.
The reason is that on the field theory side non-planar diagrams necessarily
include traces of more than two gluino superfields, and these vanish
exactly. 

Furthermore, the only way $N$ enters in the effective superpotential
is through a multiplicative factor in (\ref{wdv}). Thus, we see that, 
even if on the gauge theory side we should trust the effective 
superpotential only up to the term $S^N$, the matrix model instructs
us to go on and compute \emph{any} diagram in the same way. 
This works in this case
because the functional form of the superpotential does not
change when $N$ is varied, and thus one could also take a large value
of $N$ on the gauge theory side. This generalization must of course rely on 
considering $S$ a dynamical effective field (the glueball superfield).

We have already stressed that the perturbative
approach does not give us a systematic way to compute the pure gauge
part of the effective superpotential, namely the Veneziano-Yankielowicz
superpotential for a pure $SU(N)$ SUSY gauge theory.
This must be added to the DV term in order to obtain the full effective
superpotential. The total effective
superpotential thus reads:
\beq
     W_{\mathrm{eff}} = W_{VY} + W_{{DV}} = 
     NS \left( 1 - \log \frac{S}{{\tilde \Lambda}^3} \right) 
     + N \pd{}{S} \ff (S,g,m). \label{wtot}
\eeq

The holomorphic scale ${\tilde \Lambda}$ 
appearing in (\ref{wtot}) is the one for the low-energy pure $SU(N)$ 
theory, when the massive matter field $\Phi$ has already been integrated
out. Using the matching of scales (\ref{matching}), we see that it is
related to the high energy scale $\Lambda$ by ${\tilde \Lambda}^3
=m \Lambda^2$. Were we to be strict and use only the high energy scale
in the VY superpotential, the matching of scales would come about
from the vacuum diagram of the free matrix model, which is given by
$S^2 \log m $.

It may sound a bit unpleasant that one must add the VY piece ``by hand''.
However, a very interesting fact well emphasized already in the earlier
literature~\cite{dv,Ooguri:2002gx} is that the VY superpotential can be
fully recovered in the matrix model if one takes into account the volume
of the $U(\hat N)$ group rotating the hermitian matrix $M$.
Essentially, one can show that:
\beq
    \langle 1 \rangle \propto \left(Vol(U(\hat N))\right)^{-1} 
    \simeq e^{\frac{{\hat N}^2}{ 2}
    \log \hat N + \dots}, \label{volsun}
\eeq
so that after substituting for $S$, the leading term looks like a
contribution to the planar free energy, and then the typical $S\log S$ term
also arises from (\ref{wdv})\footnote{After the first version of this 
review appeared, the matrix model prescription which yields exactly 
the VY superpotential was spelled out in \cite{Hailu:2004mq}.}.
Although this observation can be justified by string duality it is difficult 
to find a field theoretical argument for it.
In the framework we have been working, we have consistently
taken $\langle 1 \rangle = 1$ and added the VY piece by hand.

\section{The chiral ring}

One of the uses of the effective superpotential is that of 
determining the vacuum structure of the theory, namely the expectation
values of certain gauge invariant operators. It thus comes natural
to define an equivalence class of such operators identifying
those that have the same vacuum expectation
value in a SUSY vacuum. This leads to
the concept of the chiral ring that we now turn to discuss.

What we have seen in Section 2 is that a chiral superfield is defined by 
$\bar{D}_{\dot{\alpha}} \Phi = 0$ and its lowest component $\phi(x)$
is annihilated by $\bar{\mathcal{Q}}_{\dot{\alpha}}$:
\beq
     {[\bar{\mathcal{Q}}_{\dot{\alpha}}, \phi(x)]} = 0.
\eeq

The same facts are true for the gluino $\lambda_\alpha(x)$ which is the
lowest component of the chiral superfield $W_\alpha$, only now we must
switch commutators and anticommutators to account for the fact that
$\lambda_\alpha(x)$ is a fermion:
\beq
    \{\bar{\mathcal{Q}}_{\dot{\alpha}}, \lambda_\alpha(x) \} = 0.
\eeq

We call chiral operators those \emph{gauge invariant} operators
$\mathcal{O}(x)$ that are annihilated by the
$\bar{\mathcal{Q}}_{\dot{\alpha}}$. For, say, a bosonic operator: 
\beq
    {[\bar{\mathcal{Q}}_{\dot{\alpha}}, \mathcal{O}(x)]} = 0. 
    \label{chirallo}
\eeq
As we have
seen before, given such an operator one can always generate its
superpartners by acting on it with the $\mathcal{Q}_\alpha$ to fill up
a chiral multiplet. Vice-versa, given a gauge invariant chiral
superfield its lowest component will be a chiral operator.

Before going on studying the properties of such operators, we
should discuss how to construct them
in a gauge theory. Here we have at our disposal the gauge
\emph{variant} objects $\phi(x)$ and $\lambda_\alpha(x)$ that are
annihilated by $\bar{\mathcal{Q}}_{\dot{\alpha}}$. One could thus
construct gauge invariant composite operators, such as $\tr \lambda^2(x)$,
that are naively annihilated by $\bar{\mathcal{Q}}_{\dot{\alpha}}$ by
using the Leibniz rule. But one must be careful when defining
composite operators in quantum field theory. 
In particular, when using point-splitting
regularization, one must introduce a Wilson line to preserve gauge
invariance, as we will discuss shortly. 
One then has to check that such 
regularization does not spoil eq.~(\ref{chirallo}).
In this case it turns out that the naive expectation is correct.

Consider, for concreteness, two fields $q$ and
$\tilde q$, the lowest components of the quark chiral superfields
in SQCD, transforming in the fundamental and antifundamental
representations of the gauge group. Define the gauge invariant
composite operator 
\beqs
   \tilde q q(x) &=& 
    \lim_{\epsilon\to 0}\tilde q(x+\epsilon/2)
    \mathrm{P} \exp\left(-i\int_{x-\epsilon/2}^{x+\epsilon/2}
    \! A_\mu(x)d x^\mu\right) q(x-\epsilon/2) \nonumber \\
    &=&  \lim_{\epsilon\to 0}\tilde q(x+\epsilon/2)
   \left(\mathbf{1} - i \epsilon^\mu A_\mu(x) +
    \dots\right) q(x-\epsilon/2),
   \label{split1}
\eeqs
where the explicit insertion of the Wilson line 
guarantees gauge invariance. 
The dots in the second expression represent the high order
terms in the Wilson line needed to preserve gauge invariance to all
orders but they will not be needed here.

When acting with $\bar{\mathcal{Q}}_{\dot{\alpha}}$ on the LHS of
(\ref{split1}) we get zero by definition when hitting $\tilde q$ and
$q$ but the SUSY variation of $A_\mu$ gives rise to a gluino (cfr. eqs. 
(\ref{SUSYYM}) and (\ref{qoper})):
\beq
    {[\bar{\mathcal{Q}}_{\dot{\alpha}}, \tilde q q(x) ]} = - 
    \frac{1}{\sqrt{2}}\lim_{\epsilon\to 0}\tilde q(x+\epsilon/2) \epsilon_\mu
    \sigma^\mu_{\alpha\dot{\alpha}} \lambda^\alpha(x) q(x-\epsilon/2).
\eeq
The quantity in the limit is naively linear in $\epsilon$ but we could
get into trouble if the OPE between $\tilde q$ and
$q$ had a singularity in $\epsilon$ so that a non-zero result could
arise. Fortunately, no such singularity is present in this case, as
can be seen by noticing, for instance, that the propagator between 
$\tilde q$ and $q$ is zero (there is a non-zero propagator
between $\tilde q$ and $\tilde q^\dagger$ and between 
$q$ and $q^\dagger$). 

It is a general fact that there are no singularities in the OPE of
lowest components of
chiral fields and thus the gauge invariant composite operators made out
of them obey the naive equation 
${[ \bar{\mathcal{Q}}_{\dot{\alpha}}, \mathcal{O}(x) ]} = 0$ and are
thus chiral operators. We will see later, when discussing the 
Konishi anomaly, that
there are other cases (not just involving $\phi(x)$ and
$\lambda_\alpha(x)$) where such singularity arises and the naive SUSY
transformations are modified. 

Now that we have constructed chiral operators, we can immediately derive
the important consequence~\cite{Novikov:ee}
that the v.e.v. of an arbitrary (time ordered) product of such
operators is totally independent of their spacetime position. Consider
for instance the product of two bosonic 
chiral operators $\oone$ and $\otwo$ and
take the derivative with respect to $x^\mu$:
\beqs
  &&\frac{\partial}{\partial x^\mu} 
   \langle 0|T\Big(\oone\otwo\Big)|0\rangle  =  \label{timeor}\\
  && \langle 0|T\Big(\frac{\partial}{\partial x^\mu}\oone\otwo\Big)|0\rangle
  + \delta_\mu^0 \langle 0|{[\oone,\otwo]}|0\rangle
    \delta(x^0-y^0). \nonumber
\eeqs
Both terms vanish separately. The first because (for, say, $x^0 > y^0$):
\beqs
     \langle 0|\frac{\partial}{\partial x^\mu}\oone\otwo|0\rangle
     &=& - i \langle 0|{[\mathcal{P}_\mu, \oone]}\otwo|0\rangle \label{tor}\\
     \phantom{\frac{d}{d}}  
     &=& \frac{i}{2} {\bar\sigma}_\mu^{\dot{\alpha}\alpha} 
     \langle 0|{[\{\mathcal{Q}_\alpha, \bar{\mathcal{Q}}_{\dot\alpha}\},
     \oone]}\otwo|0\rangle \nonumber\\ 
     \phantom{\frac{d}{d}} 
     &=& \frac{i}{2} {\bar\sigma}_\mu^{\dot{\alpha}\alpha} 
     \langle 0|\{\bar{\mathcal{Q}}_{\dot\alpha},{[\mathcal{Q}_\alpha,\oone]}\}
     \otwo|0\rangle \nonumber \\  
     \phantom{\frac{d}{d}} 
     &=& \frac{i}{2} {\bar\sigma}_\mu^{\dot{\alpha}\alpha} 
     \langle 0|\{\bar{\mathcal{Q}}_{\dot\alpha},{[\mathcal{Q}_\alpha,\oone]}
     \otwo\}|0\rangle \nonumber \\
      \phantom{\frac{d}{d}}  &=& 0 \nonumber.
\eeqs
The idea behind the manipulations above is to use the SUSY algebra,
the Jacobi identity and the chirality of the operators to bring 
$\bar{\mathcal{Q}}_{\dot\alpha}$ to act on the vacuum state, where it
vanishes assuming that the vacuum is supersymmetric.

The second term in (\ref{timeor}) is also zero because the equal time 
commutator vanishes since the same arguments leading to (\ref{tor})
apply more generally to the OPE of the two operators.

Given that the v.e.v. of chiral operators is independent of their position,
one can go to the limit of large separation and apply cluster
decomposition to obtain:
\beq
    \langle 0|T\Big(\mathcal{O}_1(x_1)\dots
     \mathcal{O}_n(x_n)\Big)|0\rangle = \langle \mathcal{O}_1 \rangle
    \dots \langle \mathcal{O}_n \rangle,
\eeq
where there is no longer any need to specify the  positions.
This is the important property of factorization of correlation functions
involving chiral operators.

Since objects of the type 
$\{\bar{\mathcal{Q}}_{\dot\alpha}, ... \}$ do not contribute to the
expectation values in a SUSY vacuum, 
it is natural to define an equivalence relation between chiral
operators. Two chiral operators $\oone$ and
$\mathcal{O}_2(x)$ are equivalent if there exist a \emph{gauge
invariant} operator $X_{\dot{\alpha}}(x)$ such that 
\beq
    \oone = \mathcal{O}_2(x) + 
    \{\bar{\mathcal{Q}}^{\dot\alpha},X_{\dot{\alpha}}(x) \}, \label{ringdef}
\eeq
where the generalization to operators carrying Lorentz indices should
be obvious -- the extra indices just appear in the same way in 
$\oone$,   $\mathcal{O}_2(x)$ and $X_{\dot{\alpha}}(x)$ and the
anticommutator in (\ref{ringdef}) becomes a commutator if $\oone$ and 
$\mathcal{O}_2(x)$ are fermionic.

Note that for (\ref{ringdef}) to be consistent, $X_{\dot{\alpha}}(x)$
must satisfy:
\beq
   {[\bar{\mathcal{Q}}_{\dot\beta}, 
   \{\bar{\mathcal{Q}}^{\dot\alpha},X_{\dot{\alpha}}(x) \} ]}=0
\eeq
which is not automatic from the SUSY algebra.

The set of equivalence classes of chiral operators under
(\ref{ringdef})
forms a ring, known as the \emph{chiral ring}. It is easy to check
that the product of equivalence classes is well defined (i.e. it is
independent of the representative) by applying the Leibniz rule a few
times.

The equivalence (\ref{ringdef}) can also be formulated in superspace
by saying that, if $\mathcal{O}_1(x)$ and $\mathcal{O}_2(x)$ are the
lowest components of a chiral superfield $\Sigma_1$ and $\Sigma_2$
respectively, there exist a (gauge invariant) 
superfield $Y_{\dot{\alpha}}$ such that:
\beq
     \Sigma_1 = \Sigma_2 + \bar{D}^{\dot{\alpha}} Y_{\dot{\alpha}}.
     \label{superdef}
\eeq
We will also refer to two such superfields as equivalent in the chiral ring.
Again, the superfield  $Y_{\dot{\alpha}}$ must obey 
$\bar{D}_{\dot{\beta}}\bar{D}^{\dot{\alpha}} Y_{\dot{\alpha}} = 0$ for
consistency. Furthermore, the relation 
$\bar{D}_{\dot{\alpha}} Y_{\dot{\beta}} + 
\bar{D}_{\dot{\beta}} Y_{\dot{\alpha}}= 0$ can
always be imposed since it disappears from (\ref{superdef}).
These two conditions together allow one
to ``solve'' the constraint as $ Y_{\dot{\alpha}} =
\bar{D}_{\dot{\alpha}} Z$, for some superfield $Z$ in terms of which
(\ref{superdef}) now reads
\beq
     \Sigma_1 = \Sigma_2 + \bar{D}^{\dot{\alpha}}\bar{D}_{\dot{\alpha}}Z 
     = \Sigma_2 + 2 \bar{D}^2 Z.
     \label{superdefnew}
\eeq
Thus, two superfields are equivalent if they differ by a ``chirally
exact'' term. We have also shown, in passing, that the lowest
component of a chirally exact object is the SUSY variation of some
other operator and, thus, its v.e.v. is identically zero in a SUSY
vacuum. 

Let us make all of these statements concrete and apply the above
formalism to gauge theories. 
Let $\Phi$ be any chiral superfield transforming under the 
representation $T^a$ of the gauge group. We write 
$W_\alpha = W_\alpha^a T^a$. As usual, we denote the lowest components
of $\Phi$ and $ W_\alpha$ (in the WZ gauge) by $\phi$ and
$-\sqrt{2} i\lambda_\alpha$ respectively. 
The basic relation that we will extensively use is:
\beq
      \lambda_\alpha \phi = -\frac{1}{2\sqrt{2}}
      {[\bar{\mathcal{Q}}^{\dot\alpha},
      \nabla_{\alpha\dot\alpha} \phi]},
\eeq
or, equivalently, in terms of superfields 
\beq
    W_\alpha \Phi = \bar{\nabla}^2 \nabla_\alpha \Phi. \label{fundrel}
\eeq
Since the fields appearing in the Lagrangian are gauge variant we must
use covariant derivatives. Of course, when acting on gauge invariant
combinations, $\bar{\nabla}^2$ becomes $\bar{D}^2$.

The proof of these relations is simple: e.g.
\beq
    W_\alpha \Phi = \frac{i}{2}{[ \bar{\nabla}^{\dot\alpha}, 
    \nabla_{\alpha\dot\alpha} ]} \Phi 
    = \frac{i}{2}\bar{\nabla}^{\dot\alpha} 
    \nabla_{\alpha\dot\alpha} \Phi 
    = \frac{1}{2}\bar{\nabla}^{\dot\alpha}
    \{\bar{\nabla}_{\dot\alpha}, \nabla_{\alpha} \} \Phi =
    \bar{\nabla}^2 \nabla_\alpha \Phi.
\eeq  
(In the case where $\Phi$ is in the Adjoint representation, it is
customary to think of it as a matrix and write ${[W_\alpha, \Phi]}$
instead.)

We now use (\ref{fundrel}) to obtain relations in the chiral
ring. We call these relations ``classical'', and similarly for the
ring generated by them, since they receive quantum corrections as
we will discuss later on. 

Let us specialize, for concreteness, to pure gauge theories, where the
only chiral superfield is $W_\alpha$ itself, transforming in the
adjoint representation. Eq. (\ref{fundrel}) reads:
\beq
    \{ W_\alpha, W_\beta \} = \bar{\nabla}^2 \nabla_\alpha W_\beta.
    \label{alluneed}
\eeq
Note that the RHS of (\ref{alluneed}) is symmetric in $\alpha$ and
$\beta$ because of the relation $\nabla^\alpha W_\alpha = 
\bar{\nabla}_{\dot\alpha} \bar{W}^{\dot\alpha}$, which can be 
easily checked using the definitions of 
$W_\alpha$ and $\bar W_{\dot\alpha}$.

Consider now taking the trace of a string of $W_\alpha$. 
Because of (\ref{alluneed}), anticommuting any two $W_\alpha$ will 
contribute only a trivial element in the chiral ring, e.g.:
\beq
    \tr(W_\alpha W_\beta W_\gamma) = - \tr(W_\alpha W_\gamma W_\beta) 
    + \bar{D}^2 \tr( W_\alpha \nabla_\beta W_\gamma).
\eeq
Using again  (\ref{alluneed}), we see that the only non-trivial
piece of the trace is when all the indices are antisymmetrized, 
thus excluding all traces of three or more $W_\alpha$.

Hence, of these type of invariants, the only one
that is non-trivial in the chiral ring is the glueball
superfield introduced in (\ref{glueS}):
\beq
      S = - \frac{1}{32\pi^2} \tr W^\alpha W_\alpha. \label{endofstory}
\eeq
Notice that, in spite of the fact that 
$S \propto \bar{D}^2\tr( W^\alpha e^{-V} (D_\alpha e^{V}))$, 
the glueball superfield is not chirally exact because 
$\bar{D}_{\dot\alpha}\tr( W^\alpha e^{-V}(D_\alpha e^{V}))$ 
is not gauge invariant.

Let us now be more exhaustive and discuss the full chiral ring of all 
pure gauge theories based on simple Lie groups.

For a pure $SU(N)$ gauge theory, eq.~(\ref{endofstory}) 
is the end of the story because 
traces of $W_\alpha$ exhaust all the invariants one
can construct. In fact, the only primitive invariants at one's disposal to
make a singlet are $\delta^i_j$, $\epsilon_{i_1,\dots i_N}$ 
and $\epsilon^{i_1,\dots i_N}$, but $\epsilon_{j_1,\dots j_N}\times
\epsilon^{i_1,\dots i_N}$ is proportional to products of 
$\delta^i_j$ and one is only left with traces. 

For other gauge groups one has to be more careful. 
For instance, for the remaining classical groups
one could use the extra primitive
invariants ($\delta^{ij}=\delta^{ji}$ for the orthogonal and
$\eta^{ij}=-\eta^{ji}$ for symplectic groups) to raise one index of
the $W_\alpha$ and then try to use \emph{one} $\epsilon_{i_1,\dots i_N}$ to
make a singlet. But even this will not work: For $SO(2k+1)$ the
$\epsilon$-tensor has an odd number of indices whereas one needs an even
one. For $Sp(k)$ $W^{ij}_\alpha$ is symmetric and thus vanishes when
contracted with $\epsilon$ and, finally, for $SO(2k)$ ($k>2$)
the contraction of two $W_1$  or of two $W_2$ with $\epsilon$ 
vanishes because of their Grassmann nature. 

Thus, for pure gauge theories based on classical groups the chiral
ring is generated by the glueball superfield~\cite{Witten:2003ye}. 
This is conjectured to be true even for exceptional 
groups although the presence of further
primitive invariants (e.g. $\sigma^{ijk}$ for $G_2$) makes this
statement non-trivial. Amongst the exceptional groups, so far the 
conjecture has been proven only for $G_2$ \cite{Etingof:2003dd}.

Having established the generators of the chiral ring, we now proceed to 
see that it has a further structure.
An important and non-trivial fact that
one can show using group theory is that relation (\ref{alluneed}) also
implies 
\beq
    S^h = 0 \qquad \hbox{in the classical chiral ring.} \label{sheqalzero}
\eeq 
The integer $h$ is the dual Coxeter
number of the group, i.e. half of the index of the
adjoint representation. Eq. (\ref{sheqalzero}) has been proven 
for the classical groups \cite{Witten:2003ye}
and for $G_2$ \cite{Etingof:2003dd} and it is believed to be true for 
the other exceptional groups as well. 
The list of dual Coxeter numbers is given by Table~\ref{tab:2}.
\begin{table}[hbt]
  \begin{center}
    \begin{tabular}{|c|c|c|c|c|c|c|c|c|}
      \hline
      $SU(k)$ & $SO(2k)$ & $Sp(k)$ & $SO(2k+1)$ & $G_2$ & $F_4$ & 
      $E_6$ & $E_7$ & $E_8$\\
      \hline
      $k$ & $2k-2$ & $k+1$ & $2k-1$ & $4$ & $9$ & $12$ & $18$ & $30$ \\
      \hline
    \end{tabular}
    \label{tab:2}
    \caption{Different groups and their dual Coxeter numbers.}
  \end{center}
\end{table}
We will not go through the proof of this fact in general but just
notice that for $SU(2)$ it can be easily obtained from the relation
(valid for any four $SU(2)$ generators)
\beq
    \tr (A B C D) = \frac{1}{2}\Big(\tr(A B) \tr(C D) +  \tr(A D) \tr(B C) -
                 \tr(A C) \tr(D B)\Big).
\eeq
Substituting $A=B=W_1$ and $C=D=W_2$ and using Fermi statistics one gets
\beq
     S^2 \propto \Big(\tr(W_1 W_2)\Big)^2 = \tr(W_1 W_1 W_2 W_2),
\eeq 
the last term being zero in the classical chiral ring.

What we have obtained so far is very similar to what was obtained in
the previous section where 
imposing the condition (\ref{bkgnd}) on the background has set 
to zero precisely all the operators that now are shown to be trivial in the 
classical chiral ring.

However, if eq. (\ref{sheqalzero}) were correct at the full quantum level, 
it would imply that in a SUSY vacuum, 
$\langle S^h\rangle = 0$.\footnote{By
factorization one would also have $\langle S\rangle = 0$.}
However, that $\langle S^h \rangle \not= 0$ can be shown explicitly 
by doing a one instanton calculation \cite{instant}. 

We thus face two choices, either
$\langle S^h \rangle \not= 0$ is a signal of the fact that the vacuum
breaks SUSY or the relation $S^h = 0$ is modified by quantum
corrections. There is overwhelming evidence that the second option is
the right one. In particular the Witten index~\cite{wittenindex}
for pure gauge theories
is never zero implying that SUSY cannot be broken even by
non-perturbative effects.

The fact that the chiral ring relation $S^h = 0 $ is modified to $S^h =
\mathrm{const.}$ is actually not that surprising. Remember that quantum
corrections cannot bring new operators into existence but can modify
the relations between existing ones by the introduction of the
holomorphic scale $\Lambda$. The classical relation should be
recovered by letting $\Lambda\to 0$. 
Simple dimensional analysis fixes 
$S^h\propto \Lambda^{3h}$ and the instanton computation
reviewed in \cite{instant} (see also \cite{Finnell:1995dr}) allows also 
the determination of the proportionality constant $C(G)$. We recognize 
in $3h$ the coefficient $\beta$ of the one loop beta-function
of pure SYM. From (\ref{biglambda}) we can see that $\langle S^h \rangle$
is thus proportional to the exponential of minus the one instanton action
$8\pi^2/g^2$. Hence eq. (\ref{sheqalzero}) is changed into
\beq
    S^h = C(G) \Lambda^{3h} \qquad \hbox{in the quantum chiral ring.} 
\eeq 

So far we considered pure gauge theories. As it is going to be important
later, let us consider an example of gauge theory 
with matter, say, a $U(N)$ gauge theory
with fields $\Phi$ in the adjoint and $Q, \tilde Q$ in the fundamental
and anti-fundamental. Relation (\ref{fundrel}) implies that, as far as the
classical chiral ring is concerned, the fields $W_\alpha$ and $\Phi$ commute
inside various singlets and that $W_\alpha Q = \tilde Q W_\alpha =
0$. Thus the generators of the chiral ring are of the following 
type~\cite{earlych,cdsw, crfla}:
\beq
    \tr \Phi^k,\;\;\; \tr W_\alpha \Phi^k, \;\;\; 
    \tr W^\alpha W_\alpha\Phi^k, \;\;\; \tilde Q \Phi^k Q.
\eeq
Classically, they are subjected to the relation 
analogous to (\ref{sheqalzero}) for $S$ and to the 
usual relations stemming from the possibility of rewriting
higher traces as functions of the lower ones. 
For instance, $\tr \Phi^{N+1}$ can be expressed as a
polynomial in $\tr \Phi, \dots  ,\tr \Phi^N$.

When matter fields are present, there are also 
relations that use the dynamics of the theory
and depend on the particular choice of superpotential. 
Let $\Phi_r$ denote a generic field content, with $r$ labeling the
various irreducible representations of the gauge group. Let
$F_r(W_\alpha, \Phi)$ be a combination of the fields that also
transforms in the representation $r$. (Since we are working in the
chiral ring, two such expressions that differ by a chirally exact
operator can be identified.)
The classical equations of motion for the matter fields are
\beq
   \bar{\nabla}^2\left( \bar{\Phi} e^V \right)^r 
    = -\frac{\partial W(\Phi)}{\partial \Phi_r}.
\eeq
Contracting with $F_r$ to make a singlet this implies 
the \emph{classical} relation
\beq
    \frac{\partial W(\Phi)}{\partial \Phi_r}F_r(W_\alpha, \Phi) = 0
    \qquad \hbox{in the classical chiral ring.} 
\eeq
By now, it should not be surprising that also this classical
relation is modified quantum mechanically. This is due to 
the so called Konishi anomaly which we now turn to discuss.

\section{The Konishi anomaly}

The Konishi anomaly is the SUSY extension of the usual chiral anomaly
and it proves extremely useful in solving for the F-terms of the
theory.

In its original form~\cite{konishiorig}, it is based on the 
observation that (contrary to
what we saw for chiral operators) the SUSY transformations of a
generic composite operator are not always the ones that we would
naively obtain applying the transformation to each component
field. Consider again the example of SQCD, with its two chiral superfields 
$Q$ and $\tilde Q$ transforming in the fundamental and
antifundamental representation of the gauge group. We saw in the
previous section that the composite (chiral) operator $\tilde q q $
constructed from the lowest components obeys 
$[{\bar {\mathcal Q}}_{\dot\alpha}, \tilde q q] = 0$ as one 
would naively expect. 

If we drop the requirement that the composite 
operator be chiral, there are many other ways to build a gauge singlet
even using just the components of one superfield instead of two. 
Of particular interest is the composite operator
$\bar{\psi}^{\dot\alpha} q$ constructed only with the first and
second components of $Q$. Naively, one would write:
\beqs
     \{{\bar {\mathcal Q}}_{\dot\alpha}, \bar{\psi}^{\dot\alpha} q\} &=&
     \{{\bar {\mathcal Q}}_{\dot\alpha}, \bar{\psi}^{\dot\alpha}\} q
     - \bar{\psi}^{\dot\alpha} {[ {\bar {\mathcal Q}}_{\dot\alpha},
     q]} \nonumber \\ 
     &=& 2 \bar{f} q + 0 = - 2\frac{\partial W}{\partial q} q
     \qquad\hbox{classical} \label{naive}
\eeqs
If (\ref{naive}) was true, the RHS could not get a v.e.v. in a SUSY
vacuum. For instance, for $W=m\tilde Q Q$ one would deduce that 
$\langle \tilde q q\rangle = 0$, contradicting many well
established facts about SQCD. 

To obtain the right formula, we must once again regularize:
\beq
     \bar{\psi}^{\dot\alpha}q (x) = 
     \lim_{\epsilon\to 0} \bar{\psi}^{\dot\alpha}
     (x +\epsilon/2)(1 - i \epsilon^\mu A_\mu(x)+ \dots)
     q (x -\epsilon/2).
\eeq
Now in the SUSY variation of $\bar{\psi}^{\dot\alpha} q $ 
there will be an extra
piece coming from the SUSY variation of $A_\mu$:
\beq
    \{{\bar {\mathcal Q}}_{\dot\alpha}, \bar{\psi}^{\dot\alpha} q\} = 
    - 2 \frac{\partial W}{\partial q}q - 
     \frac{1}{\sqrt{2}}\lim_{\epsilon\to 0} \bar{\psi}_{\dot\alpha}
     (x +\epsilon/2) 
     \epsilon^\mu\bar{\sigma}_\mu^{\dot\alpha\alpha}
     \lambda_\alpha(x) q(x-\epsilon/2)
\eeq
For our purposes it is enough to think of the gluino $\lambda$ as a
background field, although this is not necessary. The extra piece
seems to vanish as $\epsilon\to 0$ but now we have a singularity in
the OPE coming from the presence of a term of the type 
$\sqrt{2}\int d^4x \; \bar q \lambda \psi $ in the Lagrangian.
If one ``brings down" such an interaction in the path integral     
one obtains the contribution (written for simplicity in the Abelian   
case)
\beqs
   &&\!\!\!\!\!\!\!\!\!\!\!\!\lim_{\epsilon\to 0}
   \epsilon^\mu\bar{\sigma}_\mu^{\dot\alpha\alpha} \lambda_\alpha(x)\!\!
   \int\!\! \mathrm{d}^4 x^\prime \langle 0|T\big(\psi_\beta(x^\prime)
   {\bar\psi}_{\dot\alpha} (x+\epsilon/2)\big)|0\rangle
   \lambda^\beta(x^\prime)  
   \langle 0|T\big(q(x-\epsilon/2)\bar q(x^\prime)\big)|0\rangle 
   \nonumber \\
   &&\!\!\!\!\!\!\!\!\!\!\!\!\propto \lim_{\epsilon\to 0}
   \epsilon^\mu\bar{\sigma}_\mu^{\dot\alpha\alpha} \lambda_\alpha(x)
   \epsilon^4 \frac{\epsilon_\nu}{\epsilon^4} \sigma^\nu_{\beta\dot\alpha}
   \lambda^\beta(x)\frac{1}{\epsilon^2} \propto \lambda^2(x).
   \label{koniabel}
\eeqs

More precisely, keeping track of all numerical factors and generalizing 
to the non-Abelian case one gets \cite{konishiorig}
\beq
     \{{\bar {\mathcal Q}}_{\dot\alpha}, \bar{\psi}^{\dot\alpha} q\} =
    2 \left( \bar{f} q + \frac{1}{16\pi^2} 
    \tr \lambda^\alpha  \lambda_\alpha \right).
    \label{konishianomaly}
\eeq

As usual, the same relation can be expressed in terms of superfields
as
\beq
    \bar{D}^2 \Big(\bar{Q} e^V Q\Big) 
    = -\frac{\partial W}{\partial Q} Q -
    \frac{1}{32\pi^2} \tr W^\alpha W_\alpha. \label{konishianomalysuper}
\eeq
This is so because we require the composite object 
$\bar{Q} e^V Q$ to be a superfield, i.e. to contain a
supermultiplet of fields properly transforming into each other by a
SUSY transformation. For instance:
\beq
   \bar q q  \stackrel{\bar{\mathcal{Q}}}{\rightarrow}
   \bar{\psi}^{\dot\alpha} q \stackrel{\bar{\mathcal{Q}}}{\rightarrow}
   \bar{f} q + \frac{1}{16\pi^2} \tr \lambda^\alpha \lambda_\alpha,
\eeq
where the rightmost entry is thus the lowest component of 
$\bar{D}^2 \bar{Q} e^V Q$.

In the application of the Konishi anomaly to the recent developments,
it is sometimes necessary to consider a generalization of formulas
(\ref{konishianomaly}) and (\ref{konishianomalysuper})~\cite{cdsw}
that we now turn to discuss. Consider chiral matter $\Phi$ in an arbitrary 
representation of the gauge group.
Eq. (\ref{konishianomalysuper}) can be understood as the anomalous
Schwinger-Dyson equation coming from varying 
$\Phi \to \Phi + \eta\Phi$  ($\eta \ll 1$) in the path integral.
The classical piece comes from the variation of the classical action
whereas the anomaly comes from the non-invariance of the functional
measure.

One can generalize the transformation to
\beq
   \Phi \to \Phi + \eta F(\Phi, W_\alpha),
\eeq
where $F$ is a holomorphic function transforming in the same 
representation as $\Phi$. The classical piece is straightforward:
\beq
     \bar{D}^2 \Big(\bar{\Phi} e^V F\Big) 
     = -\frac{\partial W}{\partial \Phi} F 
     \qquad\hbox{classical}.
\eeq
As for the anomalous piece, the only additional complication is in
keeping track of the color indices. Let $r, s, \dots$ denote the
indices of the representation in which $\Phi$ transforms and let us
write the gluino field as a matrix in that representation. The anomaly
is, schematically\footnote{There can also be corrections from chirally
exact terms that would contribute to the D-terms, but the
non-renormalization theorem prevents additional perturbative
contribution to the F-terms.}:
\beq
    \lim \epsilon \bar{\psi}^r \lambda^s_r F_s \propto
    \lim \epsilon \int \langle \bar{\psi}^r \psi_p\rangle \lambda^s_r 
    \lambda^p_q \langle F_s \bar{\phi}^q \rangle,
\eeq
where the spacetime dependence is as in~(\ref{koniabel}). Using 
\beq
    \langle \bar{\psi}^r \psi_p\rangle \propto \delta^r_p
    \qquad\hbox{and}\qquad
    \langle F_s \bar{\phi}^q \rangle \propto \frac{\partial
    F_s}{\partial\phi_l}\langle \phi_l \bar{\phi}^q \rangle \propto
    \frac{\partial F_s}{\partial\phi_l} \delta^q_l,
\eeq 
yields
\beq
    \lim \epsilon \bar{\psi}^r \lambda^s_r F_s \propto \lambda^s_r 
    \lambda^r_q \frac{\partial F_s}{\partial \phi_q}.
\eeq

The generalized Konishi anomaly is thus:
\beq
    \bar{D}^2\Big( \bar{\Phi}^q (e^V)_q^r F_r\Big) = 
    -\frac{\partial W}{\partial \Phi_r} F_r -
    \frac{1}{32\pi^2} {W^\alpha}^s_r {W_\alpha}^r_q 
    \frac{\partial F_s}{\partial \Phi_q}.
     \label{generalkonishi}
\eeq
The use of (\ref{generalkonishi}) is mainly in the context of the
chiral ring, where we may set all chirally exact terms to zero,
including the LHS of (\ref{generalkonishi}). In this context 
(\ref{generalkonishi}) reads:
\beq
   \frac{\partial W}{\partial \Phi_r} F_r =
    - \frac{1}{32\pi^2} {W^\alpha}^s_r {W_\alpha}^r_q
   \frac{\partial F_s}{\partial \Phi_q}. 
   \quad\hbox{in the quantum chiral ring} 
   \label{ringkonishi}    
\eeq
In particular,
these relations hold for v.e.v.s in 
a SUSY vacuum, and they
are often enough to solve for the F-terms of the effective theory.
Specifically, they can be used to prove the
Dijkgraaf - Vafa conjecture as we now discuss. 

\section{\!Glue\-ball super\-po\-ten\-tial via the Koni\-shi ano\-maly}

In this section, we will exploit a relationship which exists between
the effective superpotential and the SUSY vacuum expectation values
of chiral operators to determine completely the 
former. The relation
was discussed in Section 3 and reads:
\beq
     \pd{}{\lambda_k}W_{\mathrm{eff}}= \langle X_k \rangle.
     \label{dwisx}
\eeq
This relation was derived when we realized that integrating out and back in
degrees of freedom was technically a Legendre transform. Alternatively, 
one can derive the above expression directly by expressing $W_{\mathrm{eff}}$
in terms of the path integral over the elementary fields, weighted by
the action including the 
tree level superpotential $W_{\mathrm{tree}}=\sum\lambda_k X_k$ . 
Then considering the coupling $\lambda_k$ as a
source term for the operator $X_k$ the relation (\ref{dwisx}) follows.

The usefulness of eq.~(\ref{dwisx}) is that if we are able to determine
$ \langle X_k \rangle$ as a function of the couplings $\lambda_k$ and
of $S$ we can then obtain $W_{\mathrm{eff}}$ up to a function
independent of the coupling.
The Konishi anomaly and its generalizations provide us with just
a way to determine such expectation values.~\footnote{That the Konishi 
anomaly plays an important role
in the Dijkgraaf - Vafa conjecture was first emphasized 
in~\cite{gorsky}.}

Let us see how this works in a simple example, where the non-generalized
Konishi anomaly is sufficient to solve for a non-trivial $W_{\mathrm{eff}}$
\cite{binor}.
We take SQCD with gauge group $U(N)$ and, for simplicity, only one 
flavor~\footnote{The $U(1)$ factor is unimportant as it decouples in 
the IR.}.

The only gauge invariant we can build with this matter content is a 
single meson superfield, $M=\tilde Q_a Q^a$. A tree level superpotential
that one can write is the following:
\beq
     W_{\mathrm{tree}}=m M +\lambda M^2.  \label{wtreemsq}
\eeq
The quarks are thus massive, and the second term introduces a non-trivial
interaction. The latter is non-renormalizable, but
it is straightforward to see that (\ref{wtreemsq}) can be obtained
by writing the renormalizable superpotential:
\beq
      W_{\mathrm{tree}}=m \tilde Q Q + \frac{\mu}{ 2} \tr \Phi^2 + 
      g  \tilde Q\Phi Q,  \label{wtreeqphi}
\eeq 
and then integrating out the adjoint field $\Phi$, assuming that 
$|\mu| \gg |m|$ and identifying $\lambda= -\frac{g^2}{2\mu}$.

Let us now write down the Konishi anomaly relation for, say, the $U(1)_Q$
rotating the superfield $Q$. This is basically the computation
leading to (\ref{konishianomalysuper}). Using (\ref{wtreemsq}),
the relation in a SUSY vacuum becomes:
\beq
     m\langle M\rangle + 2 \lambda \langle M^2 \rangle = S. \label{sqcdkon}
\eeq 
By using factorization, we can reexpress $\langle M^2 \rangle =
\langle M\rangle^2$ and then solve (\ref{sqcdkon}) for $\langle M\rangle$:
\beq
     \langle M\rangle = -\frac{m}{ 4\lambda} \pm \sqrt{\frac{m^2}{16 \lambda^2}
     +\frac{S}{ 2\lambda}}. \label{mvev}
\eeq
The two signs in (\ref{mvev}) have a clear meaning in the classical limit, 
where $S \rightarrow 0$. For the positive sign, the v.e.v. tends to zero, 
and we are thus in a situation where classically the $U(N)$ 
gauge group in unbroken.
For the negative sign, the v.e.v. tends to a finite value, and thus $U(N)$
is broken to $U(N-1)$. Note that when the coupling $\lambda$ is sent
to zero this second classical vacuum is sent to infinity and disappears.

We thus have two partial differential equations for $W_{\mathrm{eff}}$:
\beq
     \pd{W_{\mathrm{eff}}}{m} 
     = \langle M(S,m,\lambda)\rangle, \qquad
     \pd {W_{\mathrm{eff}}}{\lambda}  = 
     \langle M(S,m,\lambda)\rangle^2. \label{pdeweff}
\eeq
Using the expression (\ref{mvev}), we obtain the solution:
\beq
      W_{\mathrm{eff}}= -\frac{m^2}{8 \lambda} \pm \frac{m^2}{8 \lambda}
      \sqrt{1+\frac{8\lambda}{m^2}S} + S \log \frac{m}{\Lambda}
      + S\log \left(1 \pm \sqrt{1+\frac{8\lambda}{ m^2}S}\right) +C(S),
      \label{weffsqcd} 
\eeq
where of course the coupling independent piece of $W_{\mathrm{eff}}$ is 
still undetermined, and we have used the holomorphic scale $\Lambda$
to compensate for the dimension of $m$. This can be done because a rescaling 
in $\Lambda$ can always be absorbed in $C(S)$. 

The coupling independent piece $C(S)$ can be determined by using low-energy
information on the theory. Consider simply the positive branch
of (\ref{mvev}), and take
the coupling $\lambda$ to zero. Here we should recover the low energy physics
of SQCD with a massive flavor. We know what this is, namely pure $SU(N)$
SYM, and we also know it is described by the Veneziano-Yankielowicz
superpotential (\ref{WVY}). By thus imposing that (\ref{weffsqcd})
with the positive signs reduces to (\ref{WVY}) when $\lambda\rightarrow 0$
(for this, we have to match the holomorphic scale $\tilde \Lambda$ 
of the pure SYM theory
to the scale $\Lambda$ of the theory with one flavor, $\tilde \Lambda^{3N}
=m\Lambda^{3N-1}$) we obtain:
\beqs
     W_{\mathrm{eff}} & = & NS\left(1-\log \frac{S}{\Lambda^3}\right)
      + S \log \frac{m}{ \Lambda} - \frac{S}{2} \label{wefffinal} \\
     & & -\frac{m^2}{8 \lambda} \pm \frac{m^2}{8 \lambda}
     \sqrt{1+\frac{8\lambda}{m^2}S}  + S\log \left(\frac{1}{2} \pm 
     \frac{1}{2}\sqrt{1+\frac{8\lambda}{m^2}S}\right). \nonumber 
\eeqs
We have thus determined the full effective superpotential by using
the Konishi anomaly (\ref{sqcdkon}) to determine the v.e.v.s, which
have allowed us to solve for the coupling dependence of $W_{\mathrm{eff}}$.
The remaining piece is then determined by taking specific values of the
coupling constants which make the low-energy physics more familiar.

It is worth noting that the same result (\ref{wefffinal}) can be 
obtained through a matrix-vector model based on the tree level superpotential
(\ref{wtreeqphi}) \cite{acfh1}. There one can show that only the 
graphs with the topology of a disk give a non-trivial contribution, the
boundary of the disk being composed of quark propagators. The result
of this perturbative computation is a power series in $S$,
which sums to the exact expression (\ref{wefffinal}). 
In this latter approach the VY piece of $W_{\mathrm{eff}}$ has to be added
to the result.

We now turn to the example discussed 
in \cite{cdsw}, where the 
generalized Konishi anomaly relations will be needed. We consider a
$U(N)$ gauge theory with a matter field in the adjoint representation
and a generic tree level superpotential:
\beq
     W_{\mathrm{tree}} = \sum_{k=0}^n \frac{g_k}{k+1}\tr \Phi^{k+1}.
\eeq
The classical vacuum structure is determined by the extrema of
$W_{\mathrm{tree}}(z)$, seen as a holomorphic function of a complex
variable. If we assume that the $n$ extrema are all
distinct:
\beq
     W_{\mathrm{tree}}'(z)= g_n \prod_{i=1}^n(z-a_i), \qquad
     a_i \neq a_j \quad \mbox{for} \quad i\neq j,
\eeq
then in the classical vacuum the eigenvalues of $\Phi$ will partition into
groups corresponding to every extremum and the fluctuations around 
these vacua will all be massive. If 
we have $N_i$ eigenvalues equal to $a_i$, the classical gauge symmetry
is spontaneously broken in the following way:
\beq
     U(N) \; \rightarrow \; \prod_{i=1}^n U(N_i) . \label{uni}
\eeq
At the quantum level, we expect for every $U(N_i)$ factor that
at low enough energies the $SU(N_i)$ piece will confine, while
the $U(1)$ piece will be in a free Coulomb phase. For every factor, 
we thus define the effective fields:
\beqs
         S_i & = & -\frac{1}{ 32 \pi^2} \tr W^\alpha_i W_{\alpha i},
          \label{upperdefi} \\
         w^\alpha_i & = & \frac{1}{4 \pi} \tr W^\alpha_i.
\eeqs
The quantity that we want to determine is $W_{\mathrm{eff}}$ as a
function of the couplings $g_k$ and the effective fields $S_i$ and
$w^\alpha_i$. Notice that the trace in (\ref{upperdefi}) is over the
full $U(N_i)$ subgroup and not $SU(N_i)$ for later convenience.

We now make a crucial remark that simplifies the discussion. The overall
$U(1)$ of the full $U(N)$ gauge group is free, since the matter field
is in the adjoint representation. This also means that there are no
Yukawa couplings involving the gaugino relative to
this $U(1)$, and thus its action is also one of a free fermion.
We thus have the possibility of shifting the $U(1)$ gaugino 
by a constant Weyl spinor
without changing anything in the theory. In superfield terms, this
symmetry of the theory is implemented as:
\beq
      W_\alpha\, \rightarrow\, W_\alpha -4\pi \chi_\alpha {\bf 1}_{N\times N}.
      \label{invarw}
\eeq
When the gauge symmetry is broken as in (\ref{uni}), the symmetry
still acts as an overall simultaneous shift of all the gaugino superfields, 
$W_{\alpha i}\rightarrow W_{\alpha i} -4\pi \chi_\alpha 
{\bf 1}_{N_i\times N_i}$, so that the effective fields transform as:
\beqs 
        S_i & \rightarrow  & S_i + \chi^\alpha w_{\alpha i} 
        -\frac{1}{2} N_i\, \chi^\alpha \chi_\alpha,\nonumber \\
        w^\alpha_i & \rightarrow  & w^\alpha_i -N_i\, \chi^\alpha. 
        \label{invariance}
\eeqs
The effective superpotential $W_{\mathrm{eff}}(S_i,w^\alpha_i,g_k)$
has to be invariant under the above transformation, since it cannot
depend on the decoupled overall $U(1)$.

A clever way to constrain the form of $W_{\mathrm{eff}}$ using the
invariance under (\ref{invariance}) is to apply to this fermionic
symmetry the same tricks used with supersymmetry in Section 2. Namely, consider
the invariance as a translation in an auxiliary Grassmann variable
$\psi^\alpha$. Then build a ``superfield'' using $\psi^\alpha$
such that the translation $\psi^\alpha \rightarrow \psi^\alpha +\chi^\alpha$
implements (\ref{invariance}). Finally, an invariant function is trivially
built by writing the integral over $\psi$-superspace of any function
of the $\psi$-superfield. Hence, we define:
\beqs
      \mathcal{S}_i & = & - \frac{1}{2} \tr \left(\frac{W^\alpha_i}{4\pi}
       -\psi^\alpha \right) \left(\frac{W_{\alpha i}}{ 4\pi}
       -\psi_\alpha \right) \label{cals}\\
      &= & S_i + w^\alpha_i \psi_\alpha - \frac{1}{2}N_i \psi^\alpha
       \psi_\alpha . \nonumber 
\eeqs
A manifestly invariant effective superpotential is then given by:
\beq
     W_{DV} = - \int\! \mathrm{d}^2\psi
     \mathcal{F} (\mathcal{S}_i,g_k),  
     \label{weffcalf}
\eeq
which, after performing the Grassmann integral, becomes:
\beq
     W_{DV} = \sum_i N_i \pd{\mathcal{F}}{S_i}
      + \frac{1}{2}  \sum_{i, j} \frac{\partial^2 \mathcal{F}}{\partial
       S_i \partial S_j} w^\alpha_i w_{\alpha j}. \label{weffquad} 
\eeq
To compute the derivatives we can consider $\mathcal{F}$ 
as a function of the $S_i$, that is the lowest components of the 
$\mathcal{S}_i$.

Note that (\ref{weffquad}) is quadratic in $w^\alpha_i$, a result which
could also be obtained in the perturbative set up of Section 4 by
observing that only planar diagrams contribute. We should stress here that
this feature appears only when $S_i$ is the glueball field obtained by tracing
over $U(N_i)$ and not over its $SU(N_i)$ subgroup. If we were to consider
the latter, then $W_{\mathrm{eff}}$ would contain higher powers
of $w^\alpha_i$.

Let us anticipate now that the main result of this section will be 
to identify the function $\mathcal{F}(S_i, g_k)$ with the 
planar free energy
of the related matrix model, as eq.~(\ref{wdv}) at the end of Section 4 
suggests. To see how this comes about, let us 
derive equations for this function $\mathcal{F}$.

First of all, the coupling dependence of the effective superpotential
is fixed by the equations:
\beq
     \pd{ W_{\mathrm{eff}} }{g_k} = 
     \left\langle \frac{1}{k+1 } \tr \Phi^{k+1}\right\rangle.
\eeq
Using eq.~(\ref{weffcalf}), the above equation reads:
\beq
     \int d^2\psi \frac{\partial}{\partial g_k}\mathcal{F}
     (\mathcal{S}_i,g_l) = -
     \left\langle \frac{1}{k+1 } \tr \Phi^{k+1}\right\rangle.
\eeq
We can write an equation directly for the integrand $\mathcal{F}$ if we
write the RHS as the highest component of a $\psi$-superfield,
generalizing (\ref{cals}). We thus have:
\beq
    \frac{\partial}{\partial g_k}\mathcal{F}(\mathcal{S}_i,g_l) 
    = - \left\langle \frac{1}{2( k+1) }
     \tr \left(\frac{W^\alpha}{4\pi}
     -\psi^\alpha \right) \left(\frac{W_{\alpha}}{4\pi}
     -\psi_\alpha \right)  \Phi^{k+1}\right\rangle.
\eeq
For convenience, the equations can be written setting $\psi^\alpha=0$
without loss of information, since from the lowest components the higher
ones are generated acting as in (\ref{invarw}). The equation we will
thus take as defining the coupling dependent part of 
the function $\mathcal{F}$ is:
\beq
     \frac{\partial}{\partial g_k} \mathcal{F}(S_i,g_l) = -
     \left\langle \frac{1}{k+1 } \tr
     \frac{W^\alpha W_{\alpha}}{32 \pi^2}  \Phi^{k+1}\right\rangle.
\eeq
The next step is to introduce a generating function for the 
operators on the RHS of the above equation. It reads:
\beq
     R(z) = - \frac{1}{32 \pi^2}  \left\langle\tr \frac{W^\alpha
         W_{\alpha}}{z-\Phi} \right\rangle 
     \equiv - \frac{1}{32 \pi^2} \sum_{k=0}^\infty \frac{1}{z^{k+1}}
      \langle \tr W^\alpha W_{\alpha} \Phi^k\rangle .
     \label{gaugeresol}
\eeq
Note that in the above we are not really paying attention to the
ordering of the $W^\alpha$ with respect to the $\Phi$, the reason 
being the relation in the chiral ring $[W^\alpha,\Phi]=0$.

The expression (\ref{gaugeresol}) also gives us a more precise, gauge
invariant, definition of the gaugino condensates $S_i$. Indeed, suppose
that the analytic structure of the function $R(z)$ is such that it has $n$
cuts over the $z$-plane (we will shortly see how this arises). The cuts
can be seen as a sort of quantum resolution of simple poles. By defining
the contour $C_i$ as the one going around the $i$-th cut, we define:
\beq
      S_i= \frac{1}{2\pi i} \oint_{C_i}\!\mathrm{d}z\; R(z). \label{sidef}
\eeq
Classically, the above relation is understood as follows. If the contour
$C_i$ encircles the eigenvalue $a_i$, then for any matrix $M$ the integral
projects on the eigenspace corresponding to this eigenvalue. Indeed, 
having diagonalized $\Phi$, we have:
\beqs
     \frac{1}{2\pi i} \oint_{C_i}\!\mathrm{d}z\; \tr \frac{M}{z-\Phi} &=& 
     \frac{1}{2\pi i} \oint_{C_i}\!\mathrm{d}z\; \sum_{m=1}^N
     \frac{M_{mm}}{z-\phi_m} 
     \nonumber \\
      &=&  \sum_{\phi_m=a_i}  M_{mm} 
     =  \tr P_i M \equiv \tr M_i. \label{projector}
\eeqs
Note that we also have $\tr P_i =N_i$. Since projectors should not receive
quantum corrections, the result of the manipulation above carries over to 
the quantum level.

We are now after a way to solve for the generating function (also called
resolvent, in reference to the matrix model analogy) $R(z)$. We are going
to see that it is precisely the generalized Konishi anomaly which provides
us with a closed equation for it.

Let us first of all specialize eq.~(\ref{ringkonishi}) to the case of
a matter field in the adjoint. The LHS simply becomes a trace.
As for the RHS, we have to make explicit the adjoint action of
the gaugino superfield by writing ${W^\alpha}^r_s\equiv i {W^\alpha}^p 
{{f^p}^r}_s$, with all indices in the adjoint representation. 
Then the two structure constants are traded for two
commutators, and the generalized Konishi anomaly reads:
\beq
     \left \langle\tr \left(W^\prime_\mathrm{tree}(\Phi) 
     F(W^\alpha, \Phi) \right) \right\rangle
     = - \frac{1}{32 \pi^2}\sum_{i,j=1}^N\left\langle 
     \left\{ W^\alpha, \left[ W_\alpha , 
     \pd{F}{\Phi^i_j }\right] \right\}^i_j \right\rangle.
     \label{commkon}
\eeq 
We now specialize to a particular variation of $\Phi$, namely we take:
\beq
     \delta \Phi
     = - \eta \frac{1}{32 \pi^2} \frac{W^\alpha W_{\alpha}}{z-\Phi}.
\eeq
To see what the RHS of (\ref{commkon}) looks like with this variation, 
we develop in a power series as in (\ref{gaugeresol}):
\beqs
     \pd{}{\Phi^i_j}\left(\frac{W^\alpha
         W_{\alpha}}{z-\Phi}\right)^k_l &=& \sum_{n=0}^\infty
     \frac{1}{z^{n+1}} \nonumber  
     \pd{}{\Phi^i_j} (W^\alpha W_{\alpha}\Phi^n)^k_l \\
     & = & \sum_{n=1}^\infty \sum_{p=0}^{n-1}\frac{1}{z^{n+1}} 
     (W^\alpha W_{\alpha}\Phi^p)^k_i (\Phi^{n-p-1})^j_l \nonumber \\
     & = &  \sum_{p=0}^\infty \sum_{n=p+1}^\infty \frac{1}{z^{n+1}} 
     (W^\alpha W_{\alpha}\Phi^p)^k_i (\Phi^{n-p-1})^j_l \nonumber \\
     &=& \sum_{p=0}^\infty \sum_{m=0}^\infty \frac{1}{ z^{p+1}z^{m+1}}
     (W^\alpha W_{\alpha}\Phi^p)^k_i (\Phi^{m})^j_l \nonumber \\
     & =& \left(\frac{W^\alpha W_{\alpha}}{ z-\Phi}\right)^k_i 
     \left(\frac{1}{z-\Phi}\right)^j_l . \label{trick}
\eeqs
Implementing now the chiral ring relations, which forbid traces with
more than two $W^\alpha$, we see that only one term of the (anti)commutators
in (\ref{commkon}) is non-vanishing, namely when the four $W^\alpha$
distribute themselves by two in each trace. We thus get the following basic
relation in a SUSY vacuum:
\beq
      - \frac{1}{32 \pi^2} \left \langle \tr 
     \left( W_\mathrm{tree}'(\Phi) \frac{W^\alpha W_{\alpha}
     }{ z-\Phi} \right)\right\rangle  = \frac{1}{(32 \pi^2)^2} 
     \left \langle \left( \tr \frac{W^\alpha W_{\alpha}}{z-\Phi} \right)^2
     \right\rangle.
\eeq
At this point, we can crucially use the factorization property of the
correlation functions of chiral operators to write the RHS as
$R(z)^2$. Concerning the LHS, we can add and subtract 
$W_\mathrm{tree}'(z)$ to $W_\mathrm{tree}'(\Phi)$ so that we can write:
\beq
    - \frac{1}{32 \pi^2} \left \langle \tr \left(W_\mathrm{tree}'(\Phi)
     \frac{W^\alpha W_{\alpha}}{z-\Phi} \right)\right\rangle
     = W_\mathrm{tree}'(z) R(z) + \frac{1}{4} f_{n-1} (z), \label{defoff}
\eeq
where we have defined:
\beq
      f_{n-1} (z) = \frac{1}{ 8 \pi^2} \left \langle \tr \left(
      (W_\mathrm{tree}'(z) - W_\mathrm{tree}'(\Phi) ) 
      \frac{W^\alpha W_{\alpha}}{ z-\Phi} \right)\right\rangle.
\eeq
It is easy to convince oneself that $f_{n-1} (z)$ is a polynomial of
degree $n-1$: $f_{n-1}(z)\equiv \sum_{k=0}^{n-1} f_k z^k$.
Indeed, by expanding $W_\mathrm{tree}'(z)$ in the numerator
around any eigenvalue of $\Phi$, one can see that $f_{n-1} (z)$ 
has no singularities.
Moreover, we see that at large $z$, the LHS of (\ref{defoff}) goes
like $1/z$. Thus $f_{n-1} (z)$ must cancel all the non-negative
powers in the product $W_\mathrm{tree}'(z) R(z)$, which go at most
as $z^{n-1}$. At this point the $n$ coefficients of $f_{n-1} (z)$ 
are unknown complex parameters, to be related to more physically
significant quantities shortly.

We have thus finally obtained the equation for $R(z)$. It reads:
\beq
     R(z)^2 = W_\mathrm{tree}'(z) R(z) + \frac{1}{4} f_{n-1} (z).
     \label{finalR}
\eeq
Solving the above quadratic equation yields
\beq
     R(z) = \frac{1}{2} \left( W_\mathrm{tree}'(z) - \sqrt{ 
     W_\mathrm{tree}'(z)^2 +  f_{n-1} (z)}\right),
\eeq
where the sign has been chosen in order to get the right $1/z$ behavior
at infinity.
The above solution gives us an expression of $R(z)$ (and thus, expanding, 
of all the relevant v.e.v.s) in terms of the couplings $g_k$ and the
coefficients $f_k$ of the polynomial.

In order to have a useful solution for $R(z)$, namely one which is a
function of $g_k$ and the $S_i$, we only have to use the relations
(\ref{sidef}) to get expressions of the $S_i$ in terms of the $f_k$
and $g_k$.
By inverting these relations, we finally obtain an expression $R(z)$
in terms of $g_k$ and $S_i$
that can be used to integrate for $\mathcal{F}(S_i,g_k)$.
Fortunately, in this case we do not need to perform this computation 
explicitly because the implicit expression (\ref{sidef}) is all we need 
to show the correspondence with the matrix model.

We now pause to mention what should be done in order to perform a 
more complete
treatment of this theory. One defines two more generating functions:
\beq
     w_\alpha(z) = \frac{1}{4\pi} \left\langle \tr \frac{W_\alpha}{
     z-\Phi} \right\rangle, \qquad T(z) =  \left\langle \tr \frac{1}{
     z-\Phi} \right\rangle,
\eeq
which can be seen as higher components of $R(z)$ in a $\psi$-superfield.
Anomaly equations for them can also be written, which can be solved linearly.
New polynomials with undetermined coefficients like $f_{n-1}(z)$
arise, but again these coefficients can be solved for the physical
quantities:
\beq     
     w^i_\alpha = \frac{1}{2\pi i} \oint_{C_i} \mathrm{d}z\; w_\alpha(z),
     \qquad
     N_i = \frac{1}{2\pi i} \oint_{C_i} \mathrm{d}z\; T(z).
\eeq
In principle, it is the expansion of $T(z)$ that is needed to obtain
$W_\mathrm{eff}$, but as we have seen, the $\psi$ translational
symmetry makes it possible to
consider only the resolvent $R(z)$ to fully solve the problem. This
is no longer true for some theories with a different gauge group and/or
matter content, for which one cannot write the fermionic $\psi$ symmetry.
One can still write generalized Konishi anomaly relations that can
be solved, but in this case one has eventually to get the expression of 
the analog of $T(z)$ in order to obtain $W_\mathrm{eff}$.

What we are going to show in the following is that there is an even
more direct way to compute $\mathcal{F}(S_i,g_k)$
through a bosonic matrix model computation. To prove this, we
are going to show that exactly the same equations for $R(z)$ and  
$\mathcal{F}$ arise within the matrix model.

Let us define the free energy $\hat{\mathcal{F}}$ 
of the matrix model whose potential is given by $W_\mathrm{tree}$:
\beq
     e^{-\frac{\hat N^2}{ \hat S^2}\hat{\mathcal{F}}} =
     \int\!\mathrm{d}M\; e^{-\frac{\hat N}{\hat S}W_\mathrm{tree}(M)},
     \label{fmm}
\eeq
where $M$ are hermitian $\hat N\times \hat N$ matrices and we take $\hat N$
to be large. Note that $\hat N$ is {\em not} related to the $N$ of 
the gauge theory. For the moment we may also think $\hat S$ 
as unrelated to $S$ although a relation between these last quantities 
will arise in the following.
 
First of all we derive differential equations for $\hat{\mathcal{F}}$
by taking derivatives with respect to the couplings $g_k$ on both sides:
\beq
     \pd{\hat{\mathcal{F}}}{g_k} = 
     \frac{\hat S}{\hat N} \left\langle \frac{1}{ k+1} \tr M^{k+1} 
     \right\rangle. \label{matrixpde}
\eeq
In the large $\hat N$ limit, both sides of the above equation will
have a genus expansion, with the planar contribution being the leading one.
(With this normalization they both tend to a finite limit.)

As in (\ref{gaugeresol}), we introduce a generating function for the v.e.v.s
on the RHS of (\ref{matrixpde}). This is called the matrix model resolvent:
\beq
     \hat R (z) = \frac{\hat S}{ \hat N} \left\langle \tr \frac{1}{
     z - M } \right\rangle. \label{mresol}
\eeq
We are now left with the task of deriving an equation for the resolvent.
Similarly to the Konishi anomaly, we can obtain such an equation
from the Ward identities implied by reparameterization invariance of
the matrix model free energy. Namely, the integral in (\ref{fmm})
should be invariant under a simultaneous reparameterization of the
matrix $M\rightarrow M + \delta M$ both in the measure and in the 
potential. Let us take the following variation:
\beq
     \delta M = \eta \frac{1}{z - M}, \qquad \eta \ll 1. \label{mvariat}
\eeq
The measure changes as:
\beq
     d M \rightarrow d M \left(1+ \pd{\delta
         M_i^j}{M_i^j}\right). 
\eeq
Now, using the same trick as in (\ref{trick}), we have that:
\beq
     \pd{}{M_i^j}\left( \frac{1}{ z - M}\right)_i^j
     = \left(\tr  \frac{1}{z - M} \right)^2.
\eeq
On the other hand, the variation (\ref{mvariat}) acts on the potential
by bringing down its derivative, so that the Ward identity reads:
\beq
     \frac{\hat S}{\hat N} \left\langle \left(\tr \frac {1}{z - M} \right)^2
     \right\rangle = \left\langle \tr \frac{W_\mathrm{tree}'(M)} {z-M}
     \right\rangle. \label{ward} 
\eeq
This expression can be turned into an equation for $\hat R(z)$ again
using factorization, but this time the reason we can factorize is 
very different from the one in the gauge theory context. Indeed, we can
write:
\beq
     \frac{\hat S^2}{\hat N^2} \left\langle \left(\tr  \frac{1}{z - M} 
     \right)^2 \right\rangle \rightarrow \frac{\hat S^2}{\hat N^2} 
     \left( \left\langle\tr  \frac{1}{z - M}\right\rangle \right)^2
     = \hat R(z)^2,
     \label{mfacto}
\eeq
only in the planar limit, that is, keeping only the leading order
in $\hat N$ for both expressions. 

\begin{figure}[htbp]
  \begin{center}
    \setlength{\unitlength}{3pt}
    {\renewcommand{\dashlinestretch}{30}
      \begin{picture}(100,24)(0,0)
        \put(10,16.5){\arc{11.5}{1.9656}{7.4592}}
        \put(10,6){\arc{11.5}{5.1072}{10.6008}}
        \put(10,6){\ellipse{7.50}{7.50}}
        \put(10,16.5){\ellipse{7.50}{7.50}}
        \thicklines
        \spline(0,11.25)(.5,12)(1,11.25)(1.5,10.5)(2,11.25)(2.5,12)(3,11.25)(3.5,10.5)(4,11.25)(4.5,12)(5,11.25)(5.5,10.5)(6,11.25)(6.5,12)(7,11.25)(7.5,10.5)(8,11.25)(8.5,12)(9,11.25)(9.5,10.5)(10,11.25)
        \put(10,11.25){\circle*{1}}
        \thinlines
        \put(30,6){\arc{11.5}{5.1072}{10.6008}}
        \put(30,16.5){\arc{11.5}{1.9656}{7.4592}}
        \put(30,6){\ellipse{7.50}{7.50}}
        \put(30,16.5){\ellipse{7.50}{7.50}}
        \thicklines
        \spline(30.0,11.25)(30.5,12)(31,11.25)(31.5,10.5)(32,11.25)(32.5,12)(33,11.25)(33.5,10.5)(34,11.25)(34.5,12)(35,11.25)(35.5,10.5)(36,11.25)(36.5,12)(37,11.25)(37.5,10.5)(38,11.25)(38.5,12)(39,11.25)(39.5,10.5)(40,11.25)
        \put(30,11.25){\circle*{1}}
        \thinlines
        \put(70,11.25){\arc{18.5}{3.3229}{6.1019}}
        \put(70,11.25){\arc{18.5}{0.1813}{2.9603}}
        
        \put(70,0){\arc{31.8}{4.09690}{5.32787}}
        \put(70,22.5){\arc{31.8}{0.95531}{2.18627}}
        
        \put(70,0){\arc{27.8}{4.09690}{5.32787}}
        \put(70,22.5){\arc{27.8}{0.95531}{2.18627}}
        
        \put(70,11.25){\ellipse{22.50}{22.50}}
        \thicklines
        \spline(50,11.25)(50.5,12)(51,11.25)(51.5,10.5)(52,11.25)(52.5,12)(53,11.25)(53.5,10.5)(54,11.25)(54.5,12)(55,11.25)(55.5,10.5)(56,11.25)(56.5,12)(57,11.25)(57.5,10.5)(58,11.25)(58.5,12)(59,11.25)(59.5,10.5)(60,11.25)
        \put(60,11.25){\circle*{1}}
        \spline(80,11.25)(80.5,12)(81,11.25)(81.5,10.5)(82,11.25)(82.5,12)(83,11.25)(83.5,10.5)(84,11.25)(84.5,12)(85,11.25)(85.5,10.5)(86,11.25)(86.5,12)(87,11.25)(87.5,10.5)(88,11.25)(88.5,12)(89,11.25)(89.5,10.5)(90,11.25)
        \put(80,11.25){\circle*{1}}
    \end{picture} 
    }
  \end{center}
  \caption{Factorization - the disconnected two-point function of 
the operator $ (1/\hat N)\tr M^4$ is finite while the connected one
is suppressed by a factor $ 1/{\hat N}^2$.}
  \label{fig:fac2}
\end{figure}
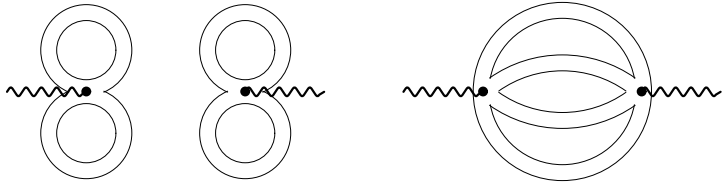

This can be understood as follows. Consider the correlation of two 
operators $(1/\hat N) \tr M^4$ as depicted in Fig.~\ref{fig:fac2}.
The diagrams contributing to the correlator have the same form of vacuum 
diagrams, only this time at the point where the operator is inserted we 
have a factor $(1/\hat N)$ instead of $\hat N$ as there would be for an
ordinary vertex. Thus we see that connected diagrams with one insertion
are finite while those with two insertions are suppressed by $(1/\hat N^2)$
leading to the abovementioned factorization property.
The mechanism leading to
factorization here is thus totally different from the gauge theory context,
where supersymmetry and the properties of chiral operators are crucially used.

As for the RHS of (\ref{ward}), we apply the same addition and subtraction
of $W_\mathrm{tree}'(z)$ 
to write the final equation for the resolvent, known as the 
loop equation, as:
\beq
     \hat R(z)^2 = W_\mathrm{tree}'(z) \hat R(z) +\frac{1}{4} \hat f_{n-1}(z),
     \label{mmquad}
\eeq
with:
\beq
     \hat f_{n-1}(z) = 4 \frac{\hat S}{ \hat N} \left\langle \tr \frac{
     W_\mathrm{tree}'(M)-W_\mathrm{tree}'(z)} {z - M } \right\rangle,
\eeq
also an unknown polynomial of degree $n-1$.
We thus see that the loop equation for $\hat R(z)$ is \emph{the same} as
eq.~(\ref{finalR}) for $R(z)$ and thus that
$\hat R(z)$ has the same analytic structure as $R(z)$
in the gauge theory, namely it will display a total of $n$ cuts. 

As before, the coefficients of the function 
$\hat f_{n-1}(z)$ can be reexpressed in terms of the following
variables:
\beq
     \hat S_i = \frac{1}{ 2\pi i} \oint_{C_i} \!\mathrm{d}z\; \hat R(z),
\eeq
where the contour $C_i$ goes around the $i$-th cut\footnote{Note 
that in the hermitian matrix model context, 
the cuts are on the real axis and thus the $\hat S_i$ are real, unlike
the gauge theory context where both can be generally complex. The relation
between the quantities on the two sides will be done by analytic continuation.
}. Using the definition
(\ref{mresol}) of the resolvent, and applying the same reasoning as in
(\ref{projector}), we obtain:
\beq
     \hat S_i = \frac{1}{2\pi i} \oint_{C_i}\!\mathrm{d}z\; \frac{\hat
       S }{\hat N} 
     \left\langle \tr \frac{1}{z - M } \right\rangle = 
     \frac{\hat S}{\hat N} \hat N_i .
\eeq
The $\hat S_i$ are proportional to the fraction of eigenvalues 
populating the $i$-th cut\footnote{This last relation has no analogue in
the gauge theory.}.

We are now finally ready to state the correspondence between the matrix
model and the gauge theory computation. We already have the same 
potential $W_\mathrm{tree}(z)$ entering the equation for the resolvent.
We now identify also the two polynomials of degree $n-1$, that is
we impose $\hat f_{n-1}(z) =  f_{n-1}(z)$ so that the coefficients
entering the loop equations are all the same. We can then trivially
identify the two resolvents: $\hat R(z) = R(z)$.
This in turn implies $\hat S_i  = S_i$. 
Lastly, expanding the resolvents
we also obtain identical equations for the free energy on one side and
for the function $\mathcal{F}$ on the other, so that we identify
$\hat{\mathcal{F}} (S_i, g_k) = \mathcal{F}(S_i, g_k)$.

This is thus the core of the correspondence between the computation
of the effective superpotential on the gauge theory side, and the
free energy of the matrix model. Using a mixed notation, and recalling
the relation between $\mathcal{F}$ and $W_{DV}$, we have
that:
\beq
     W_{DV} = \sum_i N_i \pd{\hat{\mathcal{F}}}{S_i}
     \label{fullsplendor} 
\eeq
where we concentrate on the piece of $W_{DV}$ relevant purely
to the glueball superfields.

There are some important remarks one has to make on (\ref{fullsplendor}).
The first is to state once again that the $N_i$ are not related
to the $\hat N_i$, the former being finite and the latter being taken
to infinity in order to extract the planar contribution to the free energy
(and to be able to use the factorization in (\ref{mfacto})).

The second remark is that once the correspondence 
is derived as before, one can turn to any other alternative derivation
of $\hat{\mathcal{F}}$ on the matrix model side in order to compute it,
as for instance presented in Appendix~B.
Then, when the result is stated in terms of the filling fractions
$\hat S_i$ and the couplings $g_k$, the gauge theory effective superpotential
is obtained using (\ref{fullsplendor}).

The last comment on (\ref{fullsplendor}) is that one could take it at face
value and apply it even to any part of the matrix model free energy
$\hat{\mathcal{F}}$ which depends only on the filling fractions $\hat S_i$
and not on the couplings $g_k$. That would imply the possibility of deriving
on the gauge theory side the effective superpotential also for the pure
gauge low-energy dynamics. Keeping in mind that all the arguments of
this section do {\em not} apply to this case, one can still perform
a simple computation and find, very similarly to (\ref{volsun}),
that the Veneziano-Yankielowicz superpotential is reproduced.
This fact still needs a better understanding in gauge theoretic terms.

\section*{Acknowledgments}

We thank Vanicson L. Campos for collaborating with us on work
related to this review. One of us (G.F.) wishes to thank the
organizers and the participants of the schools at Nordita and 
Parma in which part of the material in this review was tried out. 
R.A. would like to thank the ITP in G\"oteborg for support and warm
hospitality during most of the work leading up to this review.
We also thank Matteo
Bertolini, Massimo Bianchi, Paolo di Vecchia,
Alessandra Feo, Francesco Fucito, Kumar Narain, Kostas Sfetsos and 
Daniela Zanon for discussions.

This work is partly supported by EU contracts HPRN-CT-2000-00122 
and HPRN-CT-00131, by the ``Actions de Recherche
Concert{\'e}es" of the ``Direction de la Recherche Scientifique -
Communaut{\'e} Fran{\c c}aise de Belgique", by a ``P\^ole
d'Attraction Interuniversitaire" (Belgium) and by IISN-Belgium
(convention 4.4505.86). R.A. is a Postdoctoral Researcher of
the Fonds National de la Recherche Scientifique (Belgium).
The research of G.F. is supported by the Swedish Research Council
(Vetenskapsr\aa det) contract 622-2003-1124.

\appendix

\section{Superspace notation}

In this appendix we collect the notation and conventions used throughout 
this paper. 

The Minkowski metric is $\eta^{\mu\nu}=\mathrm{diag}(+1,-1,-1,-1)$.

The gamma matrices $\gamma^\mu$ satisfy the Clifford algebra 
$\{\gamma^\mu, \gamma^\nu\}=2\eta^{\mu\nu}$, and we always use the
Weyl representation
\beq
   \gamma^\mu=
   \begin{pmatrix}
     0 & \sigma^\mu \\
     \bar\sigma^\mu & 0 
   \end{pmatrix},
\eeq
where 
\beqs
\sigma^0= \bar\sigma^0= \begin{pmatrix}
     -1 & 0\\
     0 & -1 
   \end{pmatrix}&,&\quad
\sigma^1= -\bar\sigma^1= \begin{pmatrix}
     0 & 1\\
     1 & 0 
   \end{pmatrix}, \nonumber \\
\sigma^2= -\bar\sigma^2= \begin{pmatrix}
     0 & -i\\
     i & 0 
   \end{pmatrix}&,&\quad
\sigma^3= -\bar\sigma^3= \begin{pmatrix}
     1 & 0\\
     0 & -1 
   \end{pmatrix}
\eeqs
and the indices are read, row by column, as $(\sigma^\mu)_{\alpha\dot\alpha}$
and $(\bar\sigma^\mu)^{\dot\alpha\alpha}$. For instance, 
$(\sigma^2)_{1\dot 2} = -i$.

The generators of the Lorentz transformations are
\beqs
    \sigma^{\mu\nu\beta}_{\phantom{\mu\nu}\alpha} & = & 
    \frac{1}{4}\left(\sigma^\mu_{\alpha\dot\alpha}
    {\bar\sigma}^{\nu\dot\alpha\beta} -
    \sigma^\nu_{\alpha\dot\alpha}
    {\bar\sigma}^{\mu\dot\alpha\beta} \right) \\
    \bar\sigma^{\mu\nu\dot\alpha}_{\phantom{\mu\nu}\dot\beta} & = & 
    \frac{1}{4}\left({\bar\sigma}^{\mu\dot\alpha\alpha}
    {\sigma}^\nu_{\alpha\dot\beta} -
    {\bar\sigma}^{\nu\dot\alpha\alpha}
    {\sigma}^\mu_{\alpha\dot\beta} \right).
\eeqs

Dotted and undotted indices describe two component 
Weyl spinors of opposite chirality and
we raise them and lower them using the totally antisymmetric tensors
$\epsilon_{21}= \epsilon^{12}=1$ and 
$\epsilon_{\dot 2 \dot1}= \epsilon^{\dot 1 \dot 2}=1$:
\beq
     \psi^\alpha=\epsilon^{\alpha\beta}\psi_\beta, \quad
     \psi_\alpha=\epsilon_{\alpha\beta}\psi^\beta, \quad
     \bar\psi^{\dot\alpha}=\epsilon^{\dot\alpha\dot\beta}
           \bar\psi_{\dot\beta}, \quad
     \bar\psi_{\dot\alpha}=\epsilon_{\dot\alpha\dot\beta}
           \bar\psi^{\dot\beta}.  \label{raiselower}
\eeq

Complex conjugation changes chirality: 
$(\psi_\alpha)^* = \bar\psi_{\dot\alpha}$ and when two pairs of
indices are not explicitly written, the following summation
conventions are understood:
\beq
    \psi\chi = \psi^\alpha\chi_\alpha\quad \hbox{and} \quad
    \bar\psi\bar\chi = \bar\psi_{\dot\alpha} \bar\chi^{\dot\alpha}.
\eeq
So far everything is as in \cite{wb} except for the signature of the
metric. We now define the \emph{square} of a spinor as
\beq
    \psi^2 = \frac{1}{2} \psi^\alpha\psi_\alpha \quad \hbox{and} \quad
    \bar\psi^2 = \frac{1}{2} \bar\psi_{\dot\alpha}
    \bar\psi^{\dot\alpha}.  \label{fermisq}
\eeq
(This is done as to eliminate many of the powers of two appearing
explicitly otherwise.)

Superspace has coordinates $x^\mu, \theta^\alpha,
\bar\theta^{\dot\alpha}$ and the fermionic derivatives are defined as
\beq
     \partial_\alpha = \frac{\partial}{\partial\theta^\alpha}
     \quad \hbox{and} \quad
     \bar\partial_{\dot\alpha} = 
         \frac{\partial}{\partial\bar\theta^{\dot\alpha}},
\eeq
so that $\partial_\alpha \theta^\beta = \delta_\alpha^\beta$ and 
$\bar\partial_{\dot\alpha} \bar\theta^{\dot\beta} = 
\delta_{\dot\alpha}^{\dot\beta}$. 

The supercharges and superspace derivatives are defined as follows: 
\beqs
     Q_\alpha = \partial_\alpha - 
     \frac{i}{2}\sigma^\mu_{\alpha\dot\alpha}
     \bar\theta^{\dot\alpha}\partial_\mu \quad &\hbox{and}& \quad
      \bar Q_{\dot\alpha} = \bar\partial_{\dot\alpha} - 
     \frac{i}{2}\theta^\alpha
     \sigma^\mu_{\alpha\dot\alpha}\partial_\mu \\
     D_\alpha = \partial_\alpha + 
     \frac{i}{2}\sigma^\mu_{\alpha\dot\alpha}
     \bar\theta^{\dot\alpha}\partial_\mu \quad &\hbox{and}& \quad
      \bar D_{\dot\alpha} = \bar\partial_{\dot\alpha} + 
     \frac{i}{2}\theta^\alpha
     \sigma^\mu_{\alpha\dot\alpha}\partial_\mu. \nonumber
\eeqs
The only two non-vanishing anticommutators are
\beq
      \{ Q_\alpha, \bar Q_{\dot\beta} \} =
     -\{ D_\alpha, \bar D_{\dot\beta} \} =
     -i\sigma^\mu_{\alpha\dot\beta} \partial_\mu =
     - \sigma^\mu_{\alpha\dot\beta} P_\mu.
\eeq

Indices on the derivatives $D_\alpha$ and $\bar D_{\dot\beta}$ 
are raised and lowered using the same
conventions as in (\ref{raiselower}) but we 
define the square of the fermionic derivatives with an extra minus
sign as compared to (\ref{fermisq}):
\beq
    D^2 = -\frac{1}{2} D^\alpha D_\alpha 
   \quad \hbox{and} \quad
    \bar D^2 = -\frac{1}{2} \bar D_{\dot\alpha}
    \bar D^{\dot\alpha}, 
\eeq
so that: $ D^2 \theta^2 =  \bar D^2  \bar\theta^2 = + 1$.
The fermionic integrals are just another notation for the derivative:
$ \int \;d^2\theta \;\theta^2 =  \int\; d^2\bar\theta \; \bar\theta^2 = 1$.
It is straightforward to check that these integrals are 
invariant under the translations $\theta \to \theta + \xi$ and
$\bar\theta \to \bar\theta + \bar\xi$.

All the identities involving the $\sigma$ matrices described in the 
Appendices A and B of Wess and Bagger \cite{wb} are still valid by letting
$\eta_{\mu\nu} \to - \eta_{\mu\nu}$ and recalling that 
$\theta\theta = 2 \theta^2$ and $\bar\theta\bar\theta = 2
\bar\theta^2$ in our notation.

\section{One cut solution to the cubic matrix model}

In this appendix we show how to compute the leading (planar) 
contribution to the free energy $\ff$ of the one cut cubic matrix model
\cite{bipz}, thus completing the computation of Section~4.  
To be slightly more general, we present the computation for a generic 
potential  $W$ and only at the end we substitute 
$W(z) = \frac{m}{2} z^2 + \frac{g}{3} z^3$.

We consider the following matrix integral for a generic potential $W(M)$
of a random hermitian matrix $M$ (here we write the trace explicitly)
\beq
    e^{-\frac{\hat N^2}{\sh^2}\ff + \dots} = \measure{\hat N^2}{M}
    e^{-\frac{\hat N}{\sh} \tr W(M)}\,,
    \label{eqw}
\eeq
where $\ff$ is the planar free energy, the dots denote non-planar 
(sub-leading) contributions and with respect to Section~4 we 
have already substituted for $\epsilon = \sh / \hat N$.

Since $M$ is hermitian, there exists a unitary matrix $\Omega$ such that
\beq
    M=\Omega^\dagger \Lambda \Omega\,,
\eeq
where $\Lambda = diag \left(\lambda_1, \ldots , \lambda_{\hat N} 
\right)$ is a diagonal matrix with the real eigenvalues $\lambda_i$ of
$M$ as entries.  
The measure in the new variables becomes $\mathrm{d}^{\hat N^2}\! M =
\mathrm{d}^{\hat N}\!\Lambda\, \mathrm{d}^{\hat N^2-\hat N}\!\Omega\,
\Delta(\lambda)^2$  
with $\Delta(\lambda)^2$ being the Jacobian of the transformation.
  
This Jacobian can be determined in the following way.
The metric on the matrix space can be written as 
\beqs
    \mathrm{d} s^2&=&\tr\,{\mathrm{d} M \mathrm{d} M} \nn \\ 
    &=&\tr(\mathrm{d}\Omega^\dagger\Lambda\Omega+\Omega^\dagger
    \mathrm{d}\Lambda\Omega+\Omega^\dagger\Lambda 
    \mathrm{d}\Omega)^2\nn\\  
    &=&\tr(\mathrm{d}\Lambda+[\Lambda,\mathrm{d}\Omega\Omega^\dagger])^2\nn\\
    &=&\tr(\mathrm{d}\Lambda^2+[\Lambda,\mathrm{d}\Omega\Omega^\dagger]^2)\nn\\
    &=&\mathrm{d}\lambda_1^2 + \ldots + \mathrm{d}\lambda_{\hat N}^2 +
    \sum_{i\neq j}(\lambda_i-\lambda_j)^2 
    |\mathrm{d}\Omega_{i k}\Omega_{k j}^\dagger|^2. 
\eeqs
The Jacobian is just the square root of the determinant of this metric
\beq
    \Delta(\lambda)=\sqrt{\det\,G}=\displaystyle \prod_{i<j}
    (\lambda_i-\lambda_j)^2.
\eeq
This result can be inserted into eq. \refe{eqw}
\beq
    \measure{\hat N^2}{M}  e^{-\frac{\hat N}{\sh}\tr W(M)} =
    \measure{\hat N^2-\hat N}{\Omega} \mathrm{d}^{\hat N}\lambda\;
    \prod_{i<j} (\lambda_i-\lambda_j)^2 e^{-\frac{\hat N}{\sh} 
    \sum_k W(\lambda_k)}. 
\eeq
The integral over the unitary matrices gives an overall constant 
that we absorb into the measure leaving
\beq
    \measure{\hat N^2}{M}  e^{-\frac{\hat N}{\sh}\tr W(M)} =
    \measure{\hat N}{\lambda}  e^{-\frac{\hat N}{\sh} \sum_k
      W(\lambda_k) + \sum_{i\neq j} \log|\lambda_i-\lambda_j|}. 
\eeq
Solving this integral is impossible for finite $\hat N$. 
Instead, we use the steepest descent method for the evaluation, which 
corresponds to taking the large $\hat N$ limit, thus singling out the
planar diagrams. 

To leading order in $\hat N$, the free energy is given by
\beq
    - \frac{\hat N^2}{\sh^2}\ff = 
    - \frac{\hat N}{\sh} \sum_k W(\hat\lambda_k)
    + \sum_{i \neq j}
    \log|\hat{\lambda}_i-\hat{\lambda}_j |,  
\eeq
where $\hat\lambda_k$ are the $\hat N$ solutions to the extremization problem
given by the simultaneous solution of the following $\hat N$ equations, 
labeled by $k=1, \dots \hat N$.
\beq
    0=-\frac{\hat N}{\sh} W'(\lambda_k) + 2 \sum_{k\neq j}
    \frac{1}{\lambda_k-\lambda_j}.  
\eeq

To solve this set of equations for large $\hat N$ one goes 
to a continuous description of the eigenvalues by defining
\beqs
   \frac{i}{\hat N}&\to& x\in[0,1]\nn \\
   \sum_i&\to& \hat N\int_0^1\!\! \mathrm{d}x\\
   \lambda_i&\to&\lambda(x) \nn
\eeqs
With these identifications the free energy and the extremum equation
become
\beqs
    \ff&=& \sh \int_0^1\!\!\mathrm{d}x\, W(\lambda(x))
      - \sh^2 \int_0^1 \!\!\mathrm{d}x\int_0^1 
      \!\!\mathrm{d}y\,\log|\lambda(x)-\lambda(y)|
    \label{eqex1} \\
    0&=&- W'(\lambda(x)) +2 \sh\int_0^1\!\!\mathrm{d}y\,
    \frac{1}{\lambda(x)-\lambda(y)}\,, 
    \label{eqex2}
\eeqs
where the integrals are to be thought of as principal values. 
Note that all dependence on $\hat N$ has disappeared and that we traded 
a set of $\hat N$ algebraic equations for one integral equation. 

To proceed, introduce the density of eigenvalues $\rho(\lambda)$ as
\beq
    \sh \mathrm{d}x=\mathrm{d}\lambda\;\rho(\lambda)
\eeq
which is non-negative,  has support within an interval $[a,b]$ 
(to be determined)\footnote{We focus here on the so called one cut
solution, i.e. the one with all eigenvalues grouped around an extremum,
corresponding to an unbroken gauge group.} and is normalized such that
\beq
    \int_a^b\mathrm{d}\lambda\,\rho(\lambda)\ =\ \sh. \label{nooorm}
\eeq
In terms of the eigenvalue density eqs. \refe{eqex1} and \refe{eqex2}
read
\beqs
     \ff&=& \int_a^b\!\!\mathrm{d}\lambda\, \rho(\lambda) W(\lambda)
     - \int_a^b\!\!\mathrm{d}\lambda' \int_a^b\!\!\mathrm{d}\lambda\,
     \rho(\lambda) \rho(\lambda') \log|\lambda-\lambda'|
     \label{eqf2} \\
     0&=&-  W'(\lambda)+2 \int_a^b\!\! \mathrm{d}\lambda'\,
     \frac{\rho(\lambda')}{\lambda-\lambda'}\,. 
     \label{eqexx2}
\eeqs
Integrating \refe{eqexx2} in the two intervals $[a,\lambda]$
and $[\lambda,b]$ and combining the results yields 
\beqs
     && \int_a^b\!\!\mathrm{d}\lambda'\,\rho(\lambda')
     \log|\lambda-\lambda'| =\frac{1}{4}\Big(2\ W(\lambda)-W(b)-W(a)\Big) 
     \nonumber \\
     && +\frac{1}{2} \int_a^b\!\!\mathrm{d} \lambda'\, \rho(\lambda')
       \log(b-\lambda')(\lambda'-a)
\eeqs
Eq. \refe{eqf2} then becomes, using the normalization of the eigenvalue
density (\ref{nooorm})
\beqs
     \ff &=& \frac{1}{2} \int_a^b\!\!\mathrm{d}\lambda\, \rho(\lambda)\Big(
     W(\lambda) - \sh \log(b-\lambda)(\lambda-a)\Big) \nonumber \\ 
     &+& \frac{\sh}{4}\Big(W(b)+W(a)\Big). 
     \label{eqf3}
\eeqs

To find the eigenvalue density define the resolvent 
\beq
    R(z)=\int_{-\infty}^{\infty}\!\!\mathrm{d}\lambda\,
    \frac{\rho(\lambda)} {z-\lambda}\, , \label{reso}
\eeq
which is an analytic function of $z$. 
The singularity of the resolvent is a cut along the real interval $[a,b]$. 
Across the cut the resolvent has the following behavior
\beqs
    R(\lambda+i\epsilon) + R(\lambda-i\epsilon) &=& 2\int_a^b
    \!\!\mathrm{d} \lambda'\, \frac{\rho(\lambda')} {\lambda-\lambda'}
    = W'(\lambda)\label{eqrho1} \\   
    R(\lambda+i\epsilon) - R(\lambda-i\epsilon) &=& \oint \!\mathrm{d}
    \lambda'\, \frac{\rho(\lambda')} {\lambda-\lambda'} 
    = -2 \pi i \rho(\lambda)\,.
    \label{eqrho2}
\eeqs
The integral in (\ref{eqrho1}) is a principal value and the one in
(\ref{eqrho2}) is a residue as can be easily seen by drawing the contours.
The idea now is to use (\ref{eqrho1}) to find the resolvent and then 
(\ref{eqrho2}) to extract the eigenvalue density $\rho$.

One particular solution to eq.~(\ref{eqrho1}) is, of course, the regular 
solution $R_{reg}(z)=\frac{1}{2}W'(z)$, but this cannot be the full story 
because it does not have the right asymptotic behavior at infinity. To
$R_{reg}(z)$ we must add a generic solution to the homogeneous equation
\beq
     R_{sing}(\lambda+i\epsilon) + R_{sing}(\lambda-i\epsilon) = 0, 
\eeq
necessarily singular along the cut and 
chosen so that the full solution has the 
correct large $z$ behavior following from (\ref{reso}):
\beq
    R(z) = R_{sing}(z) + R_{reg}(z) \sim \frac{\sh}{z} + {\mathcal O}
    \left(\frac{1}{z^2}\right), \qquad z\to\infty\,. 
    \label{eqpz}
\eeq
Across the cut we need a sign change which is implemented by
\beq
    R_{sing}(z) = P(z) \sqrt{(z-a)(z-b)}\,,
\eeq
where $P(z)$ is a regular polynomial with degree two less than 
the degree of $W(z)$
in order to cancel the various powers of $z$.
The coefficients in $P(z)$ are given implicitly by the condition \refe{eqpz}. 

The sought eigenvalue density $\rho(\lambda)$ is then, using eq. \refe{eqrho2}
\beqs
    \rho(\lambda)=\frac{1}{\pi i}R_{sing}(\lambda),
\eeqs
where the sign of the square root is chosen to give $\rho$ non negative.
Having determined the eigenvalue density, the free energy can be
calculated by evaluating \refe{eqf3}.

Up to this stage the potential $W(M)$ is generic. 
To illustrate the technique and to give the full solution to the problem of 
Section~4, we now specialize to a cubic matrix model.
Notice that although the integral with a cubic potential is not convergent,
each term in the perturbative expansion is, and this is all we need.
Let us thus take, as in Section~4
\beq
    W(M)=\frac{m}{2} M^2 + \frac{g}{3} M^3.
\eeq

Requiring that there are no ${\cal O}(z^2)$ and ${\cal O}(z)$
terms in $R(z)$ in the large $z$ limit, we get:
\beq
     P(z) = - \frac{1}{4}\left(2 m +  g (a + b) + 2 g z\right).
\eeq
Furthermore, by canceling the constant term ${\cal O}(z^0)$ and
requiring the pole to have residue $\sh$, we obtain conditions on the
end points $a$ and $b$ of the interval where the eigenvalue density has
support 
\beqs
    0 &=&2 g (a+b)^2 + g (a-b)^2 + 4 m (a+b)\label{eqconstsu} \\ 
    16 \sh &=&\left(a - b\right)^2\left(g\left(a + b\right) + m\right).
    \label{eqconst}
\eeqs
Thus, from eq.~\refe{eqrho2}:
\beqs
    \rho(\lambda)= \frac{1}{4\pi} \left( 2 g \lambda + 2 m + g
    \left(a+b\right) \right) \sqrt{(b-\lambda)(\lambda-a)}.
\eeqs

The free energy is computed by using the eigenvalue density
and evaluating eq. \refe{eqf3} 
\beqs
    \ff= \frac{1}{8\pi} \int_a^b  \!\!\mathrm{d}\lambda\, \left
      [ \left( 2 g \lambda + 2 m + g (a+b) \right) \sqrt{(b-\lambda)
        (\lambda-a)} \right. \cr 
    \left. \Big( \frac{m}{2} \lambda^2 + \frac{g}{3} \lambda^3 -\sh
      \log|(b-\lambda)(\lambda-a)| \Big) \right] \cr
    + \frac{\sh}{4} 
    \Big(\frac{m}{2} (a^2+b^2) + \frac{g}{3} (a^3 +b^3)\Big)
      \qquad\qquad\qquad\qquad\:\: 
    \label{eqfre}
\eeqs
It is convenient to introduce a parameter 
\beq
    \sigma \equiv  \frac{g}{2m}\ (a+b).
\eeq 
Then the constraints \refe{eqconstsu} and \refe{eqconst} imply
\beq
    2 \frac{g^2}{m^3} \sh + \sigma(1+\sigma)(1+2\sigma) = 0,
    \label{eqconst2}
\eeq
where we chose the branch $\sigma = 0$ for $g=0$.
Evaluating the integral \refe{eqfre} keeping only terms that depend on
$g$ (through $\sigma$) yields\footnote{A constant term arising at $g=0$
can always be absorbed into the measure as we have done many times before.}:
\beq
    \ff=-\frac{\sh^2}{3} \frac{\sigma (2+6\sigma+3\sigma^2)}
    {(1+\sigma) (1+2\sigma)^2}+\frac{\sh^2}{2}\log(1+2\sigma).
    \label{eqf4}
\eeq
The constraint \refe{eqconst2} can be solved in a perturbation
expansion in the coupling constant $g$ and substituting into $\ff$ we get: 
\beq
    \ff = - \frac{2g^2\sh^3}{3m^3} - \frac{8g^4\sh^4}{3m^6}
    - \frac{56g^6\sh^5}{3m^9} - \frac{512g^8\sh^6}{3m^{12}}
    - \frac{9152g^{10}\sh^7}{5m^{15}} + {\cal O}(g^{12}).
\eeq

The superpotential for the gauge theory is then given by the 
Dijkgraaf-Vafa formula (\ref{wdv}) of Section~4.

\section{Bibliographical note}

As mentioned in the Introduction, the work of 
Dijkgraaf and Vafa~\cite{dv} has
generated many other interesting developments that unfortunately did
not find their place in our review. In this Appendix we make an
attempt to summarize them hoping that this will help the reader in
finding his/her own way through the literature. We apologize for any
omission or imprecision that might occur. 

The most important omission is, of course, a discussion of the
geometrical aspects and implications of the 
work of Dijkgraaf and Vafa. This is
particularly embarrassing because this was the original route through
which the equivalence with the matrix model was originally discovered!
This subject deserves its own review written by someone more qualified
than us to do it justice -- here we only refer the interested reader
to some of the related 
literature~\cite{geometry} that followed~\cite{dv}.

Within the realm of gauge theories, probably the most important 
omission is the discussion of theories
with extended SUSY. The work of~\cite{dv} has a natural extension to
theories with $\mathcal{N}=2$ and $\mathcal{N}=4$ SUSY. In the
$\mathcal{N}=2$ context~\cite{Neq2} many of the results of
Seiberg-Witten theory~\cite{sw} have been reproduced using matrix
models. Various
extensions to gauge groups other than $U(N)$ and the addition of
flavor hypermultiplets have also been discussed.
In the $\mathcal{N}=4$ case~\cite{Neq4}, much attention has been
given to extracting S-duality and to the various deformations allowed
by the presence of three chiral multiplets in the adjoint
representation. In particular, one can study either the
Leigh-Strassler deformation~\cite{Leigh:1995ep} or the relevant
deformations caused by adding mass terms to some of the chiral
multiplets leading to the so called $\mathcal{N}=1^*$ or $\mathcal{N}=2^*$
theories~\cite{Polchinski:2000uf}.

The new techniques to study SUSY gauge theories immediately call
for their application to the study of various non-perturbative aspects of
their dynamics. We greatly regret not having discussed the new
results emerging from these investigations such as a much better
understanding of the parameter space of these
theories~\cite{moduliandphases}, (see
also~\cite{moremoduliandphases}), the analysis~\cite{AD} of the 
Argyres-Douglas points~\cite{Argyres:1995jj} and that~\cite{Seibergduality}
of Seiberg duality~\cite{Seiberg:1994pq}.

Another notable omission is the discussion of gravitational
corrections ~\cite{gravity}. In the presence of a non-trivial
background for the gravitino, the analysis that led to the vanishing
of non-planar diagrams is no longer valid because the gravitino can
soak up additional zero modes. The non-planar contributions have also
been checked to agree with previously known results with both a
diagrammatic analysis and a computation based on the anomaly. This has
also led to an interesting deformation of the chiral ring where
``Grassmann'' operators are no longer nilpotent even at the classical
level.

Returning to gauge theory, the matrix model has been applied to
the study of theories based on other gauge groups such as
$SO(N)$ and $Sp(N)$ without~\cite{SOSp} or with~\cite{SOSpflavor} flavors
as well as $U(N)$ with symmetric or antisymmetric
representations. (The case of $U(N)$ with ``flavor'' 
matter in the fundamental has already briefly been discussed in the main
text~\cite{acfh1}, see also~\cite{otherUNflavor}.) The
case of $SU(N)$ with a number of flavors equal to the number of colors
is particularly interesting because it allows for baryonic
deformations~\cite{Seiberg:1994bz}. Various aspects of this model 
have been investigated in~\cite{baryon}. 

Moving on to more ``exotic'' cases, orbifolds and quivers yielding
semi-simple groups have been studied in~\cite{orbifoldquiver},
whereas multi-trace deformations have been discussed
in~\cite{multitrace}. 

Regrettably, the study of chiral gauge theories using these new
techniques has not received much attention. These theories are
interesting because they can~\cite{wittenindex} and they 
do~\cite{susybreak} break SUSY dynamically. Some examples are
discussed in~\cite{binor}, see also~\cite{morechiral}.

Still within the realm of applications to four dimensional gauge
theories, a justification of the linearity principle 
of~\cite{linearityprinciple} from the matrix model has been proposed
in~\cite{linearityjust} and an attempt at deriving 
the VY term from a diagrammatic
expansion has been made in~\cite{attemptVY}. A thorough study of some
of the more formal aspects of the matrix model has been done in the
series of papers~\cite{matri}. (See~\cite{gaugedmatrix} for a discussion
of the gauged model and the role played by the ghosts 
and~\cite{supermatrix} for some
studies on supersymmetric matrix models).
Finally, although the original results are strictly four dimensional
as should be clear from the discussion of Section 4, extensions of
some results to both lower~\cite{Deq3} and higher~\cite{Deq5} dimensions
have also appeared.

\end{document}